\documentclass[11pt,a4paper,twocolumn,UKenglish]{article}
\usepackage[utf8]{inputenc}

\usepackage{pifont}
\newcommand{\cmark}{\ding{51}}%
\newcommand{\xmark}{\ding{55}}%
\usepackage{hyperref}
\usepackage[useregional]{datetime2}
\usepackage{bookmark}
\bookmarksetup{open,numbered}

\usepackage{biblatex}
\usepackage{tabularray}
\usepackage{multirow}
\usepackage{xcolor}
\usepackage{graphicx}
\usepackage{diagbox} 
\usepackage[skins]{tcolorbox}

\usepackage{geometry}
\providecommand{\keywords}[1]
{
  \small	
  \textbf{\textit{Keywords---}} #1
}

\newtcolorbox{summarybox}[2][]{
    lower separated=false,%
    colback=white,%
    colframe=white!20!gray,fonttitle=\bfseries,%
    colbacktitle=white!10!gray,enhanced,%
    attach boxed title to top left={%
        xshift=0.5cm,%
        yshift=-2mm%
    },%
    title=#2,#1%
}

\title{\textbf{Literature study on Operational Data Analytics frameworks in large-scale computing infrastructures}} %focusing on the Energy and Performance aspects

\author{
  Shekhar Suman\\
  Vrije Universiteit Amsterdam \\ Universiteit van Amsterdam\\
  \href{mailto:s.suman@student.vu.nl}{s.suman@student.vu.nl} \\ \href{mailto:shekhar.suman@student.uva.nl}{shekhar.suman@student.uva.nl}
  \and
  Xiaoyu Chu\\
  Vrije Universiteit Amsterdam \\ @Large Research\\
  \href{mailto:x.chu@vu.nl}{x.chu@vu.nl}
  \and
  Alexandru Iosup\\
  Vrije Universiteit Amsterdam \\ @Large Research\\
  \href{mailto:a.iosup@vu.nl}{a.iosup@vu.nl} \\
  \href{mailto:a.iosup@atlarge-research.com}{a.iosup@atlarge-research.com}
}

\date{\today}

\begin{document}
\maketitle

\section*{Abstract}
\addcontentsline{toc}{section}{Abstract}
By 2025, there will be zettabytes of data generated every year. The size and complexity of modern large-scale computing infrastructures like High-Performance Computing (HPC) systems continue to evolve and become complex, leaving us wondering about their manageability and sustainability concerns. Because of this reason, those complex systems are provided with fine-grained monitoring and Operational Data Analytics (ODA) capabilities to optimise their efficiency. In this literature study, we list the fundamental pillars of the large-scale computing infrastructures which enable its ODA capabilities, and conduct a study of the popular ODA frameworks operating in various such environments (predominantly HPC). Based on that, we propose a more holistic ODA framework matching the various layers of a large-scale graph-processing distributed ecosystem proposed by Sherif Sak et al, that extends the ODA functionalities presented in an existing novel ODA framework proposed by Netti et al. We compare the holistic ODA framework proposed by us to some of the state-of-the-art frameworks that we study as part of this literature to highlight the novelty, which would hopefully draw more attention to perform extensive research in this field. As part of creating awareness, we highlight the significant operational efficiencies observed as a result of the implementation of the state-of-the-art ODA frameworks to make the study appear beneficial for the readers, and lastly, discuss the trending research work ongoing in this field.\\

\keywords{Operational Data Analytics (ODA), Monitoring, Energy efficiency, Performance, High-Performance Computing (HPC), Data centre}

\section{Introduction}
The complexity of modern large-scale computing infrastructures (e.g., HPC clusters, supercomputers, and cloud systems) has grown to an extreme level, at the verge of exascale, which introduces operational challenges. The complexity of these systems originates due to their adoption of heterogeneous architectures, ability to provide support to modern workflows and other applications, novel cooling  mechanisms, modern infrastructure facilities, and several other components \cite{DBLP:conf/icppw/BourassaJBCJVS19}. These systems typically have thousands to millions of CPU cores running up to a billion threads involved in complex computation. Moreover, the dynamic nature of workloads involved in these environments adds to the complexity. All these complexities result in understanding the state of such systems significantly challenging, before even talking about the optimal decisions involved in their operation. It is, unfortunately, common that these systems generate various kinds of malfunctions many times per day, resulting in process crashes ranging to even halting of operations on the compute nodes. Resilience has been studied for HPC executions on these future exascale systems, and several technical options have been established \cite{cappello2014toward}. But, this exascale resilience problem is far from solved, and thus the associated research problems pose a critical challenge for the HPC community. Also, as the architecture of HPC has evolved over the years, so has the energy consumption of these supercomputers \cite{villa2014scaling}. Due to the expanding use of HPC systems, energy consumption is also expected to expand, causing sustainability-related concerns (for example, CO2 emissions). This needs to be taken into consideration, otherwise, it will impact heavily on climate change. Safeguarding the interests of the community is a journey, which requires a multidisciplinary team of scientists, engineers and technologists to collaborate on this research. Several initiatives are being taken in this regard to meet the organisational energy efficiency goals, and the supercomputing community is promoting the same and publishing the greenest HPC supercomputers to create awareness\cite{green500}.\\~\\
\begin{figure}
\centering
\includegraphics[width=0.5\textwidth]{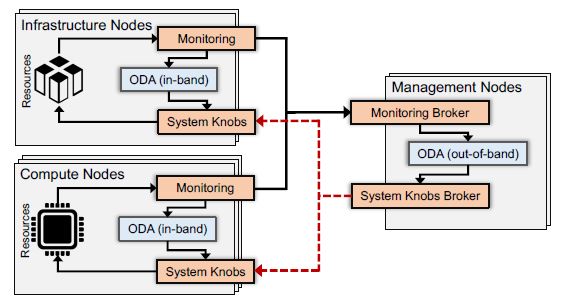}
\caption{Typical data flow involved in a distributed system using ODA framework. \cite{DBLP:conf/ipps/NettiTO021}}
\label{oda-data-flow}
% \vspace{-4mm}%Put here to reduce too much white space after your table 
\end{figure}
We set the preface of this literature by asking a pair of hypothetical questions: "If we were to extract efficiently and quickly a significant bit of information in a large-scale computing system, what kind of decisions would we be able to make? Would adding this intelligence to the system change the way how we design software operating on large-scale infrastructures like HPC or cloud?". Of course, some of the possible answers would preface an imagination questioning whether this capability has existed before, and if so, what were the limitations. This literature study aims to deeply dive into the journey to answer such questions. The first known practical usage of performance dashboards in IT operations began around 2000 when the field of business intelligence (BI) converged with performance management, resulting in the creation of the term "performance dashboard" \cite{eckerson2010performance}. Dating back to the 1980s, executive information systems (EISs) were built for the company executives to serve as executive dashboards for driving the companies by their respective bedrooms, but they never gained much traction because they were available to a few people in the company and were built on mainframes or supercomputers (which gradually gave way to client/server systems in the 1990s). The dashboards were already in use in automobiles and other vehicles back then, but businesses, governments, and non-profit organisations have been known to adopt them later. \\~\\
Large-scale computing infrastructures like supercomputers, clusters, and clouds are already pervasive as most of the members of our society interact with them on a daily basis, e.g., social networks, media streaming services, government services, etc\footnote{\url{https://www.universiteitleiden.nl/en/science/computer-science/systems-and-security/large-scale-computing-infrastructure}}. The purpose of Operational Data Analytics (ODA) is to gain insight into the behaviour of such large-scale computing infrastructure (like HPC clusters) by analysing the operational data from various layers of the computing system \cite{DBLP:conf/icppw/BourassaJBCJVS19}. The ODA-related analysis yields various interesting results and behaviours about these large-scale systems which one could not even be aware of. The typical data flow involved in a large-scale system having ODA enabled is shown in Fig. \ref{oda-data-flow}. One of the most prominent large-scale infrastructures where ODA-related research has been done extensively is the HPC cluster. High-performance computing (HPC) refers to a specialised branch of computing that has the ability to solve advanced computations (which are too large to be solved using commodity computational resources) with the help of supercomputers or computer clusters. As per one of the tech reports published by a technical college, \textit{HPC integrates systems administration (including network and security knowledge) and parallel programming into a multidisciplinary field that combines digital electronics, computer architecture, system software, programming languages, algorithms, and computational techniques}\footnote{\url{https://web.archive.org/web/20100731043053/http://system.tstc.edu:80/forecasting/techbriefs/HPC.asp}}. HPC architectures couple together powerful integrated compute nodes using a high-speed interconnect. In our study, we are mainly concerned with large-scale computing infrastructures that run exclusively a variant of POSIX-compliant Linux operating systems (OS). The OS on each compute node runs various services and specialised hardware drivers that allow the applications to utilise the resources which are distributed across several nodes. \\ ~ \\
There are impacts at various layers and scales of the large-scale infrastructures ranging from facility and hardware to the software and applications running on these systems. One of the interesting studies measured the energy efficiency of 27 popular programming languages, measuring the correlation among energy, time and memory \cite{DBLP:conf/sle/Pereira0RRCFS17}. There is a lot of research work ongoing in the related fields due to arisen interest in our community to explore the topic in a holistic manner (some of which are discussed in this literature).
\begin{figure}
\centering
\includegraphics[width=0.5\textwidth]{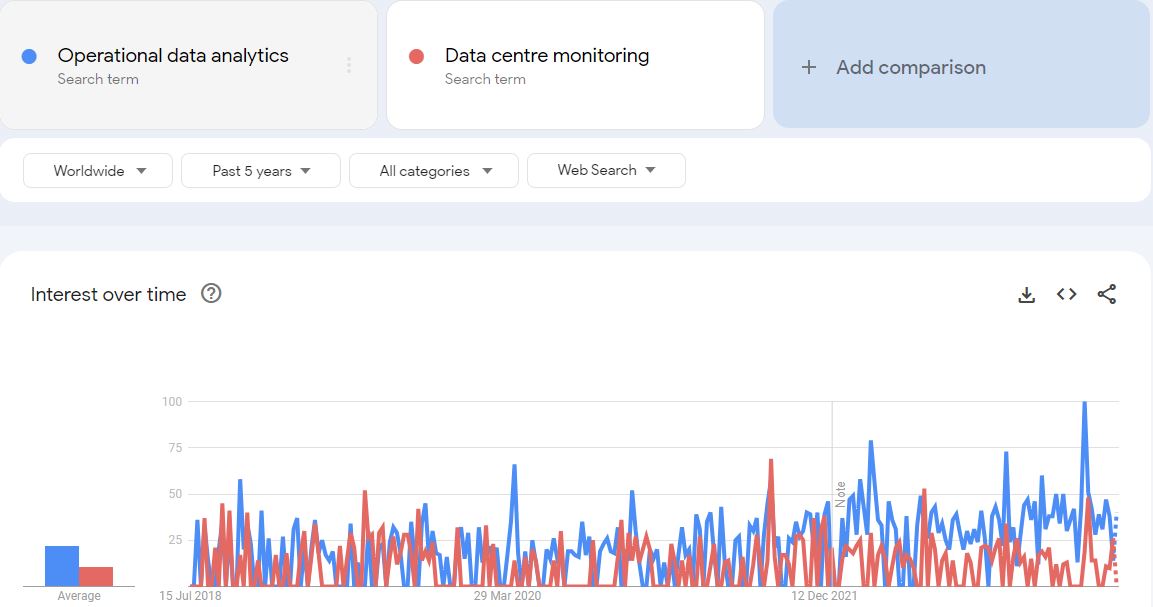}
\caption{Google Trends\protect\footnotemark for the keywords "Operational Data Analytics" and "Data centre monitoring".}
\label{monitoring-oda-trend}
% \vspace{-4mm}%Put here to reduce too much white space after your table 
\end{figure}
As reflected in Fig. \ref{monitoring-oda-trend} about the trend in the last 5 years (extracted from Google Trends), the popularity of the ODA term/keyword is increasing gradually. However, we do not see a significant increase in the keyword "Data centre monitoring", and the usage is almost constant for the past 5 years. Google Trends clarifies that the two keywords (ODA and data centre monitoring) are recognised as a search term, and not as a topic. It can be understood from the fact that the term "ODA" was officially defined for data centre operations' monitoring and analytics in 2019 by Bourassa et al. \cite{DBLP:conf/icppw/BourassaJBCJVS19}. Its popularity has been rising gradually (although slowly) as can be inferred from the increased count of published works of literature related to the topic. In contrast, data centre monitoring has been there for a long time, exhibiting almost a constant trend. Data centre monitoring might appear overall to be less actively researched compared to other trending computing terms, though the data centre components/entities are being actively researched either individually or in comparatively smaller conjunctions, and not from a holistic perspective (as what is the goal of ODA). As part of the manifesto of Future Computer Systems and Networking Research in the Netherlands, we aim to create awareness about the ODA framework within the Netherlands to promote the implementation of those capabilities across the entire information and communications technology (ICT) infrastructure \cite{DBLP:journals/corr/abs-2206-03259}.
\begin{summarybox}{ODA resurgence}
\textit{
As a large-scale computing environment (e.g., an HPC cluster, cloud) is complex due to its heterogeneity and scale of the components, gathering insights requires careful study and analysis of its components thoroughly at fine-grained levels to make the operations efficient. ODA is the process involving analysing the operation of such systems for gaining insight into their behaviour to reason about known and unknown results of the environment, which can later help in the optimisation of power usage, and maybe performance to some extent.
}
\end{summarybox}
\footnotetext{\url{https://trends.google.com/}}
\subsection{Research Questions}\label{sec:research_questions}
Here we list the research questions covered as part of the scope of this literature study. These research questions will be answered subsequently in the upcoming sections of this study. \\
\begin{itemize}
    \item \textbf{RQ1: How to lay out various components geared towards the performance and energy-efficiency of a large-scale computing infrastructure into a reference architecture for an ODA framework?} This section will be a novel contribution as part of this literature study. This research question will conceptualise a reference architecture of an ODA framework suitable for a large-scale computing data centre, conceiving from the four main pillars of an energy-efficient data centre, and that overlaps with the popular ODA reference architecture.
    \item \textbf{RQ2: What are the state-of-the-art ODA frameworks published in scientific pieces of literature for a large-scale distributed computing data centre?} This section will be the major contribution as part of this literature study. We will briefly discuss the architecture of several state-of-the-art data centre infrastructures which have deployed the ODA framework in their production environment.
    \item \textbf{RQ3: What are the various (quantitative/qualitative) energy or performance benefits that have been realised after implementing ODA techniques in a large-scale distributed data centre?} In this section, we will go through the benefits of deploying the ODA framework in a large-scale distributed data centre, as claimed in the scientific works. We will talk about the gains realised in quantitative terms (related to energy efficiency) and/or qualitative terms (in terms of performance gain), wherever applicable.
    \item \textbf{RQ4: What are the ongoing ODA-related research works in a large-scale distributed environment?} With this research question, the idea is to get an idea about what are the ongoing trends and the associated challenges in this field, as this field is gaining momentum gradually. This work will be based on the future work sections described in the articles which are scoped in this study.
\end{itemize}
\subsection{Contributions}\label{sec:contributions}
We aim to provide the following contributions to the community as part of this literature study, which would hopefully add to research in the field of Operational Data Analytics (ODA):
\begin{itemize}
    \item \textbf{C1} - We review a structured preface, setting up the proposal of a holistic ODA framework for a large-scale computing environment (e.g., requirements, fundamental blocks, modes of operation).
    \item \textbf{C2} - We propose a holistic ODA framework based on our established set of requirements, that extends a popular novel framework (laying out various ODA components into a novel system architecture).
    \item \textbf{C3} - We summarize the quantitative and qualitative gains that have been realized after enabling an ODA framework in various computing ecosystems (to promote awareness).
    \item \textbf{C4} - We discuss the ongoing research trends related to ODA in the large-scale computing ecosystem (for exploration of the research opportunities).
\end{itemize}

\section{Method for finding, selecting and characterising the relevant material}
\begin{figure}
\centering
\includegraphics[width=0.45\textwidth]{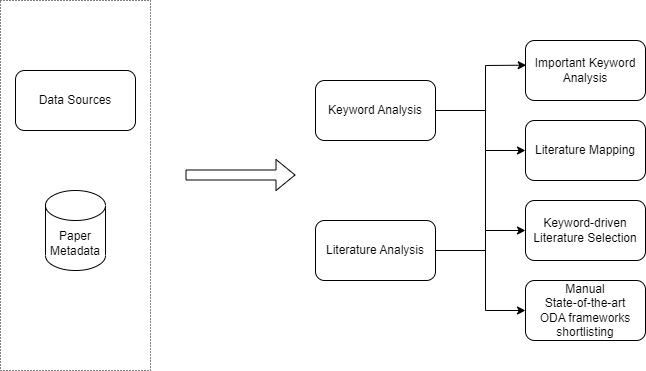}
\caption{The process followed in this literature extends from the Systematic Literature Review method by involving limited snowballing, manual search, and guided supervision.}
\label{slr-process}
% \vspace{-4mm}%Put here to reduce too much white space after your table 
\end{figure}

Various literature has been published detailing monitoring and analysing the performance and energy efficiency of high-performance computing data centres. Some have been explicitly characterised with the keyword "operational data analytics" (ODA) in a computing environment, thus associating them directly as part of the acknowledged ODA framework. Here we present our method which was used to perform the selection of literature pertaining to the usage of ODA, which is also visually presented in Fig. \ref{slr-process}.
\subsection{Analysis of commonly used methods}
Three of the most common methods adopted universally to conduct the literature surveys are: (i) unguided traversal of the related literature, (ii) snowballing \cite{DBLP:journals/misq/WebsterW02}\cite{DBLP:conf/ease/Wohlin14}, and (iii) systematic literature review method proposed by Kitchenham et al. \cite{kitchenham2007guidelines} 
Unguided traversal of the material encompasses the random traversal of the literature and related articles based on the recommendation provided either by the publications like Springer, IEEE, etc. or other scholarly websites.
Snowballing method uses a similar mechanism but it relies on some sort of limit, by imposing some restrictions such as on the criteria for finding and selecting material. Although one could select any of the methods for conducting the literature review, it is important to note that one should carefully examine the works of literature relevant to a topic so that they do not miss the useful ones, and the search is bounded and appropriate with less human bias as possible. The scientific literature review process has some laid-down processes which make it easier to follow the directions of the author to repeat the literature survey process and identify the relevant works of literature in an optimal way.
\begin{summarybox}{Literature Study Method}
\textit{
In this study, we follow the Systematic Literature Review process given by Kitchenham et al., along with the guidance of the supervisor, and with a mix of snowballing at times and manual search for works of literature published in MODA workshops for more information related to the subjects being studied.
}
\end{summarybox}
\begin{table*}[t]
  \centering
\begin{tabular}{p{1cm} | p{7cm} | p{3cm}}%\toprule
 \hline
  \multicolumn{3}{c}{\textbf{Planning the Review}}\\
  \hline
  \cmark        & Identification of the need for a review & (Section 1) \\
  \xmark        & Commissioning a review   &   \\
  \cmark        & Specifying the research question(s)   & (Section 1.1)  \\
  \cmark        & Developing a review protocol   & (Section 2.1)  \\
  \xmark        & Evaluating the review protocol   &           \\
  \hline
  \multicolumn{3}{c}{\textbf{Conducting the Review}}\\
  \hline
  \cmark        & Identification of research & (Section 2.2) \\
  \cmark        & Selection of primary studies & (Section 2.2) \\
  \xmark        & Study quality assessment   &        \\
  \cmark        & Data extraction and monitoring & (Section 2.3) \\
  \cmark        & Data synthesis   &  (Section 4 - 8) \\
  \hline
  \multicolumn{3}{c}{\textbf{Reporting the Review}}\\
  \hline
  \xmark        & Specifying dissemination mechanisms   &   \\
  \cmark        & Formatting the main report   & (Overall)  \\
  \xmark        & Evaluating the report   &       \\
  \hline
% \end{longtblr}
\end{tabular}
\caption{\label{table:literatureSelection}Overview of steps implemented (\cmark) in this paper which comprises the Systematic Literature Review method by Kitchenham et al. \cite{kitchenham2007guidelines}}
\end{table*}
\begin{table*}[t]
  \centering
\begin{tabular}{p{5cm} | p{9cm}}%\toprule
 \hline
The technique used to find literature & List of works of literature found using that technique \\ \hline
Systematic Library search (AIP, DBLP, Google Scholar) & \cite {DBLP:conf/hpdc/NettiMGOTO020, DBLP:conf/cluster/NettiSOWB21, DBLP:journals/corr/abs-2209-07164, DBLP:journals/pc/NettiOGTS22, DBLP:conf/icppw/BourassaJBCJVS19, DBLP:conf/cluster/OttSBWCRB20, DBLP:conf/icppw/BautistaRDWK19, DBLP:conf/supercomputer/TeraiYMS21, DBLP:phd/dnb/Netti22, DBLP:conf/ipps/NettiTO021, DBLP:conf/supercomputer/OzerNT020, DBLP:journals/iotj/BorghesiBB23, bautista2022omni, liu2020continuously, DBLP:conf/cluster/JakobscheLCC21, markus2021framework, shvets2021endless, DBLP:conf/sc/ShinOKEW21, kunz2022hpc, DBLP:conf/europar/AksarSALBEC21, DBLP:journals/mam/CascajoSC22, DBLP:conf/asiasim/FujitaSFNT21, moschny2021deep, DBLP:journals/fgcs/VersluisCGLPCUI23, DBLP:journals/tc/DemirbagaWNMAGZ22, DBLP:conf/cluster/TeraiSTY20, DBLP:journals/tpds/BorghesiMMB22, DBLP:conf/supercomputer/BrownNGBPCMFG22, DBLP:conf/heart/0001KSTW21, DBLP:conf/hpca/RoyPKARST21, DBLP:conf/ccgrid/ShilpikaLESVPM22, DBLP:conf/cluster/SchwallerTTAB20, demirbaga2022real, muller2019development, wood2021online, DBLP:journals/fgcs/NettiKBSBB20, DBLP:journals/corr/abs-2107-11832, DBLP:journals/corr/abs-2206-03259, DBLP:conf/supercomputer/MolanBBGB21, DBLP:conf/europar/MolanBBB22, DBLP:journals/fgcs/MolanBCBB23, DBLP:conf/europar/MolanBBB22a, DBLP:conf/cf/MolanBBB22, DBLP:phd/dnb/Ilsche20, DBLP:conf/hpdc/RajeshDGBLYKS21, wood2022scalable, DBLP:conf/sc/JhaCBXESKI20, DBLP:conf/supercomputer/ArdebiliBAB22, fernandezenergy, fischer2020metrics, } \\ \hline
% {DBLP:conf/ipps/NettiTO021, DBLP:conf/hpdc/NettiMGOTO020, DBLP:conf/sc/NettiMAGOT019, DBLP:journals/tpds/BorghesiMMB22, DBLP:journals/iotj/BorghesiBB23, DBLP:journals/fgcs/MolanBCBB23, DBLP:conf/europar/MolanBBB22, DBLP:conf/cf/MolanBBB22, DBLP:conf/wosp/MolanKBB23, DBLP:conf/europar/MolanBBB22a, DBLP:conf/supercomputer/TeraiYMS21, DBLP:conf/icppw/BourassaJBCJVS19, DBLP:conf/icppw/BautistaRDWK19, DBLP:conf/cluster/NettiSOWB21, DBLP:conf/sc/ShinOKEW21, fischer2020metrics, wood2021online, wood2022scalable, DBLP:journals/tc/DemirbagaWNMAGZ22, demirbaga2022real} \\ \hline
Manual Search (MODA Workshop) & \cite{DBLP:conf/supercomputer/SivalingamR20, DBLP:conf/supercomputer/TraceyHSE20, DBLP:conf/supercomputer/OzerNT020, DBLP:conf/supercomputer/TeraiYMS21, DBLP:conf/supercomputer/MolanBBGB21, DBLP:conf/supercomputer/EganPS22, DBLP:conf/supercomputer/ArdebiliBAB22} \\ \hline
 % Snowballing & \cite{DBLP:journals/ife/WildeAS14, DBLP:journals/cacm/SakrBVIAAAABBDV21, mqtt31, DBLP:conf/icppw/BartoliniBBCLBC19, bautista2022omni, DBLP:conf/cluster/SottileM02, DBLP:journals/corr/cs-DC-0306096, DBLP:conf/cluster/SacerdotiKMC03, DBLP:journals/pc/MassieCC04, DBLP:conf/IEEEscc/KatsarosKG11, mongkolluksamee2010strengths, DBLP:conf/lisa/Oetiker98, DBLP:conf/sc/EvansBBDFGJP14, DBLP:conf/sc/AgelastosABCEFGMNORSSTT14, DBLP:conf/cluster/AgelastosABGLMO15, DBLP:journals/fgcs/FortiGB21a, DBLP:conf/uic/SukhijaB19, DBLP:conf/medes/SukhijaBJGDLQL20, DBLP:conf/ucc/GraciaRBA16, DBLP:conf/xsede/Chan19, DBLP:conf/jcsse/RattanatamrongB20, naqvi2017time, DBLP:journals/fgcs/WangXZGZ18, simmonds2009scf, barhate2018hybrid} \\ \hline
Snowballing & Rest of the references (omitted for better readability of the report) \\ \hline
\end{tabular}
\caption{Categorisation of literature sources used in this study along with the technique of their identification.}
\end{table*}
\subsection{Systematic Literature Review process}
% Below in Table \ref{table:literatureSelection} we list the identification and selection criteria, which have been referenced from the famous \cite{kitchenham2007guidelines}.
In this subsection, we mainly list the methods followed in our literature study, from the overall systematic literature review process directed by Kitchenham et al. We further briefly explain below those methods we have followed under different review processes for this study, also shown in Table \ref{table:literatureSelection}. We use the literature resources retrieved from AIP\footnote{\url{https://github.com/atlarge-research/AIP}}, DBLP\footnote{\url{https://dblp.org/}} and Google Scholar\footnote{\url{https://scholar.google.com/}}. Additionally, we also consider the works of literature published especially in all the 3 ISC-HPC International Workshops on Monitoring and Operational Data Analytics (MODA) \footnote{\url{https://moda20.sciencesconf.org/page/scope}}\footnote{\url{https://moda21.sciencesconf.org/page/customizable\_page}}\footnote{\url{https://moda.dmi.unibas.ch/wp-content/uploads/2023/02/index22.html}} up till March 2023, although several of them already overlap with the Google Scholar results as mentioned in Table \ref{table:literatureSelectionOverview}. In addition to the scientific literature review process, we also used snowballing at times (to explore some topics in detail, wherever deemed necessary) and the supervisor's guidance to limit the search scope and refine the results, and enhance the quality of the literature study.
\begin{itemize}
    \item Planning the Review
    \begin{itemize}
        \item \textbf{Identification of the need for a review}: The need for this review arises to know about the topic being researched, ODA in this case. The need is to summarise the main information about the topic thoroughly and unbiasedly, to serve as a prelude to further research.
        \item \textbf{Specifying the research question(s)}: This is the most crucial activity during the planning phase. The important point to note is that the research questions should be of primary interest to the researchers from the same field. The right research questions in the field of software engineering are those that either result in change in the current practice, or strengthen the confidence in existing practice, and is meaningful to the researchers as well as the practitioners. We have presented our research questions in Section \ref{sec:research_questions}.
        \item \textbf{Developing a review protocol}: A review protocol consists of the methods used to undertake the systematic review and some planning information. The key components of our review protocol are the rationale, research questions, literature search strategy, study selection criteria and procedure, and project timetable.
    \end{itemize}
    \item Conducting the Review
    \begin{itemize}
        \item \textbf{Identification of research}: A systematic review process aims to find various primary works of literature related to the research question, using an unbiased search strategy. In this case, we used "HPC" and "Operational Data Analytics" search strings in the beginning, and a sophisticated search string was constructed using the boolean AND and those two keywords (i.e., "operational data analytics" AND "monitoring") in the final version of literature selection. We used AIP, DBLP, and Google Scholar as the main digital libraries for querying the literature sources, amongst which Google Scholar referred to ACM Digital Library, IEEExplore and Science Direct as the supplement library to get slightly enriched results.
        \item \textbf{Selection of primary studies}: After identifying the potentially relevant primary studies, we need to follow study selection criteria to reduce the likelihood of bias. Some of the well-known criteria for inclusion and exclusion of literature are based on the research questions, language, journal, research design, and date of publication. In this case, we have considered the inclusion and exclusion of literature based on how relevant the works of literature are in answering the research questions related to the ODA framework design and implementation in HPC environments, and those which are published in the English language.
        \item \textbf{Data extraction and monitoring}: The main objective of this stage is to list details about the data extraction procedure and analysis of the retrieved data. The data extraction strategy primarily utilised Python script-based retrieval of relevant literature, but secondarily also a manual search technique while snowballing to find associated works of literature, wherever required. 
        \item \textbf{Data synthesis}: This stage deals with the collation and summarisation of the results of the included primary studies. The summary of results related to the works of literature considered in this review process is presented in Table \ref{table:literatureSelectionOverview}.
    \end{itemize}
    \item Reporting the Review
    \begin{itemize}       
        \item \textbf{Formatting the main report}: This systematic review has been formatted in a technical report (categorised as a "literature study") which would eventually be merged into a journal or a conference paper. We have tried to follow the guidelines of Kitchenham et al. as much as possible to ensure that the structure is appropriate for technical reports.
    \end{itemize}
\end{itemize}
\subsection{Analysis of selected material}
We performed the analysis of our selected material using a four-step process.
\begin{itemize}
    \item Search queries followed by manual inspection of the context of the article
    \item Scanning each selected article to summarise the details related to the ODA framework   
    \item Extraction of the HPC infrastructure overview and their ODA framework published in all such reports
    \item Manual interpretation of the ODA functionalities covered in those works of literature for the purpose of selecting the unique ones to be studied as state-of-the-art, which might have accidentally introduced human bias
\end{itemize}
\begin{table}[t]
  \centering
    \begin{tabular}{p{2.5cm} | p{1.5cm} | p{3cm}}%\toprule
 \hline
 Article Source & Selected relevant Literature & Overlap with Google Scholar result \\
 \hline
 Google Scholar & 55 & - \\
 AIP   &  5 & 5\\
 DBLP   & 8 & 8\\
 MODA Workshops   & 7 & 7\\
  \hline
\end{tabular}
\caption{\label{table:literatureSelectionOverview} Overview of articles retrieved using the Systematic Literature Review method, with Google Scholar results as the reference point.}
\end{table}
As can be seen in Table \ref{table:literatureSelectionOverview}, the reference source of article selection is Google Scholar with the highest result for our search based on the keyword \textit{"operational data analytics" AND "monitoring"}. In addition to this, we also explore some detailed information about the architecture of the large-scale computing infrastructure or the ODA framework for a given article in its referenced articles, as part of slight snowballing for providing concise information to the readers.

\subsection{Threats to validity}
Although we selected our literature selection criteria from a broad range of methods, we cannot overrule the source of biases in our selection. First, our selection criteria only include the works of literature published as part of research in the scientific world or those which have published their literature in MODA workshops (like IBM research), which means we have not explored the publications or works of commercial organisations. Second, our search criteria are limited to works of literature published only in English, which might have resulted in the exclusion of scientific work not known in the English-speaking scientific community. Third, we have mainly focused on the "monitoring" and "operational data analytics" (ODA) keywords for exploring the scientific works of literature, among which the latter is known to be defined formally for the said purpose in 2019 (however these keywords have been used in the past some years ago). Additionally, there are various works of literature similar to this topic (having the ODA keyword), which we did not include while answering some of the research questions.\\ Lastly, we have tried to select ten unique/better advanced ODA/monitoring frameworks with detailed analysis and insights (but this potentially involves human bias, as 'better' has no defined criteria). We do justify the reason why we have selected the ten state-of-the-art frameworks out of several works of literature published to date, but this could be highly influenced by the researcher's state of mind (human bias).

\section{Reference architecture for an ODA framework}
\begin{figure*}
\centering
\includegraphics[width=0.7\textwidth]{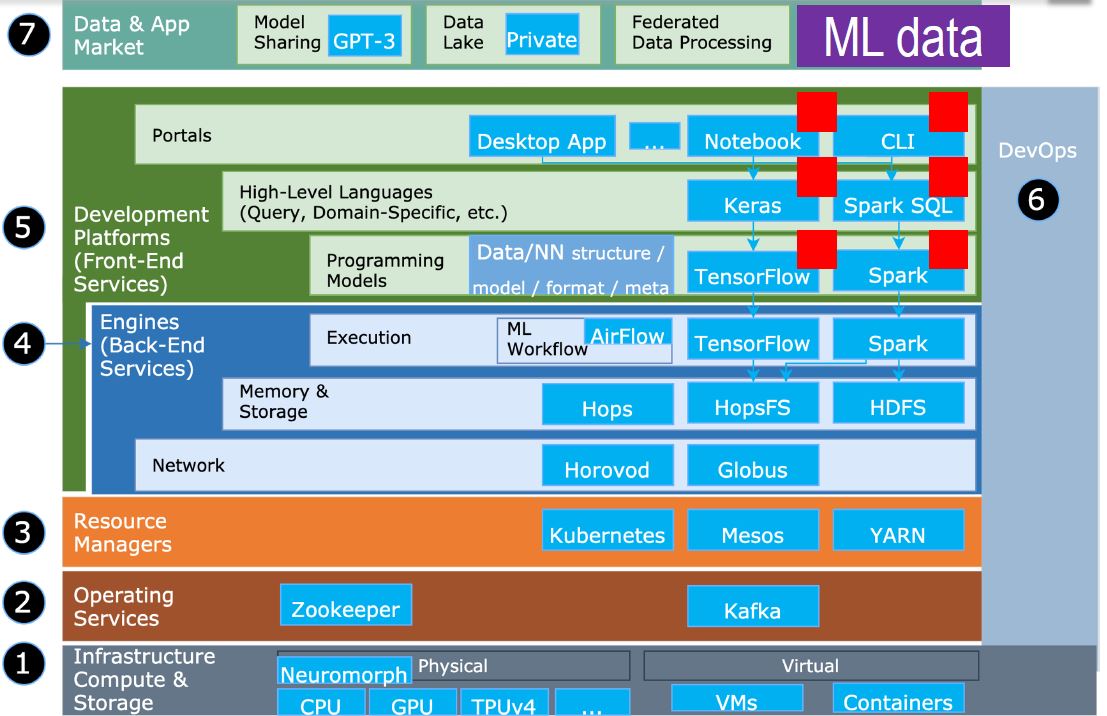}
\caption{Revised version of "A reference architecture for graph processing ecosystems that can be mapped to various global distributed ecosystems". \cite{DBLP:journals/cacm/SakrBVIAAAABBDV21}}
\label{ds-ecosystem}
\vspace{-3mm}%Put here to reduce too much white space after your table 
\end{figure*}
In this section, we answer our first research question by proposing a reference architecture for an ODA framework after doing an analysis of several existing ODA frameworks. We need to first understand various levels of a distributed ecosystem before presenting the requirements of an ODA framework reference architecture. Sherif et al. proposed a reference architecture for graph processing ecosystems which can be mapped to various distributed ecosystems \cite{DBLP:journals/cacm/SakrBVIAAAABBDV21}. Their proposed reference architecture consists of 7 distinct layers namely, infrastructure computing and storage, operating services, resource managers, back-end services (execution engines), front-end services (development platforms), DevOps, and Data/Model sharing and federated processing, as shown in Fig. \ref{ds-ecosystem}.
% \vspace{-3mm}%Put here to reduce too much white space after your table 
% We present a brief summary of the layers included in the reference architecture proposed by Sherif et al. below:
% \begin{itemize}
%     \item \textbf{Infrastructure Compute \& Storage (Layer 1)}:
%     \item \textbf{Operating Services (Layer 2)}:
%     \item \textbf{Resource Managers (Layer 3)}:
%     \item \textbf{Engines/Back-Eend Services (Layer 4)}:
%     \item \textbf{Development Platforms/Front-end Services (Layer 5)}:
%     \item \textbf{DevOps (Layer 6)}:
%     \item \textbf{Data \& App market (Layer 7)}:
% \end{itemize}
In this section, we address research question 1 (RQ1) and provide a response against the same. Before we propose our reference architecture, we go through the foundational elements that enable the implementation of an ODA framework in an HPC environment.
\subsection{Foundational elements of ODA}
HPC computing is as complex as one could imagine because of the components' scale and their heterogeneity. So, the possible scope of insights could be gained from a diverse range of sources from something as low-level as hardware counters and source code performance hotspots to as high-level as application data dependency graphs. In our literature study, we are concerned with HPC systems running an almost exclusive variant of the Linux operating system. \\

As Chad Wood mentions in his literature survey, there are three foundational elements involved in the online monitoring for HPC systems: \textit{monitoring}, \textit{analysis} and \textit{feedback} \cite{wood2021online}. The term feedback here implies the interaction with applications and the execution environments based on the analysis of monitored information. A critical point to consider is that an event or a state needs to be observable before it can be monitored, analysed or utilised online. Subsequently, the state once made observable would allow the capture and usage of associated data, which are briefly described further as extracted from Chad Wood's dissertation literature \cite{wood2022scalable}. It must be noted that this idea can be easily extended to other computing environments like cloud computing, large-scale big data computing, etc.
\subsubsection*{Observability}
Observability is the critical first step into online monitoring, and the depth and significance of the observation would vary based on the method and invasiveness of the techniques used. Observability could be achieved or enhanced using a variety of techniques as listed below:
\begin{itemize}
    \item \textbf{Application source instrumentation}: Software source code can be instrumented in a way that it reports its progress from state to state. This could be done in the form of function calls (or macros) embedded in-line between the normal application code. At the same time, the instrumentation should have an "off switch" of some kind as sometimes it is desirable to disable it (for example, maybe during code moving from active development into a production scenario).
    \item \textbf{Shared Library, Runtime, or Service Instrumentation}: HPC software consists of multiple libraries that interact based on the core logic of an application. Making observations within the code when an application makes API calls to an external library or service would enable the instrumentation within to be executed.
    \item \textbf{Sampling and Tracing}: Here, sampling means inspecting the state of an application with the help of performance counters made available by the OS, and tracing derives from sampling in the sense that each action of an application could be counted as a significant event and measured though each action might not be of interest (thus, tracing often has overhead higher in orders of magnitude). Sampling is by far considered the most efficient method for gathering observations to use for monitoring an HPC system. The basic idea revolves around exploiting binary formats, memory layout conventions and operating system APIs to apply instrumentation to a compiled application without changing its source code.
    \item \textbf{Probing and Inferences from Indirect Sources}: Observations of sources outside of the scope of applications are often required to form an intuition about the behaviour of a system and its components. The idea is to combine information from multiple external sources allowing for interesting questions to be asked and answered from a holistic perspective.
\end{itemize}
\subsubsection*{Capturing and using Data}
Something that has been rendered observable would allow the capturing and use of data, which can be characterised by one of the possible aspects listed below. Our practical interest lies in the type of data that can be accumulated to discover and react to trends and patterns through algorithms.
\begin{itemize}
    \item \textbf{Representation and meaning}: The information captured from the observations should not only include the measurements but also some standard notion of interpreting them. Any eventual application of observations would expect them to be correct, consistent, and precise. This could include "encoding the data and metadata", "encoding the expertise (users, developers, etc.)", "time, change, identity and consistency (continuous interaction)", and "combination and unit semantics (from different domains perspective)".
    \item \textbf{Patterns within HPC}: Performance observations of the application are often connected to various combinations of the operating system code, its version and configuration, the version of system libraries and vendor drivers, and the version of linked libraries among others. The permissions and user priority levels also may have a direct impact on the performance characteristics.
    \item \textbf{Exposing or exporting data}: The relevant performance metrics need to be exposed to a monitoring system, as in-memory logs and filesystem storage cannot be continuously populated over a long period of time. Moreover, because of the simple and limited nature of exposing information, we need to retain it which requires exporting, or recording and migrating that information between components of the HPC system. Some of the primary techniques for exporting information are logging, checkpoint, caching, polling and pulling, broadcast or push, hybrid pull/push, and publish/subscribe.
    \item \textbf{Introspection, Opacity, and Interface Standardisation}: It is a lamentable fact associated with online monitoring systems that many of them have opaque interfaces, protocols and data formats (due to the reason that many of them are developed keeping in mind specific research experiments). Oftentimes, the capabilities have been available but it is now emerging as a trend to utilise these abilities. A suggested solution could be to use generic performance annotation hooks and source-level instrumentation which is disabled by default but could be activated as needed at runtime. Understanding the application-level context allows a generic annotation framework to go beyond simple introspection tasks where there is a defined division between the communication and computation phases.
\end{itemize}
\subsection{Popular advanced monitoring frameworks for distributed environments}
Chad Wood has also covered some of the most advanced monitoring frameworks used in a distributed environment \cite{wood2021online}. Some of the past and present heavy hitters which have paved the way (as a major step forward) for state-of-the-art online monitoring are detailed below briefly:
\subsubsection*{SuperMon}
SuperMon, a cluster monitoring system engineered to be high-speed and to minimise overhead, was first released during the terascale era of HPC in the early 2000s \cite{DBLP:conf/cluster/SottileM02}. The design goal was to allow for low-impact monitoring of high-frequency events, thus enabling the invisible behaviour of the cluster to be open for observation. The system operated online and gathered data from all nodes (also periodically ran the \textit{ping} command to test the responsiveness of nodes) and assembled them into a coherent single set of samples that represented the state of the cluster as a whole.
\subsubsection*{MonALISA}
Monitoring Agents in A Large Integrated Services Architecture (MonALISA) project was born around 1998 to help the administrators and users to observe the grid computing system in aggregate \cite{DBLP:journals/corr/cs-DC-0306096}. It utilised forward-deployed ``station servers" positioned at the major grid system locales that ran a variety of agent-based services with the capability of deploying, starting, stopping, discovering and utilising arbitrary monitoring agents online. The focus was more towards flexibility and self-organising capabilities instead of maximising the throughput of monitoring data or concerning overhead or performance perturbation. Agents used to report whatever information they could provide, and also report the changes in their availability. The monitoring clients used to subscribe to the information streams from those agents, which propagated through the system enabling those clients to receive streams of information from all the available agents. \\
This project also included a client that projected the monitoring data (of a variety of topics like system availability, load balance, data link saturations, etc.) into a dashboard-style visualisation. It has the capability to optimise grid-based workflows based on the types of agents deployed and the sensitivity of an application. It interfaced with a variety of other monitoring tools like MRTG and Ganglia (which are discussed later). These have been some of the noteworthy achievements of the MonALISA team.
\subsubsection*{Ganglia} Ganglia is a popular distributed monitoring solution targeting both cluster and grid computing environments \cite{DBLP:conf/cluster/SacerdotiKMC03,DBLP:journals/pc/MassieCC04}. Its implementation uses XML encoding of the data and RRDTool for data storage and analysis/visualisation. Its functionality revolves around a monitoring daemon which uses TCP/IP multicast listen/announce protocols to monitor activity within a cluster, thus helping to gather a set of built-in metrics as well as allowing plugins to capture arbitrary user-defined metrics. Ganglia was reported to be in use at over 500 compute clusters worldwide by 2004.
\subsubsection*{Nagios} Nagios is an online monitoring tool having a strong emphasis on network devices and their service status \cite{DBLP:conf/IEEEscc/KatsarosKG11,mongkolluksamee2010strengths}. It has an in situ server that runs in the background local to the nodes of a cluster, and its service periodically probes the state of the machine and services. It has gained hundreds of plugins over the years, and also has been used as the core component of several different commercial monitoring solutions.
\subsubsection*{MRTG} The Multi Router Traffic Grapher (MRTG) is a tool of the past that was designed to monitor the inbound and outbound traffic on an internet gateway router \cite{DBLP:conf/lisa/Oetiker98}. This Perl script used to read the octet counters of the router every 5 minutes, and generated a graph which was represented in a web page on the same server where the script ran. MRTG embraced the idea that the less recent the monitoring data the less important it is, by allowing for a "lossy data storage".
\begin{figure*}
\centering
\includegraphics[width=0.65\textwidth]{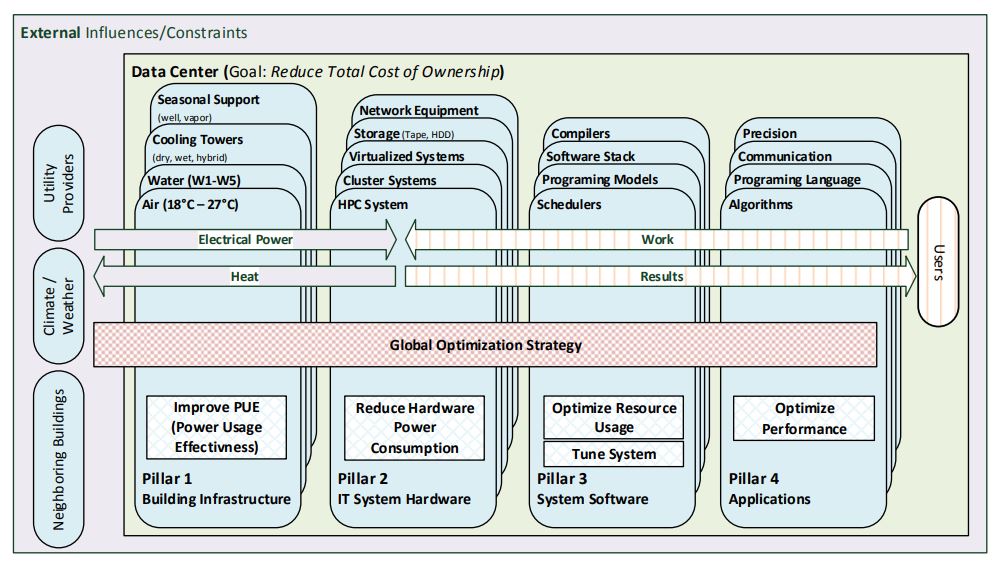}
\caption{Four pillars of energy efficient HPC. \cite{DBLP:journals/ife/WildeAS14, DBLP:conf/cluster/NettiSOWB21}}
\label{hpc-four-pillars}
% \vspace{-4mm}%Put here to reduce too much white space after your table 
\end{figure*}
\subsubsection*{TACC Stats} TACC Stats has a long history in HPC data monitoring since Texas Advanced Computing Center (TACC) introduced it in 2013 \cite{DBLP:conf/sc/EvansBBDFGJP14}. The nice part is that it is enabled and functioning by default (always-on) without any intervention from the users, developers and administrators of HPC clusters. It accumulates performance and utilisation metrics for all jobs running on the cluster, by utilising sources ranging from the filesystem to the messaging services to the job scheduler, and to operating system performance introspection APIs. It has been designed in a modular way and can be extended to incorporate additional data metrics based on the requirement and data availability.
\subsubsection*{LDMS} Lightweight Distributed Metric Service (LDMS) is a widely deployed online monitoring framework regarded as one of the most important for current petascale and future exascale HPC clusters \cite{DBLP:conf/sc/AgelastosABCEFGMNORSSTT14,DBLP:conf/cluster/AgelastosABGLMO15}. The idea behind the design of LDMS was to bridge the gap between coarse-grained system event monitoring and fine-grained application profiling tools. LDMS is highly configurable and full of samplers (used to sample data periodically), aggregators (which pull data from samplers or other aggregators), and storage components (that can write to a variety of formats), and is essentially a distributed data collection, transport and storage tool. It does only a few things but is considered to do them in a highly effective manner.
\subsubsection*{FogMon} FogMon is a lightweight self-organising distributed monitoring framework for Fog infrastructures \cite{DBLP:journals/fgcs/FortiGB21a}. The crux of fog computing is a common orchestration layer that delivers a MAPE (Monitoring, Analysis, Planning, and Execution) loop that supports the dynamic life-cycle management of multi-service data-aware Fog applications. FogMon aims to support this orchestration layer, strongly emphasising the monitoring component.
\subsubsection*{Additional advanced monitoring solutions of note}
Some of the popular dedicated monitoring solutions which may be of interest to the reader (for further exploration) are listed below:
\begin{itemize}
    \item \textbf{Prometheus + Kubernetes}: Prometheus is an open-source monitoring framework that provides out-of-the-box monitoring capabilities for the Kubernetes container orchestration platform \cite{DBLP:conf/uic/SukhijaB19,DBLP:conf/medes/SukhijaBJGDLQL20,DBLP:conf/ucc/GraciaRBA16}. It is becoming a popular monitoring tool to be used for Docker and Kubernetes monitoring, as both Prometheus and Kubernetes are Cloud Native Computing Foundation (CNCF) projects. More information can be found here\footnote{Prometheus - \url{https://prometheus.io/docs/introduction/overview/}}
    \item \textbf{Telegraf + InfluxDB}: Telegraf is an open-source plugin-driven server-based agent for collecting and reporting all metrics and events from databases, systems, and IoT sensors \cite{DBLP:conf/xsede/Chan19,DBLP:conf/jcsse/RattanatamrongB20,naqvi2017time}. It has output plugins which could send data to a variety of other datastores, services, and message queues, including InfluxDB, Graphite, OpenTSDB, Datadog, Librato, Kafka, MQTT, NSQ, and many others. More information can be found here\footnote{Telegraf - \url{https://www.influxdata.com/time-series-platform/telegraf/}}
    \item \textbf{Zabbix}: Zabbix is an open-source software tool for network monitoring and application monitoring of millions of metrics \cite{DBLP:journals/fgcs/WangXZGZ18,simmonds2009scf}. More information can be found here\footnote{Zabbix - \url{https://www.zabbix.com/features}}
\end{itemize}
\subsection{Existing ODA framework for a distributed ecosystem}
Netti et al. has presented a conceptual framework for HPC operational data analytics in their literature \cite{DBLP:conf/cluster/NettiSOWB21}. First, we review the four pillars of an energy-efficient HPC shown in Fig. \ref{hpc-four-pillars}, which is crucial to understand how the community views the domains of a data centre:
\begin{itemize}
    \item \textbf{Building Infrastructure}: comprises of the facility and infrastructure needed to run the HPC systems and support the data centre's overall operation
    \item \textbf{System Hardware}: hardware components constituting the HPC system (e.g., motherboards, firmwares, CPUs, GPUs, memory, network equipment)
    \item \textbf{System Software}: software stack at the system level which is used by the users and their applications (e.g., system management software, the resource management and scheduler, compute nodes' operating system, other tools and libraries)
    \item \textbf{Applications}: individual workloads as well as the workload mix executed on a system, and can be considered as a unit of work for an HPC system.
\end{itemize}
\begin{figure}
\centering
\includegraphics[width=0.4\textwidth]{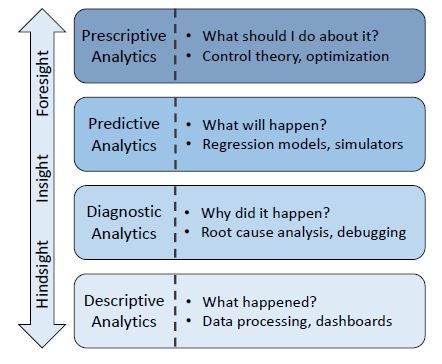}
\caption{The four types of data analytics. \cite{gartner2013}}
\label{oda-analytics-types}
% \vspace{-4mm}%Put here to reduce too much white space after your table 
\end{figure}

\begin{figure*}
\centering
\includegraphics[width=0.85\textwidth]{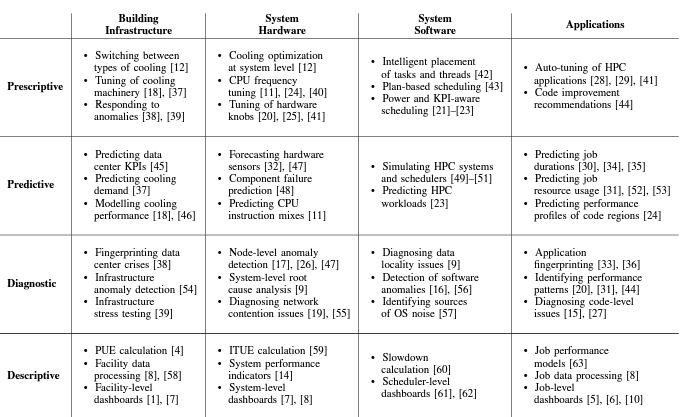}
\caption{A series of examples categorised under the generic ODA framework proposed by Netti et al. \cite{DBLP:conf/cluster/NettiSOWB21}}
\label{oda-four-pillars}
% \vspace{-4mm}%Put here to reduce too much white space after your table 
\end{figure*}

The authors in their original paper have presented the idea of types of data analytics, before integrating the pillars and the former (represented in Fig. \ref{oda-analytics-types}). The defined types of data analytics (in the context of HPC and operational data analytics) are briefly defined below:
\begin{itemize}
    \item \textbf{Descriptive}: the first degree of examination of data that centres around answering the question "what happened?". This is the simplest and the most basic of all, and does not require complex knowledge extraction.
    \item \textbf{Diagnostic}: the root-cause analysis examining the phenomenon that "why did something happen?". This analytics aims to provide systematic automation of such diagnoses.
    \item \textbf{Predictive}: aims to forecast a system's state in the near future by utilising machine learning models or other heuristic techniques.
    \item \textbf{Prescriptive}: aims to answer the question "what is the best way to manage my resources" by suggesting the best course of action towards a particular efficiency goal
\end{itemize}
While descriptive and diagnostic analytics aim for a better understanding of the past (hindsight), predictive and prescriptive analytics aim for future insights (foresight). A series of examples categorised under the generic ODA framework proposed by Netti et al. is shown in Fig. \ref{oda-four-pillars}.
%\newpage
\subsection{Reference architecture for an ODA framework}
As mentioned earlier, ODA is needed at all levels of a distributed ecosystem, which relies directly on the monitoring of various data sources at different time scales. We consider two types of data sources, as identified by the Wintermute team in their ODA framework \cite{DBLP:conf/hpdc/NettiMGOTO020}:
\begin{itemize}
    \item \textbf{In-band}: data that is sampled and consumed within a specific component at any layer of the distributed ecosystem. The underlying techniques operate at a fine temporal scale and require low analysis overhead and latency while collecting data.
    \item \textbf{Out-of-band}: data coming from any of the available sources in the system, including historical or asynchronous data. The underlying operation has to be performed at a coarse scale (in the order of minutes or higher) and must be synchronised explicitly, but latency and overhead are less of a concern in this scenario.
\end{itemize}
Additionally, we also categorise ODA techniques into two modes of operation as listed below, originally grouped by the Wintermute team \cite{DBLP:conf/hpdc/NettiMGOTO020}:
\begin{itemize}
    \item \textbf{Online}: a continuous operation resulting in output resembling a time series, which can be (re)used at various levels as a feedback loop.
    \item \textbf{On-demand}: operation triggered at specifically scheduled times to steer decisions managing the information about the system's status.
\end{itemize}
Based on these studies and carefully analysing the state-of-the-art frameworks mentioned in the literature in Section \ref{state-of-the-art}, we must first do a careful analysis of the requirements. In this subsection, we list the set of functional and non-functional requirements which have been adapted from the ODA framework presented by Netti et al. \cite{DBLP:conf/hpdc/NettiMGOTO020}.
\subsubsection*{Functional Requirements}
In light of the requirements specification, we extract a series of functional and operational requirements that must be considered before creating a reference architecture of the ODA framework:\\
\begin{itemize}
    \item \textbf{Holism}: an ODA framework must provide a holistic view of a distributed ecosystem, by exposing distinct layers of the ecosystem and generating an in-band model or an out-of-band model from analysis depending on the latency and overhead requirements.
    \item \textbf{Abstraction}: an ODA could be impractical for manual configuration in cases when a large amount of independent analytical models are required. Thus, abstraction constructs are necessary for simplification and automating the configuration of ODA models.
    \item \textbf{Modularity}: an ODA framework must be modular enough to support and integrate with a wide range of external interfaces over a set of common protocols.
    \item \textbf{Flexibility}: an ODA framework should support both online and on-demand operations to address the necessities of different techniques demanded by the various components of a computing ecosystem.
\end{itemize}
\subsubsection*{Non-functional Requirements}
Below we list the considered non-functional requirements which strengthen the basis of the ODA framework:
\begin{itemize}
    \item \textbf{Scalability}: an ODA framework should be scalable up to thousands of inputs, allowing it to be configured for operating at very fine time scales.
    \item \textbf{Portability}: an ODA framework should be based on the concepts that can be ported into another domain of computing, or even other fields for studying operational efficiency.
\end{itemize}
\subsubsection*{Actors involved in ODA}
As large-scale computing infrastructures are quite large, it involves various actors in managing the operations of the data centre. It would be impractical to figure and list out the diverse roles, thus, we restrict to the key actors presented by Borghesi et al. in their ExaMon-X literature \cite{DBLP:journals/iotj/BorghesiBB23}, and reuse them here (listed below for clarity):
\begin{itemize}
    \item \textbf{System administrators}: actors who are responsible for the correct functioning of the data centre (who manage the job scheduling policies to fixing broken components, etc.).
    \item \textbf{Facility managers}: actors who are concerned with system-wide issues (e.g., energy consumption, mortgage costs, and thermal/cooling problems).
    \item \textbf{System accountants}: actors who manage and keep track of different accounts and projects, manage the access of users, and provide reports and statistics about machine usage.
    \item \textbf{System users}: actors consisting of industrial and academic partners who submit jobs to the system, and are mainly interested in fast completion time and fair pricing.
\end{itemize}
It must be noted that each type of actor would specifically be responsible for managing specific layers of the computing infrastructure, but they could also jointly work on some of the ODA functionalities for making the operations more efficient (by sharing respective experiences from different levels). There could be several actors introduced for managing the ODA functionalities, but we do not want to dive deeper but rather keep it more generic (as roles vary from organisation to organisation).
\begin{figure*}
\centering
\includegraphics[width=0.85\textwidth]{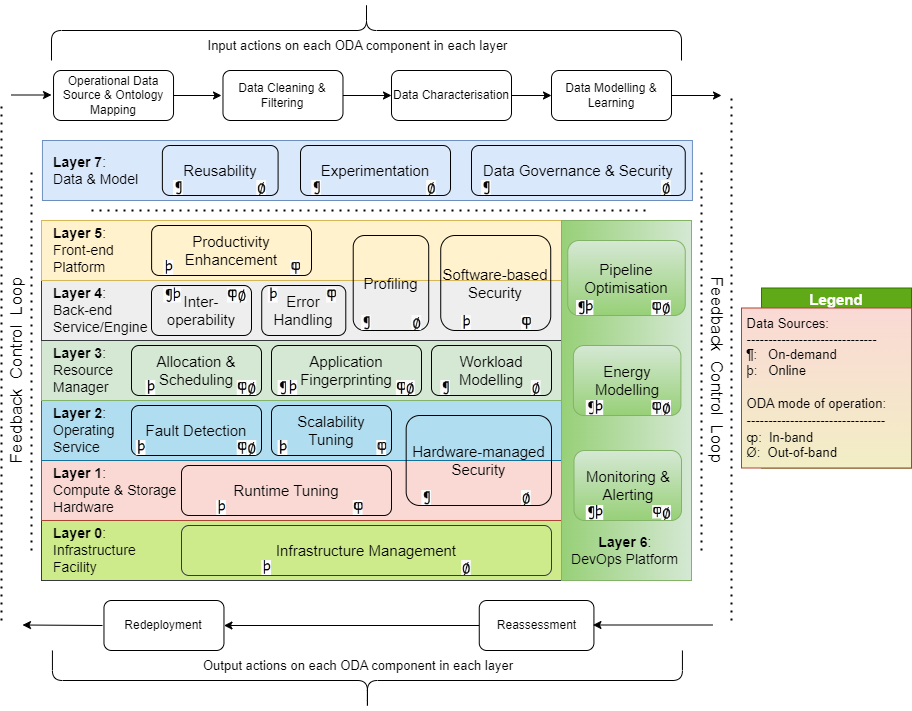}
\caption{Proposed reference architecture for the ODA framework, that is also mapped to the distinct layers of a distributed ecosystem.}
\label{oda-ref-architecture}
\vspace{-1mm}%Put here to reduce too much white space after your table 
\end{figure*}
\subsubsection*{Proposed Reference Architecture}
Below, we present in Fig. \ref{oda-ref-architecture} a reference architecture for an ODA framework which is mapped to ecosystems of distributed computing. The components in the proposed reference architecture of the ODA framework are complementary to the role handled by the respective layers. It must also be noted that although almost all of the functionalities can be automated and optimised, there might be some scenarios where manual intervention would be required (e.g., verification, validation, etc.), which we do not discuss here. It might vary on case to case basis, and the same is open for interpretation by the researchers who work in this field. The conceptual idea behind the reference architecture is already discussed in the previous subsections, and we have tried to reuse the reference architecture of the distributed ecosystem as much as possible while designing the reference framework of ODA for a large-scale computing ecosystem.\\
The relevant components of the ODA framework outlined in the reference architecture are further defined below. Please note that various components at different layers in the proposed reference architecture overlap to some extent for contributing to power usage optimisation, energy efficiency, or performance. We have tried to segregate their functionality as distinct, wherever we see the idea is broad enough to be explored in detail, or if it is complemented by other components from other layers of the computing ecosystem. An additional point to consider is that the target users responsible for managing those components must focus on selective functionalities, giving freedom for management, research and exploration in detail (they could either be distributed in different teams as described above as key actors or might be from the same team but with slightly different roles). We think that this would enable high productivity, with ease of management and focussed research in the operations, and also refer to the works of literature resulting in the identification of these key functionalities.\\~\\
The input actions on each of the ODA components first involve the understanding of various metrics available for observation, which can be modelled using \textbf{ontology mapping} of the data collected from operational data sources. The data from the data sources may not be used directly as such, and usually some sort of \textbf{data cleaning} and \textbf{filtering} is required which might be distinct for distinct ODA functionalities as per the customary requirement. Once the data is available after cleaning, \textbf{data characterisation} is needed for summarising the general characteristics, after which the designated \textbf{modelling} and/or \textbf{learning techniques} is applied with a defined objective resulting in an outcome. Each of the ODA functionalities discussed below keeps undergoing \textbf{reassessment} and \textbf{redeployment} (usually at periodic intervals but also dynamically as and when needed) which is fed back via the \textbf{feedback control loop}, for continual optimisation of the data centre operations (as shown in Fig. \ref{oda-ref-architecture}):
\vspace{-1mm}%Put here to reduce too much white space after your table
\begin{itemize}
    \item \textbf{Infrastructure Management}: Being operational at the lowermost building block (at the facility level), this ODA functionality relies on the online data source collected from various sensors from the facility and provides an out-of-band mode of service for managing the infrastructure. This component has been borrowed from the taxonomy of the ODA framework identified in the DCDB Wintermute framework literature authored by Netti et al \cite{DBLP:conf/hpdc/NettiMGOTO020}.
    \item \textbf{Runtime Tuning}: An ODA framework can be used to predict the behaviour of various components of the compute and storage hardware, and perform the dynamic tuning using system knobs. It relies on real-time data and information about specific components supplied from various layers, which is then sent to the compute nodes for dynamic adjustment of the values of those components. This component has also been borrowed from the taxonomy of the ODA framework identified in the DCDB Wintermute framework literature authored by Netti et al.
    \item \textbf{Hardware-managed security}: This is a critical security-related functionality which is a part of both compute hardware and operating service layers, and is responsible for assisting in the security of the underlying computing hardware with the help of on-demand data getting tracked at periodic intervals, thereby ensuring that the  hardware is not tampered with. This idea is slightly novel due to the fact that there is not much security-related context made available in the state-of-the-art ODA frameworks (except LBNL's newer OMNI architecture deployment which discusses some traits about the security considerations handled within the ODA framework). The need for hardware-managed security is governed by the requirements cited in certain works of literature like \cite{DBLP:conf/itrust/BaldwinS03,DBLP:books/sp/14/MavrovouniotisG14}.
    \item \textbf{Fault Detection}: This ODA component is responsible for detecting and predicting anomalous states in hardware and software components, thereby helping in improving the resiliency of the system as a whole by helping prevent catastrophic events. This monitors data from underneath as well as top layers on a real-time basis and helps detect and predict faults in both hardware and software components (continuously as well as periodically reviewing throughout the sections), preventing in turn unmasked failures. This component has also been borrowed from the taxonomy of the ODA framework identified in the DCDB Wintermute framework literature authored by Netti et al.
    \item \textbf{Scalability Tuning}: It is crucial for the operating services to closely monitor the scalability of the compute and storage resources (physical or virtual) as per the requirement of the jobs. Thus, the ODA framework must use real-time (online) data from the cluster manager and related computing layers. There are several works of literature which have highlighted similar concerns for targeting energy conservation (\cite{DBLP:journals/jcloudc/IsmaeelKM18,DBLP:conf/IEEEcloud/HauserW18}).
    \item \textbf{Allocation \& Scheduling}: As crucial as something could be, this is the backbone operation of the resource manager and is responsible for the optimal placement of jobs or allocation of resources. This needs real-time data about jobs and resources and supplies additional information to the scheduler for achieving the ODA goals, and operates in both in-band and out-of-band modes. This component has been borrowed from the taxonomy of the ODA framework identified in the DCDB Wintermute framework literature authored by Netti et al.
    \item \textbf{Application Fingerprinting}: The idea involves predicting the behaviour of user jobs and correlating this with the data from the past to characterise features such as power consumption and memory utilisation, helping to optimise management decisions. This relies on data sources from online sources as well as from the past and operates in both in-band and out-of-band modes. This component has also been borrowed from the taxonomy of the ODA framework identified in the DCDB Wintermute framework literature authored by Netti et al.
    \item \textbf{Workload Modelling}: An ODA framework uses heuristic or learning techniques to predict the properties (duration, submission pattern, etc.) of jobs submitted by users and helps in improving the effectiveness of scheduling policies. This relies on data sources from the past and operates in out-of-band mode to help reduce queuing times. This component has been derived from the "prediction of job features" functionality and borrowed from the taxonomy of the ODA framework identified in the DCDB Wintermute framework literature authored by Netti et al.
    \item \textbf{Interoperability}: The ODA framework should be able to manage and fine-tune the interoperability requirements with various layers, processing both real-time  (online) and historical data while operating in either in-band or out-of-band modes. Some of the basic prospects falling under this category which have been studied are the deployment of federated learning in cloud-edge collaborative architecture \cite{DBLP:journals/jcloudc/BaoG22}, a collaborative edge-cloud AI framework by Huawei Cloud \cite{DBLP:conf/kdd/Banitalebi-Dehkordi21}, etc.
    \item \textbf{Error Handling}: This is another crucial functionality which is responsible for monitoring the error events (online) originating from the back-end engine or as reported by other upstream layers, and acting upon them on a real-time basis to provide a reliable service. Some of the influencing works of literature guiding the inclusion of this component are \cite{peng1993software,wong2022prescriptive}, etc.
    \item \textbf{Software-based Security}: Similar to hardware-managed security, software-based security is spread across both the front-end platform and back-end engine and ensures security by monitoring and acting upon the security events collected online from multiple layers as part of near real-time operation handling. The need arises because of security concerns in a large-scale infrastructure like cloud computing as studied in \cite{DBLP:journals/jnca/SinghJP16}.
    \item \textbf{Productivity Enhancement}: The idea at the front-end layer is to enrich the user experience resulting in productivity enhancement. The component could optionally involve seeking custom input from the user or the data source could be the relevant front-end application logs analysed at regular periodic intervals, and desirable action could result in contributing to a rich user experience. Some of the associated works of literature studying the enrichment of user experience are \cite{DBLP:journals/ijcim/TzafilkouPK17,DBLP:conf/bigcom/QiuGSL017}, which highlight the significance of improvement in user experience leading to performance optimisation.
    \item \textbf{Profiling}: The ODA framework should be able to profile the code or the binary used in the front-end platform or the back-end service in an out-of-band fashion, and suggest a suitable action or perform automatic remediation for improving the performance. The potential of application profiling techniques has already been studied in various works of literature for understanding the application behaviour, and some of them dedicatedly cater to cloud computing like \cite{DBLP:conf/IEEEcloud/DoCWLZZ11,DBLP:conf/IEEEcloud/ScheunerL18} besides other forms of computing.
    \item \textbf{Monitoring \& Alerting}: The majority of the ODA framework functionality revolves around this core functionality of the DevOps layer, as evident from the state-of-the-art ODA frameworks like \cite{DBLP:conf/icppw/BourassaJBCJVS19,DBLP:conf/hpdc/NettiMGOTO020,DBLP:conf/icppw/BartoliniBBCLBC19}. Monitoring and alerting functionality relies on data from almost all other computing layers for acting on them, and includes the \textbf{visualisation} of the various operational data and/or analysis results to the different segment of users discussed previously. This functionality stands out as one of the most important ODA applications using data from both online sources and on-demand from distinct sources, and operating in both in-band and out-of-band modes.
    \item \textbf{Energy Modelling}: This is an outcome of various ODA functionalities discussed before. The idea involves modelling the energy  of a system by employing data analysis for investigating different scenarios or assumptions. This functionality is attributed to various studies based on the requirement of a reduction in power consumption and carbon emissions of a data centre infrastructure like \cite{DBLP:conf/3pgcic/PretoriusGI10,avgerinou2017trends}.
    \item \textbf{Pipeline Optimisation}: The pipelines in the DevOps layer can be optimised by cleaning up the stale data, information or logs, and applying data transformations on the speed of ingress/egress, or on the data formats for compression and type-conversion (as studied in \cite{DBLP:journals/ijcnds/OukfifOBB20,DBLP:journals/corr/abs-2202-05711}). The automatic builds defined as part of DevOps practices can be monitored and acted on efficiently to ensure their acceleration and faster delivery of results, utilising the underlying infrastructure optimally. These must be monitored continuously (on individual components) and at periodic intervals (to have a holistic overview) and should be configured to use real-time data or offline data at the DevOps layer.
    \item \textbf{Reusability}: Some of the data collected at various data sources or the developed models might be reusable in nature (for operational analysis), and it could help in avoiding reinventing the wheel at times. This intelligence could be available to an ODA framework so that it can predict, identify and reuse the data or the models as and when required. This functionality is quite intuitive as we can see the out-of-mode ODA operation reusing the data from the archival source for re-analysis.
    \item \textbf{Experimentation}: The data sources or the models should allow experimentation for enriching the learning of the entire ODA ecosystem, which can be done at periodic intervals by collecting data from various layers. Some of the similar aspects related to this are testing, measuring and benchmarking. This component has been included being part of various studies related to the experimental and numerical analysis of various metrics for forecasting or optimisation in operations (air metrics in \cite{oro2016experimental}, weather events in \cite{DBLP:journals/kbs/Chang17}, etc.).
    \item \textbf{Data Governance \& Security}: This layer is pertinent to identifying the data, characterising it, ensuring that they are of high quality, and improving its value (with archival and cleanup policies as desired). It can help organisations to respond to threats without a significant delay. This activity can be conducted periodically on-demand, and should also include the data collected from relevant layers. This topic has been researched predominantly in the past like by Boris Otto et al. in \cite{DBLP:conf/ecis/Otto11}, a dedicated literature review of data governance in \cite{DBLP:journals/puc/Al-RuitheBH19}, etc.
\end{itemize}
\begin{summarybox}{Proposed ODA Framework}
\textit{
The proposed ODA framework is aligned with the various layers of the distributed computing ecosystem, which makes it an enhanced framework richer than the existing ODA framework. It must also be noted that all four types of data analytics (descriptive, diagnostic, predictive and prescriptive) can be operated within various ODA components at all layers except the data and model layer (Layer 7) of the proposed ODA framework reference architecture, which is more of a logical layer involving provisioning of the data or the underlying models itself.
}
\end{summarybox}
As can be inferred from the above descriptions, our proposed reference ODA framework is holistic in nature covering all identified layers of a computing ecosystem. The lower-level details are abstracted at various layers and could be extended during the modelling phase. The proposed reference architecture is modular to be used at all possible layers of the distributed ecosystem and can be used with a mix of components, strengthening the interoperability of the framework. At the same time, it is flexible enough and can be adjusted to include analytics of other components introduced newly in the system. 
This ODA framework has been designed to cover all relevant forms of analytics discussed earlier, and the scalability aspects for any component can be considered during the time of its implementation, to optimise the respective operational goals. Also, we have tried to reuse the concepts from the massivising distributed processing ecosystem literature and energy-efficient pillars of HPC and have tried to build the reference architecture covering them all. We expect that this idea can be utilised in other domains of computing or even other fields to optimise operational efficiency.

\section{State-of-the-art ODA-enabled monitoring frameworks} \label{state-of-the-art}
Through this section, we answer the second research question (RQ2) by exploring the state-of-the-art ODA frameworks which have been published in the scientific literature. It must be noted that we found several works of literature which implemented advanced monitoring techniques, but there was either limited information available about the ODA-like capabilities of those frameworks or other frameworks discussed later in this section had unique reasons to be studied under the state-of-the-art framework, and thus, they have not been included here. \\
OMNI from NESRC, LBNL, Wintermute/DCDB from LRZ, and ExaMon/ExaMon-X from CINECA are the three most advanced ODA frameworks in our view (amongst the literature scoped as part of the systematic literature process) which we have studied in this literature. Fugaku systems at RIKEN is one of the powerful HPC systems with extensive research on it, and thus have been included here. AutoDiagn framework is unique in itself because this was the only framework targeting big data infrastructure, and evaluated in a cloud environment. Theta HPC system from ANL has been included in this study due to its in-depth analysis of the multi-fidelity HPC system using an end-to-end error log analysis. Summit HPC system is a 200PF pre-exascale system whose power consumption was analysed at various levels, and the authors claimed their study to be the first of its kind at that scale, which is why we included it in our study. The AI-driven EAS work published by the IBM research team is the only work publicly published by one of the IT industry leaders. Kaleidoscope is a near-realt-ime failure detection and diagnosis framework which has been deployed on Blue Waters HPC, which was once considered the largest university-based HPC infrastructure. Apollo is a unique monitoring framework as it monitors the storage subsystem within a distributed computing environment. All of these ten advanced ODA frameworks or monitoring systems/services have been discussed in detail in this section, and thus all references can be found accordingly in their dedicated subsections.\\~\\
Jakobsche et al. presented an Execution Fingerprint Dictionary (EFD) responsible for storing execution fingerprints of system metrics (keys) linked to the application and input size information (values), to help recognise the application, but it is not extensive enough to be studied independently as an ODA framework \cite{DBLP:conf/cluster/JakobscheLCC21}. Versluis et al. proposed a holistic analysis on one of the rare public datasets with an independent and detailed analysis using custom scripts, but not part of any ODA component, thus it was deemed not suitable to be considered as an ODA framework \cite{DBLP:journals/fgcs/VersluisCGLPCUI23, DBLP:journals/corr/abs-2107-11832}. A literature that aims towards the systematic assessment of node failures in production HPC systems reports that the environmental influences are not strongly correlated to the same, and hence it is not considered for deep diving further due to very limited ODA-functionality-related discussion \cite{DBLP:conf/ipps/Das0R21}. LIMITLESS is a lightweight, highly scalable monitoring tool designed to monitor performance metrics from the cluster hosted in large-scale computing infrastructures, but does not talk about the various ODA capabilities \cite{DBLP:journals/mam/CascajoSC22}. An author who worked on miniHPC, an HPC system maintained by the University of Basel's research group for research purposes, provided detailed steps for monitoring the cluster with the help of Prometheus and Grafana; however, that literature does not have much information related to the ODA capabilities enabled in the cluster \cite{kunz2022hpc}. Another author has reflected in detail in their dissertation ranging from the instrumentation to the analysis of energy measurements of HPC systems, but again lacks details about the ODA capabilities \cite{DBLP:phd/dnb/Ilsche20}. One of the thesis works conducted on RWTH Aachen University's HPC cluster "Cluster Aix-la-Chapelle" (CLAIX) in the area of performance monitoring only talks about different approaches to generating job data and evaluating job similarity, but it lacks information about other ODA capabilities in their literature \cite{fischer2020metrics}. In another literature, the authors present a case study related to the efficient use of energy in the Supercomputing Centre of Castile and Leon (SCAYLE), Spain, but that is more focussed on the design of the facility and lacks ODA-capabilities specifics \cite{fernandezenergy}. A team who studied the 10 PFlops leadership-class IBM Blue Gene/Q Mira supercomputer, located at the Argonne Leadership Computing Facility (ALCF) in Chicago (Illinois, USA), highlighted various interesting results in the context of the variability of cooling parameters and the failure rate of the cooling infrastructure, but the discussion was limited only to the cooling facility level \cite{DBLP:conf/hpca/RoyPKARST21}. Another interesting paper presented a novel multi-source, data-driven building energy management (BEM) toolkit mainly targeted to assist building operators address operational deficiencies such as energy use anomalies and inappropriate schedules \cite{markus2021framework}. Teams from LRZ (Leibniz Supercomputing Centre) and the University of Bologna introduced a fault-injection tool 'FINJ' (that enables the automation of complex experiments and helps to reproduce anomalous behaviours in a simple and deterministic way) to inject faults into an in-house experimental HPC system 'Antarex' using a machine learning approach for online fault classification \cite{DBLP:journals/fgcs/NettiKBSBB20}, but the tool itself is not comprehensive enough to be studied as part of an ODA framework. A team explored the role of workflows from the perspective of marshalling and control of urgent workloads, and at the individual HPC machine level studied the benefits of flexibility enabled due to their interoperation \cite{DBLP:conf/supercomputer/BrownNGBPCMFG22}, but that study is not significant enough to be studied as an ODA framework. The Research Computing Center of Lomonosov Moscow State University (RCCMSU) have developed TASC (Tuning Applications for SuperComputers) software which provides detailed information about different aspects of supercomputer behaviour using ODA techniques, but the capability discussed in their literature is limited to workload analysis on the Lomonosov-2 supercomputer \cite{shvets2021endless}.\\~\\
Schwaller et al. designed an architecture that enables flexible, run-time analysis and presentation capabilities for HPC monitoring data in the Sandia National Laboratories (SNL) production system environment \cite{DBLP:conf/cluster/SchwallerTTAB20}, and Aksar et al. later proposed an end-to-end machine learning framework 'E2EWatch' on top of it that diagnoses performance anomalies (job and node-level anomaly diagnosis) \cite{DBLP:conf/europar/AksarSALBEC21}, which is an ODA-enabled system but not included here due to the limited discussion of the framework. Chad Wood proposed SOSflow and Artemis (ML) frameworks for scalable observation, analysis, and tuning for parallel portability of applications in an HPC environment in their PhD dissertation which might also be interesting, but not included here because several established frameworks for HPC systems were prioritised in this study \cite{wood2022scalable}. LASSi framework developed by the ARCHER Centre of Excellence (UK National Supercomputing service) discusses application slowdown and IO usage on the shared Lustre filesystem, and possesses ODA framework capabilities \cite{DBLP:conf/supercomputer/SivalingamR20}. The National Renewable Energy Laboratory (NREL) team has explored physics-aware approaches for data centre facility monitoring like the development of a novel flexible system that identifies and visualises individual metric anomalies and component performance, and introducing a physics-informed drift and anomaly detection models to detect scale build-up in heat-exchangers, which clearly demonstrates another ODA framework with unique capabilities \cite{DBLP:conf/supercomputer/EganPS22}. There is only as much we could explore in a literature study, and thus we have limited ourselves to discussing only ten ODA-enabled monitoring frameworks (which either have comparatively succinct details or which are quite unique in comparison to others, as already discussed at the beginning of this section). Interested readers might want to explore these advanced monitoring frameworks as well.\\
The subsections below are titled in the convention of $<ODA/monitoring framework name>,<Computing system>,<Organisation, Country>$. Some of the computing infrastructures have not distinctly named their ODA framework and/or the monitoring framework, and thus, they are represented by the system level or the research organisation/laboratory in this section.
\begin{figure*}
\centering
\includegraphics[width=0.75\textwidth]{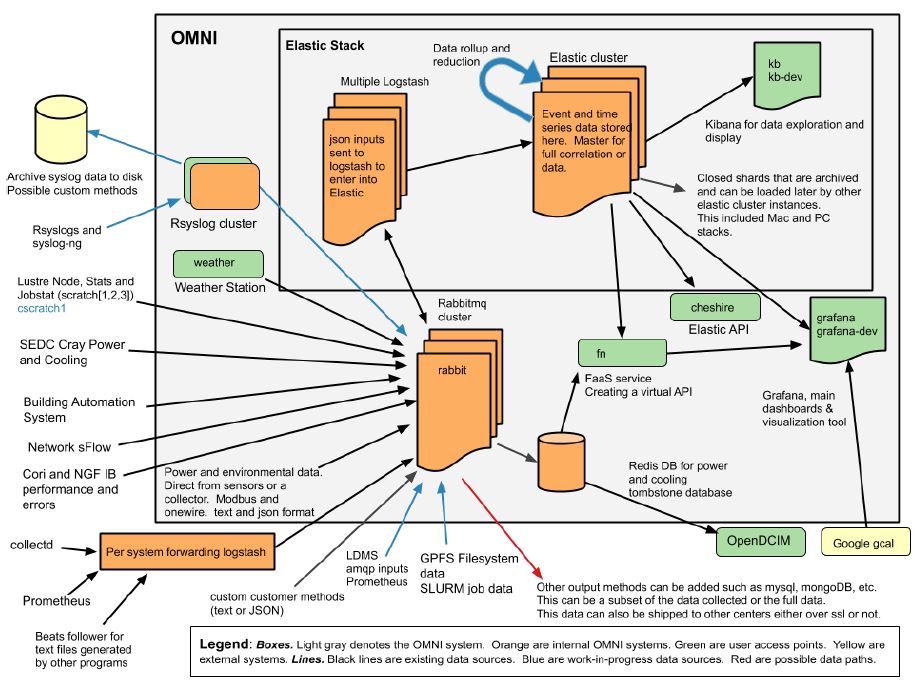}
\caption{OMNI Integrated Operational Data Collection and Analytics Architecture. (Source:~\cite{DBLP:conf/icppw/BautistaRDWK19})}
\label{omni-architecture}
\vspace{-4mm}%Put here to reduce too much white space after your table 
\end{figure*}
%%%%-------------------------------------------------------%%%%
%%%% ----------------  NERSC, LBNL, Berkeley Lab  ---------------- %%%%
%%%%-------------------------------------------------------%%%%
\subsection{Operations Monitoring and Notification Infrastructure (OMNI), National Energy Research Scientific Computing Center (NERSC), Lawrence Berkeley National Laboratory (LBNL) or, Berkeley Lab, USA}
NERSC utilises multiple monitoring systems for their operational monitoring, amongst which OMNI and SkySpark\footnote{SkySpark, SkyFoundry \url{https://skyfoundry.com/product}} are the most prominent ones \cite{DBLP:conf/icppw/BourassaJBCJVS19}. They have documented an effective use of operational data instrumentation, analysis, integration, and archiving, that aims towards effective design, commissioning, and optimisation of power usage effectiveness (PUE) in their HPC facility environments \cite{DBLP:conf/icppw/BautistaRDWK19}. It is worth mentioning that 16 of the NERSC systems appear on the TOP-500 list of fastest computing systems in the world \cite{DBLP:conf/icppw/BautistaRDWK19, bautista2022omni}. Their mission is to provide HPC and compute resources to science users guaranteeing high availability along with high utilisation of the systems to bolster the scientific research for the Office of Science in the U.S. Department of Energy (DOE).
\subsubsection*{Facility \& HPC System Overview}
\begin{figure}
\centering
\includegraphics[width=0.5\textwidth]{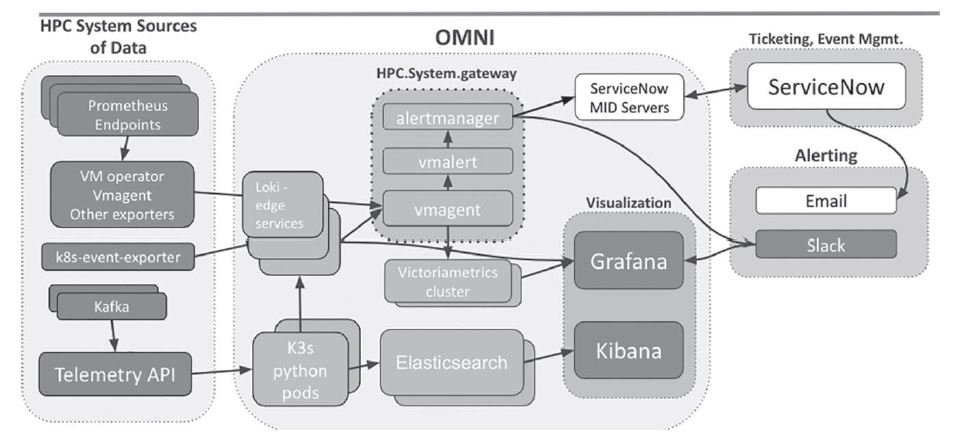}
\caption{Monitoring data pipeline for the Perlmutter HPC system to OMNI (Source:~\cite{bautista2022omni})}
\label{omni-perlmutter-architecture}
\vspace{-4mm}%Put here to reduce too much white space after your table 
\end{figure}
The main HPC systems used at the NERSC data centre (located at Shyh Wang Hall since December 2015) are the hybrid liquid and air-cooled Cori (equipped with 2,388 Intel Xeon Phi "Knight's Landing" (KNL) nodes of 68 cores each, and a large all-flash burst buffer) and Edison Cray (heaving 134,064 compute cores with a peak performance of 2.57 petaflops per second, 357 TBs of memory, and 7.56 PBs of disk) XC Series supercomputers, multiple air-cooled HPC clusters, and a high-performance storage system (HPSS) \cite{DBLP:conf/icppw/BautistaRDWK19}. Each of the Cray XC compute cabinets has "backdoor style" cooling coils to extract waste heat to the facility cooling water (CW) loop. The CW loop is connected to the closed loop side of liquid-to-liquid heat exchangers, which further connects to an open loop tower water (TW) pumping system used to reject heat to the outside air with the help of cooling towers. The cooling of entire computer room air and air-cooled systems is performed by air handling units that use air-side economisers, direct evaporative coolers, and cooling water (CW) cooling coils. The data centre building has an available power capacity of 12.5 MW, with the maximum possible power capacity with upgrades being 42 MW. The newer version Perlmutter HPC system has the architecture shown in Fig. \ref{omni-perlmutter-architecture}.
\subsubsection*{Monitoring and Optimisation tools}
The architecture of OMNI is shown in Fig. \ref{omni-architecture}. OMNI is a multifaceted platform of applications that combines a large amount of HPC and IT systems data with comprehensive cooling and facility systems performance data. The NERSC OMNI system merges the Building Management System (BMS) data (covering an array of rack-level IT sensors) with Cray syslog data, resulting in a real-time, searchable,  and easily visualised ODA system. The resulting Elasticsearch dataset then (at the time of publishing of corresponding literature) archived 25k data points/sec within the general HPSS storage system, which was further planned to expand to 100k data points/sec in late 2019. The graphical visualisation of the real-time data is managed by the Elasticsearch Grafana and Kibana web browser-based user interfaces, whereas various other analysis and visualisation needs are completed using more specific open-source software tools as needed. On the other hand, SkySpark interfaces with building systems (e.g., via BACnet) to collect and analyse available building data. The SkySpark platform gathers data through a live connection to the NERSC BMS, the Elasticsearch database, and an ION power meter database. The NERSC Energy Efficiency (EE) team have built their custom views for performance metrics which are automatically updated with the live OMNI data link.
\subsubsection*{OMNI design and implementation}
OMNI is an integrated operational data collection and analytics infrastructure. The OMNI data collection architecture is shown in Fig. \ref{omni-architecture} along with its diverse data sources. The core system requirements identified for creating OMNI system are scalability (managing the volume of systems and sensor data to provide near-real-time insights), high availability (data collection infrastructure must be always available, even in the presence of some issues at the data centre), maintainability (to apply patches, upgrades, warm hardware swaps, etc. to parts of the system without affecting the flow of data), usability (fast and easy access to the collected data for analytics, visualisation, and monitoring purposes by different actors), and lifetime data retention policy (collecting and saving the data forever theoretically, for statistical modelling and failure prediction from historical data). Accordingly, the OMNI cluster has been made independent of any other system in the facility, which becomes available as soon as the power is turned on (and the last one to be taken down before powering off). OMNI has been implemented with the help of various open-source software, on-premise hardware, and virtualisation technologies, and keeps collecting data as long as there is power to the facility. Virtualisation is implemented using oVirt and Rancher, and data ingestion and storage are managed by Elastic stack in near-real-time. The key component in the elastic stack is Elasticsearch (a distributed JSON-based RESTful search engine) that facilitates real-time ingestion and search within massive amounts of data. The Logstash component manages the server-side data processing pipeline, and forwards them to Elasticsearch for ingestion. The Kibana interface provides a web interface for data discovery, analysis, and visualisation of Elasticsearch data in addition to the monitoring information and management controls for the Elastic stack.\\
Data is collected from various systems (Systems Environment Data Collections (SEDC) data, job information from Slurm scheduler, Lustre parallel file system data, information from the Aries high-speed network, etc.) and sensors (ranging from building management systems like BACnet, Modbus, and various other facility sensors, to power distribution units (PDUs), Uninterruptible Power Supplies (UPSs) etc.) into Elastic search with the help of RabbitMQ cluster (a popular messaging broker supporting multiple messaging protocols and queuing), which sends it further to Logstash. Logstash acts as the local aggregation point by reducing the number of network connections managed by the central logging clusters, and forwards a single connection to the central logger in an encrypted manner (and could also convert collected UDP packets to TCP to ensure reliability of data). Elasticsearch indexes the data for near-real-time retrieval and querying, which could be either queries using its native RESTful APIs or by visualisation and data discovery tools such as Kibana and Grafana.\\
 \textbf{OMNI at the Edge}\cite{bautista2022omni}: The newly upgraded OMNI system is built on a lightweight Kubernetes distribution 'k3s'. It consists of multiple k3s clusters, where each of those is self-contained with master and worker nodes and its own control plane. The monitoring architecture consists of various components like a VictoriaMetrics operator (vm-operator), kube-event, kube-statemetrics, Loki/Promtail, Prometheus node exporter, a Prometheus IPMI exporter for nodes with IPMI, and a Prometheus smartctl exporter for nodes with SATA/NVME drives. In addition to these components, OMNI also uses kubernetes-mixin, a couple of Grafana dashboards, and Prometheus alerts for kubernetes to monitor the cluster visually. The vm-operator manages a VictoriaMetrics agent (vmagent) instance on each of the k3s clusters, which is responsible for efficiently scraping metrics (of various sources including Cray HPC System Environmental Data metrics, facilities information about building power, temperatures, and BACnet data – BAC: Building Automation and Control, data from the lightweight distributed management system (LDMS), and snmp data from nodes on Cori) from Prometheus-compatible exporters in VictoriaMetrics. It also supports buffered collection of metrics at the source in case the VictoriaMetrics backend is not reachable, and shipping them once the backend is made available again. The overall monitoring pipeline design with OMNI for the Perlmutter HPC system based on the edge computing services is shown in Fig. \ref{omni-perlmutter-architecture}. Further details about the new version of the OMNI system can be referenced from the associated literature \cite{bautista2022omni}.
\subsubsection*{Additional information}
The data collected from OMNI helps the NERSC team provide a holistic view of the HPC data centre along with the environmental information which together present the overall status of the data centre. NERSC engineers have been able to gain several insights from OMNI's analysis, with notable ones like the identification of incorrect voltage issues which happened after data centre relocation, and facility power planning for the next-generation Perlmutter HPC system. OMNI's analysis has highlighted a key point about the largest influence of the outside air wet bulb on the overall power usage effectiveness (PUE) of the facility. The HPC systems team have also been able to examine and optimise the queue requirements and job scheduling, thanks to the capabilities of OMNI for analysis, providing a huge scope of improvement in business decisions, capacity planning, facility planning, etc. They mention that the OMNI data collection scale is huge and the data is extremely diverse (and it belongs to NERSC), and the OMNI team is on their own to solve it. Additionally, the team uses open-source software as much as possible, which has helped them save close to \$350k yearly in licensing costs and its use is continued further.
In the newly deployed OMNI architecture (at the Edge), the deployment of gateway and other "local edge" nodes as containerised k3s pods enables monitoring of all compute nodes, services, and applications related to it. As k3s is a lightweight Kubernetes distribution it allows fine-grained scaling and reconfiguration of itself very efficiently.
%%%%-------------------------------------------------------%%%%
%%%% ------------  Wintermute, LRZ  ------------ %%%%
%%%%-------------------------------------------------------%%%%
\subsection{DCDB (Data Center Data Base) Wintermute, Leibniz Supercomputing Centre (LRZ), Germany}
\begin{figure}
\centering
\includegraphics[width=0.5\textwidth]{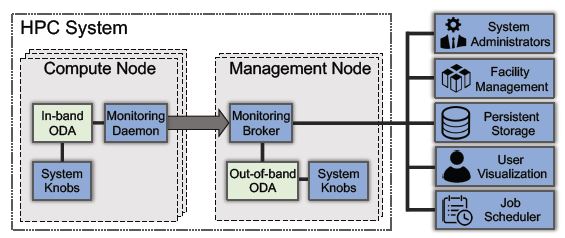}
\caption{Wintermute: A high-level overview of the suggested architecture for an online ODA framework integrated into a monitoring system, showing the main components and actors involved. (Source:~\cite{DBLP:conf/hpdc/NettiMGOTO020})}
\label{wintermute-suggested-architecture}
% \vspace{-4mm}%Put here to reduce too much white space after your table 
\end{figure}
Wintermute is an ODA framework implemented on top of the holistic DCDB monitoring system that offers a large variety of configuration options to satisfy the varying requirements of ODA applications \cite{DBLP:conf/hpdc/NettiMGOTO020}. It enables the analysis of data and granular level control at various levels of an HPC system. The authors analysed the main usage scenarios of ODA on HPC systems, many of which we have reused in our reference architecture \ref{oda-ref-architecture}. They originally suggested the use of two types of data sources (in-band and out-of-band), and the grouping of ODA techniques (into online and on-demand modes), both of them which we have borrowed in our ODA framework reference architecture. They also shared the analysis of the operational requirements while designing their online Wintermute framework for HPC systems, which we extended during our requirements analysis. Their suggested high-level overview of the online ODA framework architecture, which could be integrated into a monitoring system, is shown in Fig. \ref{wintermute-suggested-architecture}. The idea is to implement the ODA framework within an existing monitoring system, where the integration with monitoring daemons in compute nodes facilitate in-band operation and management nodes enable out-of-band operation, where the latter allows its interaction with a monitoring data broker, thus enabling access to streamed cluster-wide data as well as remote persistent storage.
\subsubsection*{Wintermute Architecture}
\begin{figure}
\centering
\includegraphics[width=0.45\textwidth]{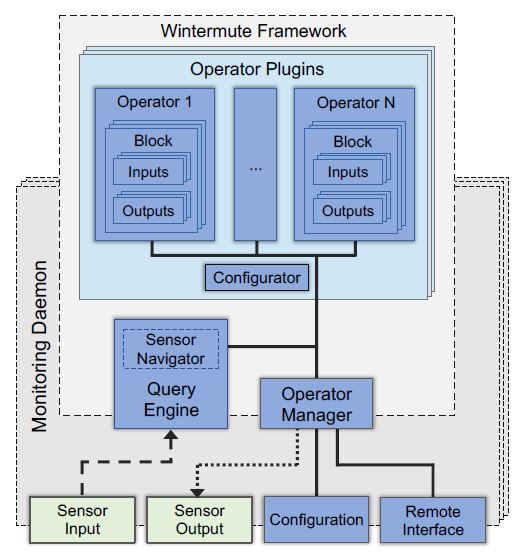}
\caption{Architecture of Wintermute (abstract-level). (Source:~\cite{DBLP:conf/hpdc/NettiMGOTO020})}
\label{wintermute-architecture}
\end{figure}
% \vspace{-3mm}%Put here to reduce too much white space after your table
Wintermute has been architected in a modular way driven by the requirements analysis, and is based on the analysis capabilities being supplied by \textit{operator plugins} that are used to instantiate \textit{operators}. Operators are the computational entities that perform all ODA tasks asynchronously, utilising a flexible local thread pool. These operators work on a set of \textit{blocks}, which are container data structures that represent physical components like compute nodes or racks, or logical entities like user jobs in an HPC system). Each block has a set of \textit{input sensors} which are used for the analysis and a set of \textit{output sensors} which are responsible for storing the results of ODA operation, that are further either consumed by the monitoring system or some other operators (output sensors). A sensor captures some system information (whose reading has a numerical value and a time-stamp) and is defined to be an atomic monitoring entity (e.g., power, temperature, CPU counter, or ODA output). The architecture of Wintermute is presented in Fig. \ref{wintermute-architecture} which shows its interaction with the external components (excluding its integration into a monitoring system). The core components can be briefly summarised as follows:
\begin{itemize}
    \item Operator Manager: It acts as the main interface between Wintermute and the monitoring system, and is responsible for loading requested operator plugins. It acts as a front-end interface responsible for exposing the available actions within the framework (e.g., start, stop or load plugins dynamically, trigger specific actions as signalled by the respective plugin).
    \item Query Engine: It is a singleton component that exposes access to a \textit{sensor navigator} object (which uses the block system to maintain a tree-like representation of the current sensor space) to operator plugins. Its uniform interface allows querying based on sensor name and time-stamp ranges.
    \item Operator Plugins: They are responsible for performing analysis by implementing a specific logic on the input sensor data alone. Job operator plugins are an extension that uses job-related data to produce output pertaining to a specific job. Plugins have two main components:
    \begin{itemize}
        \item Operator: These objects perform the required analysis tasks whenever a computation is invoked for that operator.
        \item Configurator: These components read plugins' configuration files, and instantiate the operators together with their blocks.
    \end{itemize}
    \item Configuration: It is responsible for the initialisation of Wintermute, and grants access to the designated configuration files.
    \item Remote Interface: This is the interface exposed by the monitoring daemon which is used by Wintermute in turn to expose its data retrieval and remote control features.
\end{itemize}
\subsubsection*{Integration with DCDB Monitoring system}
\begin{figure}
\centering
\includegraphics[width=0.5\textwidth]{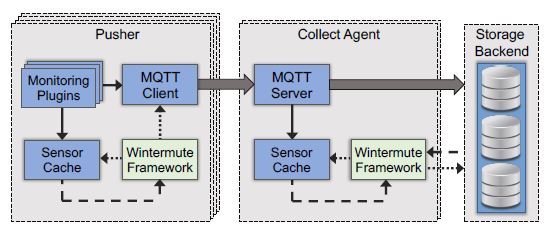}
\caption{High-level overview of the architecture of DCDB, highlighting the Wintermute framework’s integration in components and the data flow (Source:~\cite{DBLP:conf/hpdc/NettiMGOTO020})}
\label{dcdb-architecture}
\vspace{-3mm}%Put here to reduce too much white space after your table 
\end{figure}
DCDB is a holistic monitoring framework for HPC systems which comprises several components enabling it to achieve a distributed and scalable architecture \cite{DBLP:conf/sc/NettiMAGOT019}. Wintermute is implemented in C++11 and is tightly coupled with the DCDB monitoring framework, as shown in the high-level overview of DCDB in Fig. \ref{dcdb-architecture}. \textit{Pushers} use a plugin-based architecture for sampling of sensors on monitored components. \textit{Collect Agents} collect the data sent to them via the MQTT protocol \cite{mqtt31} and forward them to a \textit{storage backend} (implemented using Apache Cassandra). Also, DCDB exposes a RESTful API providing control of each of its components as well as sensor caches for faster in-memory access. The workflow of components involved is briefly discussed below (in the order presented by the respective literature authors):
\begin{itemize}
    \item Operator Location: Operators can be instantiated in both pushers and collect agents by loading the appropriate plugins. Collect Agent location is optimal for system or infrastructure-level analysis because access to the entire system's sensor space is available (or the data could even be queried from the storage backend). Whereas, the operators only have access to locally sampled sensors and their sensor cache date in the pusher locations, thus making it optimal for runtime models which require low latency, data liveness, and horizontal scalability.
    \item Operational Modes: Operators can be configured to work in both online mode (at regular time intervals, resulting in continuous analysis) as well as in on-demand mode (by explicitly invoking the RESTful API for a specific block at scheduled intervals)
    \item Block Management: The blocks of a single operator can either be arranged as \textit{sequential} when all corresponding blocks share a common operator (and are processed sequentially to avoid race conditions) or \textit{parallel} if a distinct operator is created for each block, thus allowing parallel computation to improve scalability.
    \item Analysis Pipelines: A same format is shared by the output data produced by online operators (identical to all other sensor data), which facilitates creating pipelines of other operators' input as output to another operator. This functionality is very helpful to split the computational load between multiple locations, and implementing feedback loops (via control operators at the end of a pipeline) in an HPC system.
\end{itemize}
\subsubsection*{Additional information}
The Wintermute team researched some case studies highlighting its capabilities, and showcasing its flexibility and suitability for large-scale HPC installations (which might have been extremely difficult otherwise). They conducted the experiments on the CooLMUC-3 system which is composed of 148 compute nodes, each of which has a 64-core Intel Xeon Phi 7210-F Knights Landing CPU, 96 GB of RAM and an Intel Omni Path Architecture (OPA) interconnect. DCDB continuously monitors the production system with the help of Pushers that sample data from the Perfevent, SysFS, ProcFS ad OPA plugins and a single Collect Agent that forwards the data to a dedicated storage backend. The existing Wintermute deployment is performing sensor aggregation in the CooLMUC-3 and SUPERMUC-NG\footnote{SuperMUC-NG - \url{https://doku.lrz.de/supermuc-ng-10745965.html}} systems at LRZ. The authors argued that Wintermute has a small resource footprint, despite possessing such strong ODA capabilities. They proposed an anomaly identification mechanism and implemented it in the DCDB Wintermute framework, which was later applied to the production monitoring data collected from the CooLMUC-3 HPC system at LRZ. They made several case studies at different granularity levels that captured different behaviours at different times and across different components, and also flagged suspicious behaviours as anomalies \cite{DBLP:conf/supercomputer/OzerNT020}. Additionally, DCDB has also been deployed on a modular supercomputer prototype \textbf{DEEP-EST} (“DEEP - Extreme Scale Technologies”) and Wintermute is used for an ODA use-case, which is documented in their software support report \cite{moschny2021deep}. Further discussion about the same is out of the scope of this study, but related information can be referenced from the dissertation \cite{DBLP:phd/dnb/Netti22}. Another thesis work evaluating a generic data acquisition module for the DCDB monitoring infrastructure can be referred from \cite{muller2019development}.
%%%%-------------------------------------------------------%%%%
%%%% ------------  Examon & ExaMon-X, CINECA  ------------ %%%%
%%%%-------------------------------------------------------%%%%
\subsection{ExaMon \& ExaMon-X, D.A.V.I.D.E \& Marconi HPC Systems, CINECA, Italy}
\begin{figure}
\centering
\includegraphics[width=0.5\textwidth]{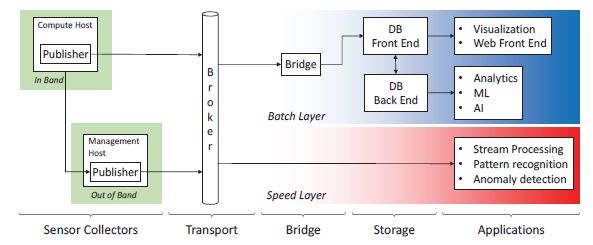}
\caption{ExaMon architecture. (Source:~\cite{DBLP:journals/tpds/BorghesiMMB22})}
\label{examon-architecture}
% \vspace{-4mm}%Put here to reduce too much white space after your table 
\end{figure}
ExaMon (also written interchangeably as Examon, and abbreviated for Exascale Monitoring
framework) is a holistic ODA framework deployed on the CINECA\footnote{The largest Italian computing centre, residing in Bologna} data centre for HPC facility monitoring and maintenance \cite{DBLP:journals/tpds/BorghesiMMB22}. It has been developed for Exascale computing, such that it could handle big data from various heterogeneous sources. ExaMon-X extends ExaMon to meet the predictive maintenance requirements on a real in-production data centre \cite{DBLP:journals/iotj/BorghesiBB23}. The term \textit{predictive maintenance} indicates a set of practices to be implemented on Industrial Internet of Things (IIoT) systems to avoid critical conditions. The advent of IIoT has already led to remarkable steps in the exploitation of several technologies (e.g., big data processing, AI) in various industries, and the complexity, scale, and cost make HPC infrastructure a perfect candidate for IIoT applications. A generic representation of how ExaMon is used in the data centre is shown in Fig. \ref{examon-architecture}. Next, we provide a synthetic summary of the two frameworks along with their implementation on two HPC systems below.
\subsubsection*{Framework overview}
\begin{figure}
\centering
\includegraphics[width=0.5\textwidth]{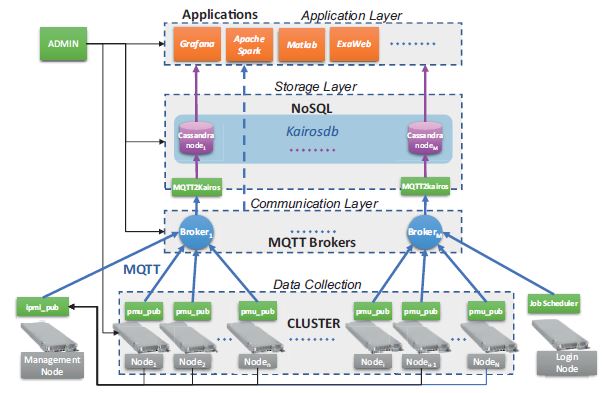}
\caption{The ExaMon monitoring framework. (Source:~\cite{DBLP:conf/icppw/BartoliniBBCLBC19})}
\label{examon-monitoring}
% \vspace{-4mm}%Put here to reduce too much white space after your table 
\end{figure}
ExaMon acts as the block responsible for low-level data collection and acts as the storage backbone, whereas ExaMon-X constitutes the mid- and high-level components used to retrieve and process the stored data and build ML/DL (Machine \& Deep Learning) models for predictive maintenance.
\begin{itemize}
    \item \textbf{ExaMon}: It is a holistic framework for HPC facility monitoring and maintenance, developed for the Exascale era, it can easily handle big data originating from very large-scale computing systems. The lowest level consists of collector components that read data from several sensors across the system and provide them in a standardised format to the upper layers of the stack. There are collectors that have direct access to hardware resources, and other collectors that sample data from applications like batch schedulers and software diagnostic tools.
    The underlying infrastructure uses the MQTT protocol (Message Queuing Telemetry Transport) that utilises the publish-subscribe messaging pattern. It involves the publisher (that sends data on a specific topic), the subscriber (that subscribes to the appropriate topic for requiring specific data), and the broker (that receives data from publishers and makes topics available to the subscribers). The ExaMon collector agents enact the role of publishers. The collected metrics are stored in KairosDB (a distributed and scalable time-series database, built on top of the NoSQL database and Apache Cassandra as back-end). MQTT2Kairos, a special MQTT subscriber, is used to bridge the MQTT protocol and the KairosDB for inserting data. The sampling rate could vary among the different metrics and is dependent on the underlying sensors. ExaMon low-level plugins aggregate the sensor measurements in a 5-second window or the data collected at a higher sampling rate. The architecture of the ExaMon monitoring framework is shown in Fig. \ref{examon-monitoring}. Some of the physical data collected using ExaMon include measurement of hardware components like PMU, IPMI, GPU, I2C, PM-BUS, and the collected data could cover a wide range of sources like CPU load of all the cores in the nodes, CPU clock, instructions per second, memory accesses (bytes written and read), fan speed, the temperature of the room that hosts the system racks, power consumption (at various levels), job dispatcher Slurm etc. Overall, there are a few hundred metrics collected on each computing node, and dozens covering the racks and rooms.
    \item \textbf{ExaMon-X}: It is an IIoT holistic framework used to monitor and maintain the HPC facility. Also developed for the Exascale era, it also has the capability to handle big data from heterogeneous sources. The authors summarise the functionalities of ExaMon-X into eight different subgroups: (i) ExaMon-client (a client interface for data collection and storage back-end, which is the role of ExaMon itself), (ii) ExaMon-dataset (responsible for data management and representation of the merged data), (iii) ExaMon imbalance (addresses the issue related to imbalanced datasets via data augmentation/oversampling or reduction/undersampling), (iv) ExaMon-normalisation (applies standardisation to make the scale of the data uniform), (v) ExaMon-model (responsible for creating the actual ML/DL models), (vi) ExaMon-training (used to train the created model using either online data or historical data sets), (vii) ExaMon-inference (trained model is put into use and inference functionalities are grouped), and (viii) ExaMon-reporting (reports the results via statistical reports, live graphs, or alerts reporting services).
    \item \textbf{Nagios}: Nagios assists ExaMon to collect information about service monitoring and node status (along with using its alarm generators). The system administrators are warned about the potentially critical conditions and need to ensure that the involved computing nodes are removed from production (if the alarm depicts a real problem in nature) and their state is registered as "DOWN+DRAIN". The data sampling frequency configured in Nagios monitoring functionalities is 15 minutes. It provides alerts and historical logs of various monitored components, and can further be extended with custom or predefined plugins targeting specific components.
\end{itemize}
\begin{figure}
\centering
\includegraphics[width=0.5\textwidth]{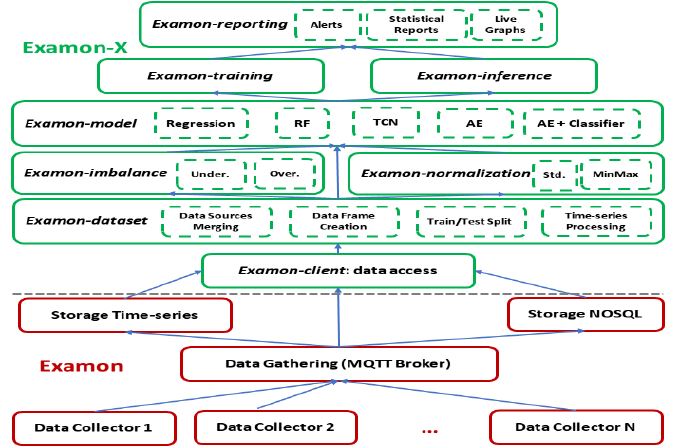}
\caption{ExaMon-X representation in the infrastructure architecture. (Source:~\cite{DBLP:journals/iotj/BorghesiBB23})}
\label{examon-x-representation}
% \vspace{-4mm}%Put here to reduce too much white space after your table 
\end{figure}
Fig. \ref{examon-x-representation} represents the overall infrastructure showcasing the data processing and analytics component.
\subsubsection*{HPC System overview}
Both ExaMon and ExaMon-X have been deployed on two supercomputers hosted at the CINECA computing centre: D.A.V.I.D.E. (where ExaMon was deployed from November 2017 until December 2019, when it was dismissed) and Marconi\footnote{MARCONI, the Tier-0 system - \url{https://www.hpc.cineca.it/hardware/marconi}} (where ExaMon has been operational since mid-2018). D.A.V.I.D.E. HPC system is based on Power architecture, composed of 45 nodes and a total peak performance of 990 TFlops, which has an estimated power consumption of less than 2 Kilowatt per node. Marconi system is the leading supercomputer since 2016 that has 3188 nodes, each of which is equipped with two 24-core Intel Xeon 8160 (Skylake) processors and 196 GB of RAM memory, and has a total peak performance of around 20 PFlops with 17PB of available storage space. These two frameworks collect physical data (measured with sensors) and workload information (retrieved from the SLURM job dispatcher) from both HPC systems with a sampling rate of 5 seconds (where an MQTT packet is dispatched to the broker). On the Marconi system, an additional class of information related to the state of the system and its services is collected using Nagios \cite{DBLP:conf/IEEEscc/KatsarosKG11,mongkolluksamee2010strengths}. The authors use Nagios to annotate collected data for the classification of the samples into normal or anomalous states.
\subsubsection*{Additional Information}
The main focus of the ExaMon \& ExaMon-X team (comprising of members from both CINECA and the University of Bologna, Unibo) has been towards anomaly detection and prediction, as HPC facilities comprise several thousands of parts, and it is critical to identify the faulty, broken or misbehaving components. As this non-trivial task has been performed by system administrators manually in the past, the research of automated methods has been sought resulting in the deployment of ExaMon and ExaMon-X on two supercomputers hosted at CINECA, which performs detection of faulty node states both on the cloud and on the edge. Some of the other use cases highlighting the usage of ExaMon-X functionalities are failure prediction of hard disks, job power prediction, and thermal prediction in HPC systems \cite{DBLP:journals/iotj/BorghesiBB23}. They have dedicated several works of literature researching anomaly detection using ExaMon and ExaMon-X frameworks, some of which are cited here. They proposed an approach to characterise the behaviour of a supercomputing node using the combination of measurements from hardware/software sensors and labels annotated by the system administrators to create a Bayesian classifier called \textit{TrueExplain} \cite{DBLP:conf/supercomputer/MolanBBGB21}. They worked on the development of a new approach for anomaly detection system based on LSTM (Long Short-Term Memory) cells \cite{DBLP:conf/europar/MolanBBB22a, DBLP:conf/cf/MolanBBB22}, that outperformed their old SoA (state-of-the-art) approach which used a semi-supervised Deep Learning model based on autoencoder networks and clustering algorithms \cite{DBLP:conf/europar/MolanBBB22}. They have published literature which proposes a set of rule-based statistical methods (flags) to explore different metrics at the room, system, sub-system, and node level of the HPC cluster and are able to identify the thermal anomalies utilising the telemetry system in the ExaMon database \cite{DBLP:conf/supercomputer/ArdebiliBAB22}.
%%%%-------------------------------------------------------%%%%
%%%% ------------  Fugaku, RIKEN  ------------ %%%%
%%%%-------------------------------------------------------%%%%
\subsection{Fugaku, RIKEN, Japan}
Fugaku\footnote{Fugaku - \url{https://en.wikipedia.org/wiki/Fugaku_(supercomputer)}}, an exascale supercomputer system, was launched officially as the successor to the K computer in March 2021, in consideration of data centre infrastructure overhauling, upgrading various system components and operational data collection/monitoring platform \cite{DBLP:conf/supercomputer/TeraiYMS21}. Additional details about the K computer (which was once ranked as the fastest supercomputer in the Top500 list) like the study of its PUE metric involving analysis of the effect of the operational impact have been discussed in another literature \cite{DBLP:conf/cluster/TeraiSTY20, DBLP:conf/asiasim/FujitaSFNT21} which we do not discuss here, as the K computer system has now been replaced with Fugaku. As a result, they are deploying their operational data collection/monitoring platform based on a three-tier pipeline architecture: (i) the first stage has the HPC system producing various types of log/metric data (including the building management system which collects data from several thousand sensors related to the power supply and cooling equipment), (ii) the second stage aggregates the data into time-series databases for further visualisation purpose, and (iii) the third stage consists of a dashboard that provides an interactive interface for the collected data (of both the HPC system and the data centre infrastructure).
\begin{figure*}
\centering
\includegraphics[width=0.75\textwidth]{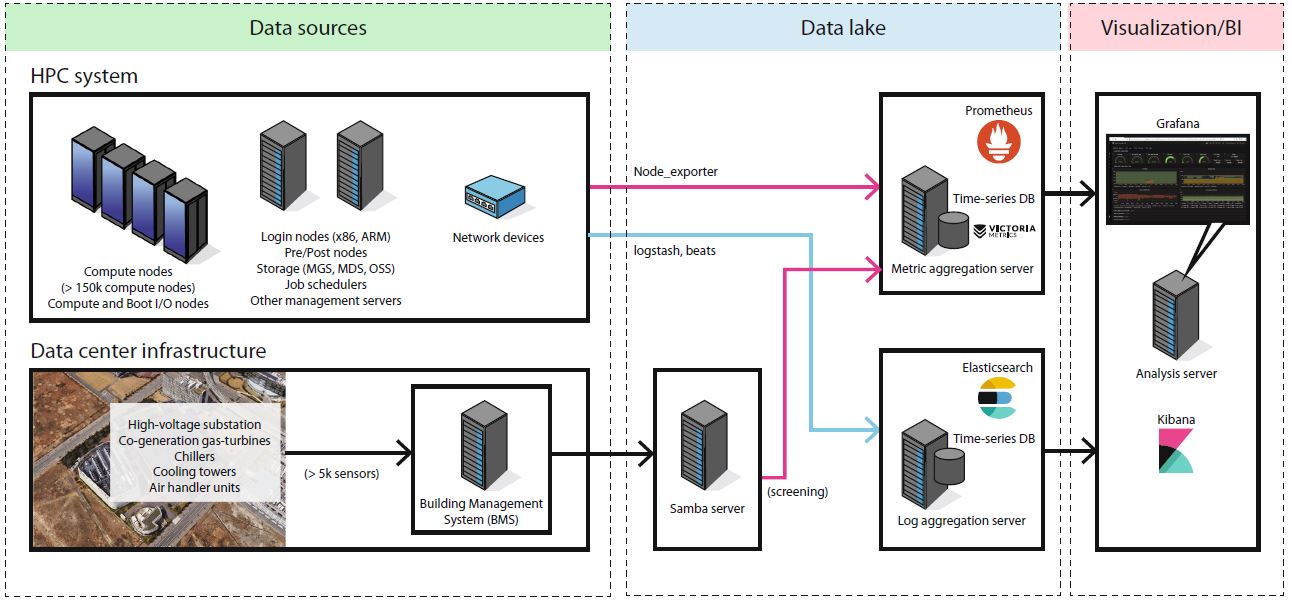}
\caption{Overview of the Fugaku data collection and system monitoring platform. (Source:~\cite{DBLP:conf/supercomputer/TeraiYMS21})}
\label{fugaku-architecture}
% \vspace{-4mm}%Put here to reduce too much white space after your table 
\end{figure*}
\subsubsection*{Environment overview}
Two types of data sources are available in the Fugaku environment: from the HPC system (which includes the compute nodes, boot I/O nodes, peripheral nodes, storage - a Lustre-based parallel file system, and network devices - switches, routers and firewalls), and from the data centre infrastructure (which includes the power supply and cooling equipment).
\begin{itemize}
    \item \textbf{HPC System}: Fugaku consists of 432 racks having 158,976 compute nodes overall, which are connected with the high-speed interconnect (Tofu interconnect D). The Fugaku HPC system employs Arm-based A64FC processors with each microprocessor having four core memory groups (CMGs), each of which integrates 13 cores that share an L2-cache, a memory controller and a ring bus network on a chip. A CMG connects a single second-generation High Bandwidth Memory (HBM2) unit directly and can access external HBM2 units via the ring bus network. The A64FX processor also has a hardware performance monitoring unit (PMU) which is used to measure dynamic program behaviour (e.g., number of CPU cycles, number of instructions, floating-point operations, and cache misses). Fugaku employs a Lustre-based parallel file system having a storage capacity of up to 150 PB available in total. The storage system is further divides into six file volumes.
    \item \textbf{Data Centre Infrastructure}: The supercomputing centre has dedicated facilities for power supply and cooling systems. Fugaku uses two types of energy sources: electricity purchased from a public utility company and energy produced by gas turbine power generators. The Fugaku HPC system is designed in a way which could consume up to 37 MW of electricity, approximately three times more than the K computer.
\end{itemize}
\subsubsection*{Data collection and monitoring system}
The HPC system produces huge log data such as syslog messages, job scheduler logs, parallel file system logs, and audit logs, most of which are recorded by servers. The servers also measure various metrics such as load average, network transfer bytes, memory usage, microprocessors and board temperature, Self-Monitoring, Analysis and Reporting Technology (S.M.A.R.T.), and electric power consumption on power supply units, having PMU metrics also classified within (most of which are time-series event data in a text-based semi-/unstructured format). Some of the network devices also generate syslog messages and specifically generate Simple Network Management Protocol (SNMP) messages. These metrics are time-series data provided in a scalar/vector numerical structured format.
Fig. \ref{fugaku-architecture} presents an overview of the collection and monitoring platform. As mentioned earlier, the first tier is data sources, which include the HPC system and the data centre infrastructure. The second tier (a data lake) stores the log/metric data in various databases (e.g., time-series databases, distributed full-text databases, traditional transactional databases, and a Portable Operating System Interface (POSIX) file system). The metric data is provided by Prometheus\footnote{Prometheus - \url{https://github.com/prometheus}} to a time-series database (TSDB). The log data is generated using Elasticsearch\footnote{Elasticsearch - \url{https://github.com/elastic/elasticsearch}} with logstash\footnote{Logstash - \url{https://github.com/elastic/logstash}}. Prometheus exporter is installed to form the pipeline between the data sources and the data lake tier, serving as a data collection tool for each compute node and the boot I/O nodes, and shipping them to the Prometheus server. Additionally, this exporter is installed on other servers, including the login nodes, pre/post nodes, and the management service (MGS) / metadata service (MDS) / object storage service (OSS) servers. Filebeat, a shipper agent in front of logstash, is installed in boot I/O nodes and the abovementioned servers. \\
Grafana\footnote{Grafana - \url{https://github.com/grafana/grafana}} and Kibana\footnote{Kibana - \url{https://github.com/elastic/kibana}} are used as front-end tools for Prometheus and Elasticsearch respectively, and used to monitor and visualise the time-series data stored in the databases. These tools focus on providing effective monitoring and reporting environments for troubleshooting. The authors also suggest that Jupyter\footnote{Jupyter - \url{https://github.com/jupyter}} might be a more appropriate tool for general users.
\subsubsection*{Additional information}
The authors highlight that it is crucial for the facility operators to monitor the power consumption of the entire centre as modern supercomputers need a complicated power supply and cooling facility management. The facility operators are also responsible for managing co-generation system (CGS) operations to ensure backup electricity is available for the IT equipment and ensure that the chillers are supplied with the necessary steam. From their perspective, workload behaviour is a crucial dependency for the amount of electricity supplied to the IT equipment (but no such control is available to balance the supply and demand between the available electricity and the workloads, up till the time of writing).\\
The combined dashboard (for both the IT equipment information and the facility information) can show the power consumption for the entire centre as well as for the Fugaku.
%%%%-------------------------------------------------------%%%%
%%%% ------------  AutoDiagn, Hadoop  ------------ %%%%
%%%%-------------------------------------------------------%%%%
\subsection{AutoDiagn, Custom Hadoop YARN Cluster, Amazon Web Services(AWS) Cloud}
\begin{figure}
\centering
\includegraphics[width=0.5\textwidth]{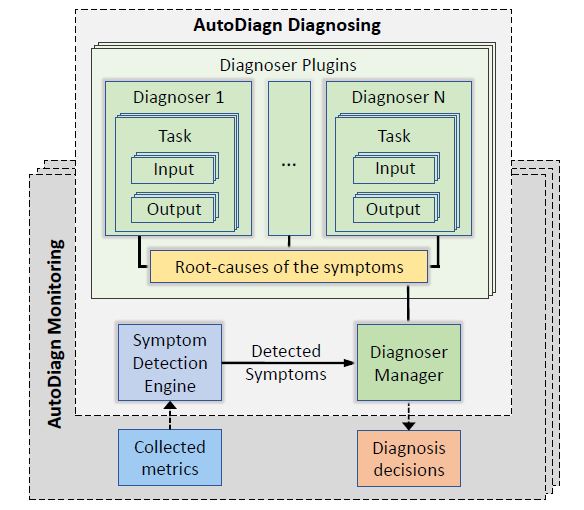}
\caption{The high-level architecture of the AutoDiagn system. (Source:~\cite{DBLP:journals/tc/DemirbagaWNMAGZ22, demirbaga2022real})}
\label{autodiagn}
\vspace{-4mm}%Put here to reduce too much white space after your table 
\end{figure}
\begin{figure*}
\centering
\includegraphics[width=0.85\textwidth]{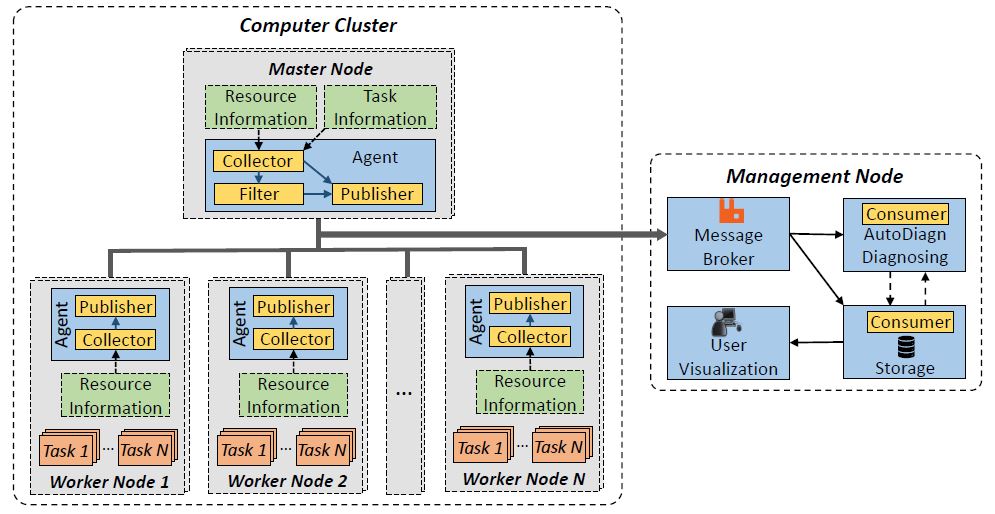}
\caption{The high-level architecture of the AutoDiagn monitoring framework. (Source:~\cite{DBLP:journals/tc/DemirbagaWNMAGZ22, demirbaga2022real})}
\label{autodiagn_monitoring}
\vspace{-4mm}%Put here to reduce too much white space after your table 
\end{figure*}
AutoDiagn is a big-data diagnosing system implemented in Java \cite{DBLP:journals/tc/DemirbagaWNMAGZ22, demirbaga2022real}. It automatically monitors the performance of big data systems and troubleshoots the issues causing performance reduction. The experimental setup uses a Hadoop YARN cluster with over 31 AWS nodes among which there is 1 master node and 30 slaves, each running an Ubuntu Server 18.04 LTS hardware virtual machine (HVM). As part of experimental requirements, the authors built two types of Hadoop clusters where one cluster has the same configuration (4 cores and 16 GB memory) for all nodes, whereas the other cluster has two distinct configurations for split-nodes (25 nodes have 4 cores and 16 GB memory, and 6 nodes have 2 cores and 4 GB memory). The high-level system architecture is shown in Fig. \ref{autodiagn}, describing the implementation of the framework for a big-data cluster.
\subsubsection*{Architecture}
AutoDiagn has two main components: \textit{AutoDiagn Monitoring} and \textit{AutoDiagn Diagnosing}. The monitoring component collects the defined metrics (logs) and feeds them to the diagnosing system in real-time, which are analysed further to troubleshoot the cause of abnormal symptoms, if any. These two components are briefly described below:
\begin{itemize}
    \item AutoDiagn monitoring: The monitoring system is a holistic solution that continuously collects information in a big data cluster. A high-level architecture of the monitoring component is shown in Fig. \ref{autodiagn_monitoring}. The key sub-components are listed below:
    \begin{itemize}
        \item Information Collection: Each compute node has an Agent deployed that provides real-time system information (e.g., usage of computing resources like CPU, memory, network bandwidth, and disk read-write speeds). A master node has an agent that also collects the job and task information in addition to the usage of compute resources. The collected information is supplied to the RabbitMQ\footnote{RabbitMQ, \url{https://www.rabbitmq.com/}} cluster via Publisher at regular time intervals.
        \item Storage: The information collected is a time-series data stored in InfluxDB. The Consumer subscribes to the desired stream topics from RabbitMQ and injects them into the InfluxDB database.
        \item Interaction with AutoDiagn Diagnosing: The symptom-related information is directly forwarded and processed in the AutoDiagn diagnosing system (with the help of a consumer). In the case of symptom detection, the diagnosing system directly queries InfluxDB for root-cause analysis, and the analysis results are stored back in the database.
        \item User visualisation: The authors use InfluxDB's client libraries to develop their RESTful APIs allowing users to query various information related to resource utilisation, job and task status, or the root cause of any performance reduction.
    \end{itemize}
    \item AutoDiagn diagnosing: It is an event-based diagnosing system which is responsible for analysing the metrics fed by the monitoring system. The key sub-components are listed below:
    \begin{itemize}
        \item Symptom Detection Engine: This engine follows a microservices architecture and subscribes to a set of metrics from the real-time streaming system (also, newer symptom detection techniques can be added easily).
        \item Diagnoser Manager: The diagnosed manager is the core entity that is responsible for identifying and selecting the right diagnosers to perform the root cause analysis. It has been developed as a front-end component that is triggered by various detected symptoms (events) using a RESTful API. It could compose a set of diagnosers for the diagnosing jobs that might require the cooperation of different diagnosers.
        \item Diagnoser Plugins: These plugins contain a set of diagnosers, where a diagnoser implements a specific logic to perform root-cause analysis of a symptom. These diagnosers perform an analysis querying the respective metrics that are stored in a time-series database.
    \end{itemize}
\end{itemize}
\subsubsection*{Additional information}
The authors have discussed some examples of big data root cause applications. They discuss the ways to detect the outlier tasks taking longer than normal to finish (when compared to similar tasks), which might prevent the subsequent tasks to make progress. There are various types of tasks in a big data system, including CPU-intensive tasks, IO-intensive tasks, and memory-intensive tasks, and the compute resource distribution could result in inefficient resource utilisation, which can also be detected using AutoDiagn. Each symptom detector and its diagnoser can be executed in parallel by partitioning the input data. AutoDiagn has a reliable centralised design for data collection due to the high availability of the RabbitMQ cluster which also takes care of the scalability. The leading author Umit also discusses BigPerf, a probabilistic performance diagnosis and prediction framework for cloud-based big data systems, which is a novel Bayesian system that diagnoses and predicts the root causes of performance degradation of Hadoop applications, in their dissertation \cite{demirbaga2022real}. Interested readers might want to explore the corresponding literature for additional insights about monitoring and diagnostics in big data systems, as we have not covered the same here for brevity.
%%%%-------------------------------------------------------%%%%
%%%% ------------  SEDC, ALCF, USA  ------------ %%%%
%%%%-------------------------------------------------------%%%%
\subsection{Theta HPC System, Argonne National Laboratory (ANL), USA}
\begin{figure*}
\centering
\includegraphics[width=0.95\textwidth]{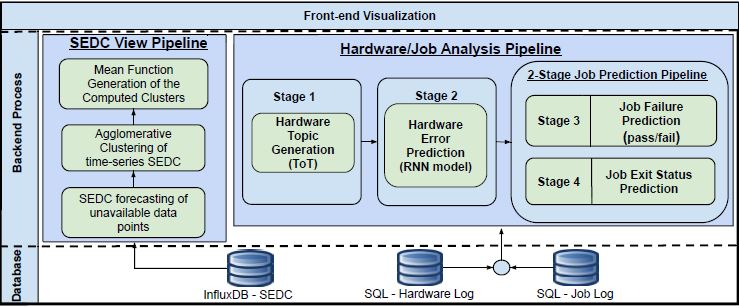}
\caption{Architectural overview of SEDC system for analysing multi-fidelity HPC system logs. (Source:~\cite{DBLP:conf/ccgrid/ShilpikaLESVPM22})}
\label{sedc-anl-architecture}
% \vspace{-4mm}%Put here to reduce too much white space after your table 
\end{figure*}
Theta is an 11.7 Petaflops Cray XC40 supercomputer system located at Argonne National Laboratory (ANL), USA having Intel Xeon Phi based compute nodes. The corresponding literature discusses an in-depth analysis of multi-fidelity HPC systems on all Cray systems, by using an end-to-end error log analysis system involving the System Environment Data Collections (SEDC) tool \cite{DBLP:conf/ccgrid/ShilpikaLESVPM22}. They extend their previous work (MELA) which was a visual analysis tool for analysing diverse log data handling scalability and interactivity, but it lacked error prediction capabilities. The authors analyse three log data types (environment/SEDC logs, hardware-error logs, and job logs) and gather insights from their interaction amongst each other at various temporal and spatial resolutions. Using the same, they build a model to predict job failure and job failure exit status. They aim to help system administrators with enriched information about the environment and provide job failure analysis to proactively perform system maintenance for reducing downtime.
\subsubsection*{System Overview}
The nodes of Theta are interconnected in a Dragonfly topology using an Aries interconnect. The various log types are essentially raw data collected from sensors, which are located at various spatial resolutions, at various temporal resolutions (thus called multi-fidelity large-scale data). The SEDC log data is recorded every 10-30 seconds, which results in the generation of a dataset of size gigabytes (GB) to terabytes (TB) every few weeks. The hardware error log constitutes data from various control systems linked with each other within the subsystem, which contributes to duplicate events. The hardware error log data range in tens of GB, and the job log data is usually hundreds of MB for a year. The system can be simply understood to be comprised of a database storage with a machine-learning back-end pipeline, that has a simple front-end visualisation. The back-end pipeline is split into an SEDC view pipeline and a hardware/job analysis pipeline, which are discussed briefly in the next sub-section.\\
An overview of the system responsible for analysing multi-fidelity HPC system logs is presented in Fig. \ref{sedc-anl-architecture}. This system operates on a machine with two 6-core, 2.4-GHz Intel E5–2620v3 processors having Intel Haswell architecture and 384 GB memory per node. The size of the datasets of SEDC/environment is 3-4 GB per day, the hardware system is 1-2 GB per month, and the job log is a few hundred MB per year. The system is implemented using Python 3.7, has a Flask server back-end, and a D3 visualisation front-end. The storage for SEDC logs is done in InfluxDB (which is optimised for time-series data), whereas the job and hardware error logs are stored in an SQL database.
\subsubsection*{System Processing}
Each node of the supercomputer has multiple SEDC measurements, which include records related to power supplies, processors, memory, and fan reading from the sensors. There could also be missing entries for the SEDC entries resulting due to the possibility of sensor failure, which is handled using a time-series forecasting method. There are approximately 700 SEDC readings with slots in the Cray XC40 supercomputer, with each slot comprising four nodes. The node reports hardware errors, and the jobs routinely run on thousands of nodes for up to 24 hours. If a job failure prediction is made, the user can select the job and the system processes the \textit{job, SEDC, and hardware error} logs for the time frame within which the job is scheduled for execution. The hardware error log is categorised as INFO messages (reporting the progress of an application), WARN messages (highlighting possibilities of hampering the normal operation), or FATAL messages (highlighting several errors that could result in application failure or aborting). And, the job log data specifically gives information about system usage by the applications. The hardware error log and job log datasets are mainly used to identify recurring patterns of errors that might result in failure.\\
Considering the back-end pipeline presented in the system architecture, it is further split into the SEDC view pipeline and the hardware/job analysis pipeline (shown in Fig. \ref{sedc-anl-architecture}). The SEDC view pipeline is responsible for processing and analysis of the environment log data to identify patterns related to anomalous HPC system behaviour. Whereas, the hardware/job analysis pipeline is split into four stages: (i) hardware topic generation, (ii) hardware error prediction, (iii) job failure prediction, and (iv) job exit status prediction (stages 3-4 form the job prediction pipeline). Stage 1 groups related hardware errors into topics for easier interpretation of data. Stage 2 utilises past hardware errors to predict the most likely future hardware errors on the respective component where a job is executed. Stage 3 is responsible for predicting if a job will fail, which then results in stage 4 handling the respective exit status. The overall system back-end has machine-learning pipelines for analysing and processing each type of log data due to large variations in their nature (from text data to numerical data). The machine-learning-related information about the system is not discussed here for brevity, but interested readers can follow the literature for further information.
\subsubsection*{Additional Information}
The ANL team analysed 91,217 jobs in the year 2018-2019, where the error in the dataset was categorised into informational, memory, transaction, and transient errors. There were some nice conclusions as part of the model analysis in the hardware/job analysis pipeline like capturing jobs with built-in fault tolerance was challenging, jobs with custom exit status to be handled separately, data inclusion from multiple sources (with cabinet/cage/slot information) improved the prediction accuracy, and environment logs could help in pinpointing anomalous system behaviours. Overall, they have developed a predictive analysis system for the comprehensive investigation of various types of logs. Additionally, the visual analytics tool analyses the job logs and provides nice insights while corresponding with other logs at various temporal and spatial resolutions.
%%%%-------------------------------------------------------%%%%
%%%% ------------  Summit, OLCF  ------------ %%%%
%%%%-------------------------------------------------------%%%%
\subsection{Summit HPC supercomputer, Oak Ridge Leadership Computing Facility (OLCF), USA}
\begin{figure}
\centering
\includegraphics[width=0.5\textwidth]{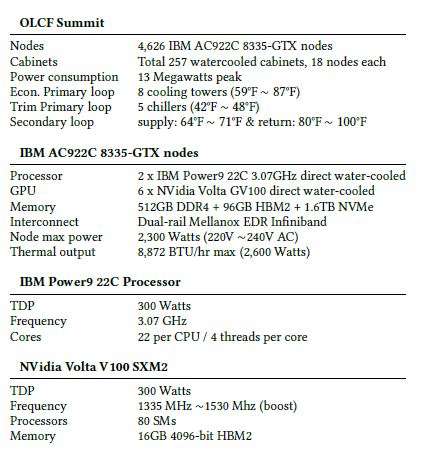}
\caption{Summit system specification. (Source:~\cite{DBLP:conf/sc/ShinOKEW21})}
\label{summit-olcf-specification}
% \vspace{-4mm}%Put here to reduce too much white space after your table 
\end{figure}
\begin{figure*}
\centering
\includegraphics[width=0.95\textwidth]{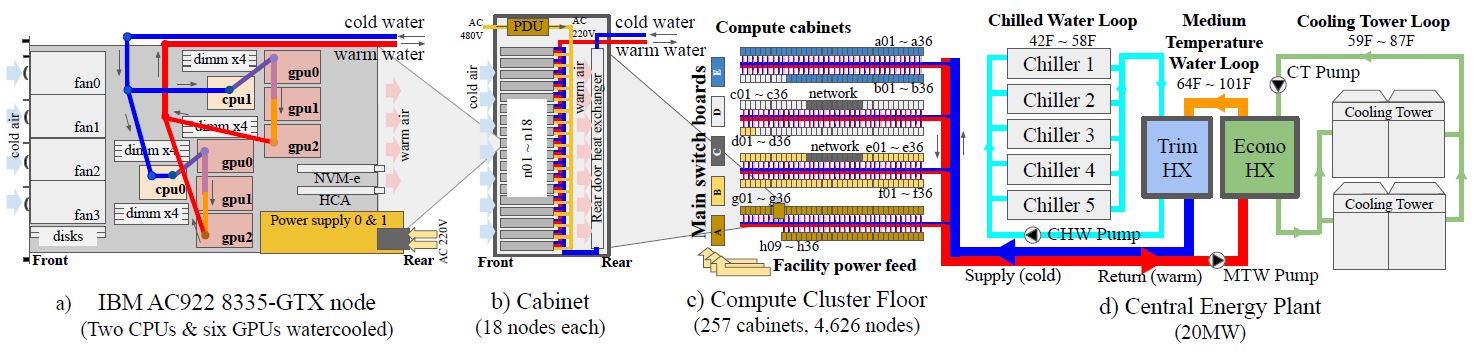}
\caption{Architectural overview of the Summit system at Oak Ridge Leadership Computing Facility. (Source:~\cite{DBLP:conf/sc/ShinOKEW21})}
\label{summit-olcf-architecture}
% \vspace{-4mm}%Put here to reduce too much white space after your table 
\end{figure*}
Summit HPC data centre has a 200PF (theoretical peak) pre-exascale system at the Oak Ridge Leadership Computing Facility, which was ranked No. 2 on the Nov. 2020 edition of the Top500 list of supercomputers \cite{DBLP:conf/sc/ShinOKEW21}. The authors mention that the OLCF team conducted an analysis of power consumption at various levels (component-level, node-level, and system-level) of the Summit supercomputer from all of the 4,626 compute nodes, each of which had over 100 metrics sampled at 1Hz frequency throughout 2020. They investigated the power characteristics and energy efficiency of over 840k jobs, 250k GPU failure logs, and power and cooling supply information from the facility for further operational insights, which they claim to be the first of its kind at that scale. The corresponding literature's contribution includes HPC data collection and workload power characterisation, cross-cutting interactions throughout the data centre beyond the traditional boundaries, and reliability and variability study of a large-scale dense GPU deployment.
\subsubsection*{Summit Architecture}
Summit caters to workloads consisting of full-system jobs related to research areas of national importance such as advanced scientific computing, basic energy, biological and environmental, fusion energy, high-energy physics, and nuclear physics. It involves Power9 CPUs, NVidia Volta V100 GPUs, and Mellanox Enhanced Data Rate (EDR) InfiniBand (IB) network technologies. Each of the 4,626 IBM AC922 nodes is powered by 2 CPUs and 6 GPUs. The system specification of Summit, OLCF is concisely presented in Fig. \ref{summit-olcf-specification}. Its peak power consumption is 13 MW being supported by a 20 MW facility (with its architectural overview shown in Fig. \ref{summit-olcf-architecture}), and was ranked 11th on the Green500 list (during the time of writing of their literature) with 14.719 GFlops/watts performance. Summit utilises medium temperature water (MTW) in the secondary loop to maximise cooling efficiency, which helps minimise chilled water use by enabling cooling towers based on evaporative cooling in the primary loop during advantageous weather conditions (Econo HX, shown in the right in Fig. \ref{summit-olcf-architecture}(d)). The facility uses chilled water only when the cooling towers cannot remove the heat sufficiently, especially during the hot and humid summer months, which ensures that the overall use of chilled water for only about 20\% of the year. Fig. \ref{summit-olcf-architecture} shows the architecture of the Summit system at OLCF depicting information about the cluster, hardware cabinets, compute cluster floor, and the central energy plant.
\begin{figure}
\centering
\includegraphics[width=0.5\textwidth]{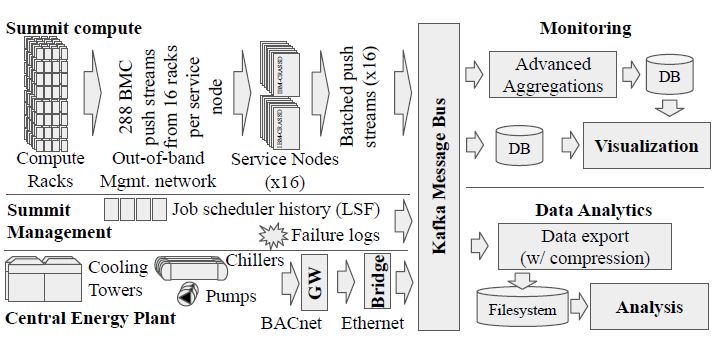}
\caption{Overview of the telemetry system for Summit's power and energy. (Source:~\cite{DBLP:conf/sc/ShinOKEW21})}
\label{summit-telemetry-system}
% \vspace{-4mm}%Put here to reduce too much white space after your table 
\end{figure}
\subsubsection*{Summit telemetry system}
The telemetry system supporting MTW operations is shown in Fig. \ref{summit-telemetry-system}. The electrical and mechanical data across the data-centre-level is aggregated with power and temperature data emitted from individual nodes, which is processed, summarised and rendered to the engineers in near-real-time. The facility cross-checks the important data-centre control parameters like MTW supply and return temperature and its flow with the temperature distribution summary of the HPC platform (27,756 GPUs and 9,252 CPUs). The telemetry system relies on out-of-band streams that are pushed at a 1Hz data rate from the baseboard management controllers (BMCs) of each compute node. The per-node \& per-component  power and temperature sensor changes at 460k metrics/sec are propagated to the point-of-analysis with an average 4.1-second delay. The primary use of such streams is to verify the data-centre state on a near-real-time basis, with the main focus on the long-term analysis of those archived streams. With the help of several lossless data compression methods throughout the telemetry data pipeline, the data stream could be compressed to a manageable 1 MB/s from those aggregated 460k metrics per second, the accumulation of which for an entire year cost 8.5TB of disk space per year.
They used Dask \cite{dask2016}, a parallel data analytics and computing tool, to aggregate the 1Hz data stream from each node into a manageable size and form, by coarsening the data into a 10-second window (but avoiding any information loss by storing statistical information of the samples). These data streams were further collapsed into a cluster-level time-series using different aggregation methods implemented by the team (depending on the analysis). As the cluster-level power consumption measurement used in this work is an aggregation of the per-node measurement, there is the possibility of various errors being induced in the measurement (e.g., variation of calculation at the huge number of sensors per node, timestamping the payloads at the aggregation point after an average 2.5-second delay). Thus, they used the 10-second window coarsening (storing min, max, mean, and standard deviation of the samples) before performing the summarisation at the cluster level.
\subsubsection*{Additional information}
The authors studied various ways to analyse the power and energy consumption at various levels - HPC system and the supporting facility, job context (HPC applications), and component-level (CPU and GPU power usage profile). They also analysed the response of the Summit system to intense change in the workloads (e.g., power and thermal response, energy efficiency value measured using PUE). They investigated the long-term impact of the overall thermal state resulting from the dynamics of power consumption (e.g., impact on GPUs at high temperatures, variability in power consumption at node-level due to compute-intense jobs at certain time intervals).
%%%%-------------------------------------------------------%%%%
%%%% ------------  EAS, IBM Research, UK  ------------ %%%%
%%%%-------------------------------------------------------%%%%
\subsection{AI-driven Energy Aware Scheduling (EAS), IBM Research, STFC Daresbury Laboratory, UK}
The IBM research team based in STFC Daresbury Laboratory, UK published a literature presenting an AI-driven holistic approach to energy and power management in data centres, which they describe as Energy Aware Scheduling (EAS) \cite{DBLP:conf/supercomputer/TraceyHSE20}. The aim of EAS is to look at energy efficiency across both hardware and software stacks of the IT infrastructure, including data centre(s), servers, network, cooling, IoT and Edge devices, to software stack ranging from firmware through to the OS, applications and workload managers. To achieve the same, EAS uses ML and AI methods for performance and power consumption modelling, and software-hardware co-design for implementing various energy/power-aware scheduling policies at different levels of the infrastructure.
\begin{figure}
\centering
\includegraphics[width=0.5\textwidth]{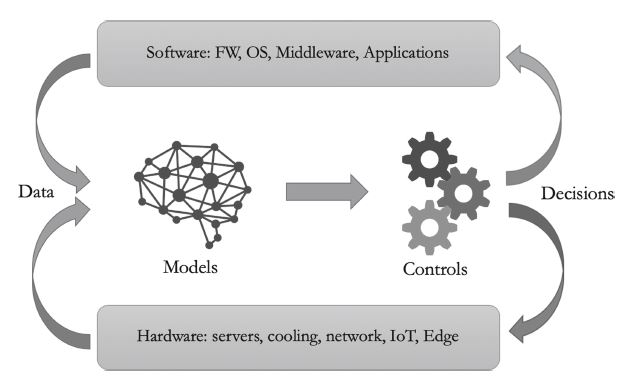}
\caption{EAS concept: data continuously collected across hardware and software stacks and fed to the AI-based model for performance and power consumption predictions, which are then used by scheduling components to send control decisions back to the respective stacks. (Source:~\cite{DBLP:conf/supercomputer/TraceyHSE20})}
\label{ibm-eas-concept}
% \vspace{-4mm}%Put here to reduce too much white space after your table 
\end{figure}
Fig. \ref{ibm-eas-concept} broadly represents the main components of the Holistic EAS, where data is collected from a broad range of hardware devices and software components and fed into an AI component which generates some optimal decisions to meet required criteria, which are then sent back to respective hardware and software system components to re-implement.
\begin{figure*}
\centering
\includegraphics[width=0.75\textwidth]{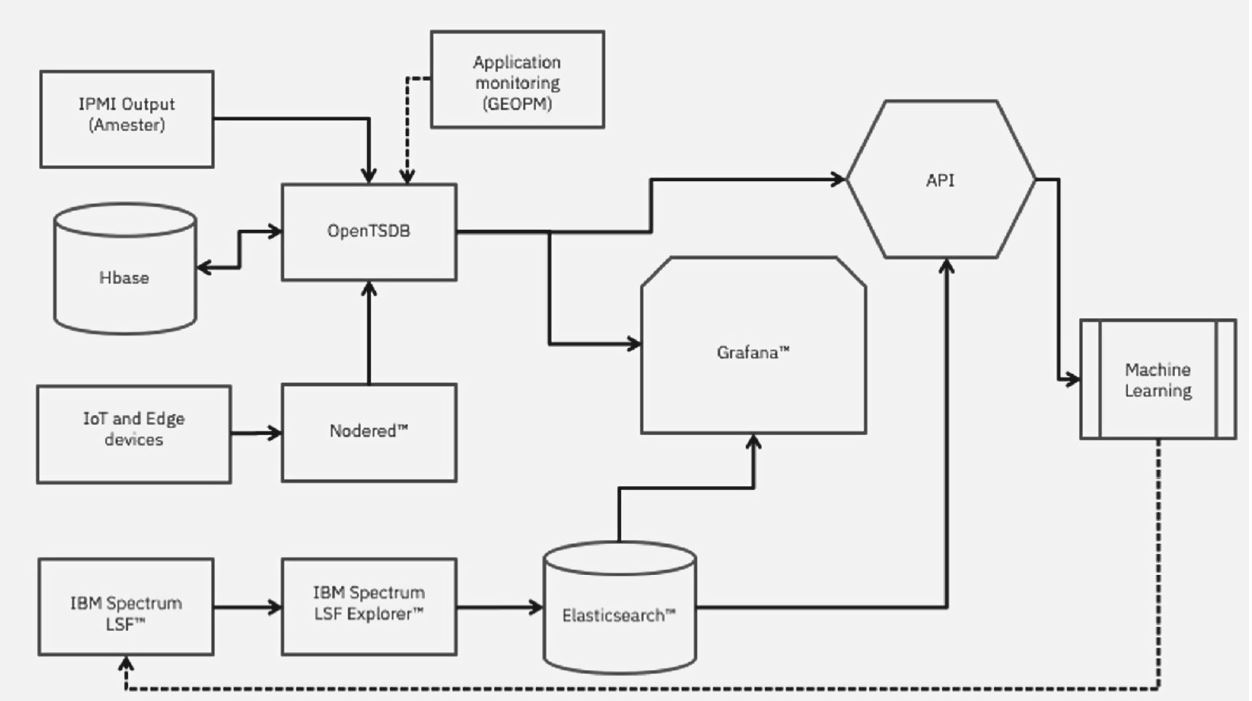}
\caption{Overview of the intelligent data collection framework architecture. (Source:~\cite{DBLP:conf/supercomputer/TraceyHSE20})}
\label{ibm-intelligent-data-collection}
% \vspace{-4mm}%Put here to reduce too much white space after your table 
\end{figure*}
\subsubsection*{Data Collection Framework}
Fig. \ref{ibm-intelligent-data-collection} represents the architecture used for data collection, where each component is housed in a Docker container resulting in numerous benefits like easier rolling upgrades independently from other components in the system, easy swapping of databases or graphing components, and easier portability to a different set of nodes (not necessarily IBM POWER servers), if required. We briefly discuss the core components of the Data Collection Framework below:
\begin{itemize}
    \item \textbf{Monitoring Tools}: The requirement for selecting a monitoring tool is to have as minimal impact on the cluster as possible, and thus clients running on the nodes are avoided for this reason. Amester tool is used to gather data from the Baseboard Management Controller (BMC) port on IBM POWER systems. BMC is a specialised service processor responsible for monitoring the physical state of the running system, and the BMC port connects to it, which helps Amester to gather out-of-band results as it will have zero impact on the running of the system and its resources. Another monitoring tool used is IBM Spectrum LSF Explorer because it integrates easily with the IBM Spectrum LSF, which is used as the workload manager. The authors are convinced that such a monitoring system could easily scale for exascale size and allows easier upgrade of the current system.
    \item \textbf{Metrics}: The requirement for metrics was driven by the need for them to be useful for general monitoring and ML, and be available on as many systems as possible. Hardware metrics considered for collection are fan power, CPU power and GPU power, and they are recorded every 250ms (the fastest the servers could read from the sensors). IBM Spectrum LSF Explorer is used for the collection of workload information which collects over 700 metrics every 10 seconds. Since this is an early analysis phase for the team, they are collecting all metrics and they can eventually identify the most useful ones for monitoring and feeding them into the ML systems. IoT-related metrics are sampled at an average rate of 5 seconds.
    \item \textbf{Databases}: Two separate databases are used given the nature of the system, to support Grafna. Elasticsearch is used to collect data from LSF, as it comes packaged with LSF Explorer and is already optimised to work with the high-frequency data sent from the former. The second database chosen is OpenTSDB, which is built on top of HBase.
    \item \textbf{Visualisation and API}: The visualisation is mainly centred around Grafana, which is very useful to generate graphs and dashboards to visualise the initial data and also allows to create live rotating dashboards. The data can also be accessed via API as the developers created a Python library (due to them using other ML libraries like Keras, PyTorch, and TensorFlow, which are all Python-based).
\end{itemize}
\subsubsection*{Experimental Setup}
The team conducted this research on a cluster of IBM POWER8/9 servers (each having 2 CPUs per node, 10/12 cores per CPU, Nominal frequency 3.6–4.2 GHz, 512/1024GB RAM, 4 NVIDIA Tesla P100/V100 GPUs per node, 16 GB/GPU, NVIDIA NVLink) with IBM Spectrum Scale storage subsystem, 100 Gb/s EDR Infiniband and 10GigE networks running RHEL 7.x operating system. This cluster is internally used as a research system in their lab for developing and testing new projects for IBM clients and for internal use, and the researchers submit jobs to it as well for real-world assessment. For gathering temperature and humidity data, ESP8266 microcontrollers are used at a room level with DHT22 sensors. Additionally, there are existing monitoring infrastructures consisting of room and rack level IoT devices, Trendpoint Enkapsis Power Management Devices, and CSIM Babel Buster edge devices (details of them can be referred from the corresponding literature) used within the room. These devices are responsible for gathering data from different parts of the data centre environment including power usage at the breaker, room temperature and current information about the main room cooling systems. Devices like Babel Buster have the advantage of communicating on multiple protocols to various IoT devices in the room, which processes this data and presents key metrics via Simple Network Management Protocol (SNMP) protocol to the Data Collection Framework.
\subsubsection*{Additional information}
There are two machine learning components of the framework that is responsible for analysing and predicting the energy consumption of the system, namely clustering and predictive management. Clustering is used to classify events according to their power consumption, fan speed, and CPU/GPU utilisation (based on k-mean classification). The authors propose to use Long Short Term Memory(LSTM) network for predictive management in this literature, which is a type of Recurrent Neural Network highly suitable for time series data. The underlying network training is based on the fan power and CPU utilisation of the cluster in three previous time steps, and output from the network is responsible for providing predictions for the cluster. The visualisations provide the temperature of the water flow in the data centre cooling systems, time series of rack power, fan power, CPU utilisation, and accordingly clustering analysis and ML prediction are done.
%%%%-------------------------------------------------------%%%%
%%%% ------------  Kaleidoscope, NCSA, USA  ------------ %%%%
%%%%-------------------------------------------------------%%%%
\subsection{Kaleidoscope, Blue Waters HPC supercomputer, National Center for Supercomputing Applications (NCSA), University of Illinois Urbana-Champaign, USA}
\begin{figure*}
\centering
\includegraphics[width=0.8\textwidth]{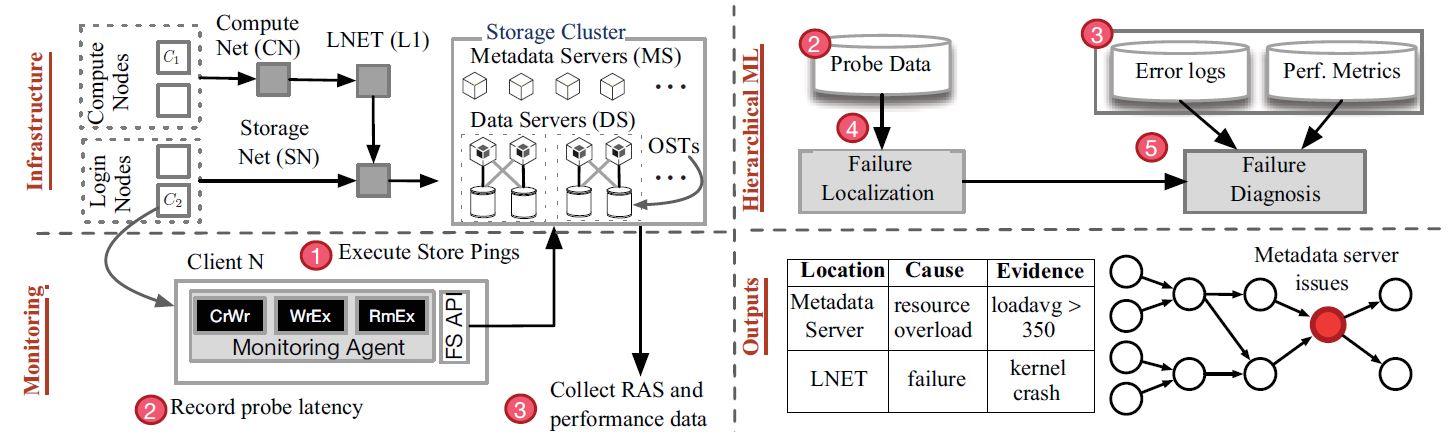}
\caption{An overview of Kaleidoscope design and implementation. (Source:~\cite{DBLP:conf/sc/JhaCBXESKI20})}
\label{kaleidoscope-design}
% \vspace{-4mm}%Put here to reduce too much white space after your table 
\end{figure*}
Kaleidoscope is a near-real-time failure detection and diagnosis framework with hierarchical domain-guided machine learning models responsible for identifying the failing components, the corresponding failure mode, and indicating the most likely cause of the failure (within one minute of failure occurrence) \cite{DBLP:conf/sc/JhaCBXESKI20}. It has been deployed on the Blue Waters supercomputer (which was once the largest university-based HPC system in terms of compute and storage nodes) and evaluated with more than two years of production telemetry data. The authors justify that the telemetry data has several problems like they are noisy due to asynchronous collection, failure propagation, and non-determinism in the system. Thus, instead of getting analysed in isolation, Kaleidoscope utilises ML methods to accurately estimate the system state. The authors focus on the analysis of high-performance storage systems after they found that those have the most failures resulting in lost compute hours.
\subsubsection*{Blue Waters storage subsystem design}
Blue Waters uses the Cray Sonexion storage subsystem, which is designed for large-scale HPC systems with I/O intensive workloads (e.g., machine learning, large simulations). Blue Waters's deployment of the Cray Sonexion involves 6 management servers, 6 metadata servers (MS), 420 data servers (DS), and 582 I/O load-balancers (LNET nodes). The storage servers are connected using an internal Infiniband network (storage network). LNET nodes connect 28,000+ computing nodes on the Cray Gemini interconnection network (compute network) to the storage network. Cray Sonexion uses the Lustre parallel distributed file system to manage 36PB of disk space across 17,820 HDD disk devices. The disks (referred to as object storage devices - OSDs) are arranged in a grid RAID, and each storage server is attached to one or more OSDs. Data servers are configured as active-active pair in Lustre to achieve load balancing and high availability, whereas metadata servers are configured as active-passive pair for the connected OSDs. The computing nodes are configured without disk, and all I/O operations are routed to the LNET nodes (via RPC), which forward the request to the storage servers. Some of the common challenges involving failure scenarios that make it hard to identify the failing component or the failure node are dataset heterogeneity and their fusion, data labelling and anticipation based on rare failures, measurement uncertainty, noise and propagation effects, timeliness of analytics for proper observability, 
\subsubsection*{Framework overview}
The design of the Kaleidoscope framework is shown in Fig. \ref{kaleidoscope-design}, and various components are separated by dotted lines into 4 parts. The upper left part corresponds to the "Infrastructure" part representing a simplified diagram of the Blue Water storage system, the lower left "Monitoring" part shows the telemetry data collected from the system, the upper right part "Hierarchical ML" represented the interconnected ML models responsible for providing the failure localisation and diagnosis capabilities, and the lower right part "Outputs" gives an interpretable set of results and dashboards which are used by the system managers. The challenges mentioned in the previous subsection are addressed by Kaleidoscope using various approaches such as using telemetry data from both the system and application views to enable fusion and comprehensive analysis, using hierarchical probabilistic ML models for analysis at different granularities and time scales, using unsupervised ML model for dealing with insufficient samples and rare failures, and low-cost automation (e.g., \textbf{Store Pings} probing monitors) for timely analytics.\\
Kaleidoscope uses end-to-end I/O probing monitors to collect telemetry data related to path-tracing for observing the health of each component on the path. \textit{Store Pings} are designed for storage systems, and are used by clients to probe a disk by means of I/O requests and record the response time. It also provides a mechanism for pinning the path of the I/O requests to a disk through specific load balancers and servers by leveraging Lustre's file system support (thus eliminating path tracing of the request). It uses direct I/O requests to avoid any caching effect, thereby ensuring that each I/O request traverses from the clients to the disks on the data servers. They are placed strategically in the system to provide both spatial and temporal differential observability in near real-time. However, it must be noted that Store Pings must be enabled only on a subset of clients to reduce its overhead and impact on other I/O requests (which can be formulated as a constraint optimisation problem and solved considering the topology, probing mechanism, and I/O request routing protocols). More details in this regard can be referred from the literature cited in the framework design image. At any given time, Store Pings are executed from all login nodes, 1 randomly chosen service node out of 64, and 1 randomly chosen I/E (import/export) node out of 25; another client can be chosen as a monitor in case of a client failure. They are executed every minute for each OSD, data server, and metadata server.
\subsubsection*{Additional information}
Kaleidoscope uses a comprehensive monitoring system to collect performance measurements and RAS (reliability, availability, and serviceability) logs for each system component in real-time. It uses the Light-weight Distributed Metric Service (LDMS) to collect performance measurements from compute nodes, load-balancers (LNETs), switches, and the ISC (Integrated System Console) to collect performance measurements on storage components (e.g., disks and servers), LDMS data, and RAS logs on a centralised server. It uses hierarchical domain-guided unsupervised ML models like the failure localisation model (for identifying the failed nodes) and the failure diagnosis model (for identifying the failure mode of the failed mode) to provide live forensics capabilities.
%%%%-------------------------------------------------------%%%%
%%%% ------------  Apollo, IIT, USA  ------------ %%%%
%%%%-------------------------------------------------------%%%%
\subsection{Apollo (Storage Resource Observer), Illinois Institute of Technology, USA}
\textit{Apollo} is a low-latency near-real-time monitoring service assisted with machine-learning capabilities ('Delphi') which monitors the storage subsystem of a distributed computing environment, and provides telemetry data (which describes the state of a remote resource for a given time window) to applications and middleware services \cite{DBLP:conf/hpdc/RajeshDGBLYKS21}. Real-time access to telemetry data is critical for application and middleware library developers, as it could be used for performance optimisation and behaviour correctness. Apollo supports fast ingestion and low-latency access to metrics using a custom distributed data structure called Storage Condition Report (SCoRe), which in turn leverages data streaming and publish-subscribe delivery mechanism.
\subsubsection*{Architecture}
\begin{figure*}
\centering
\includegraphics[width=0.75\textwidth]{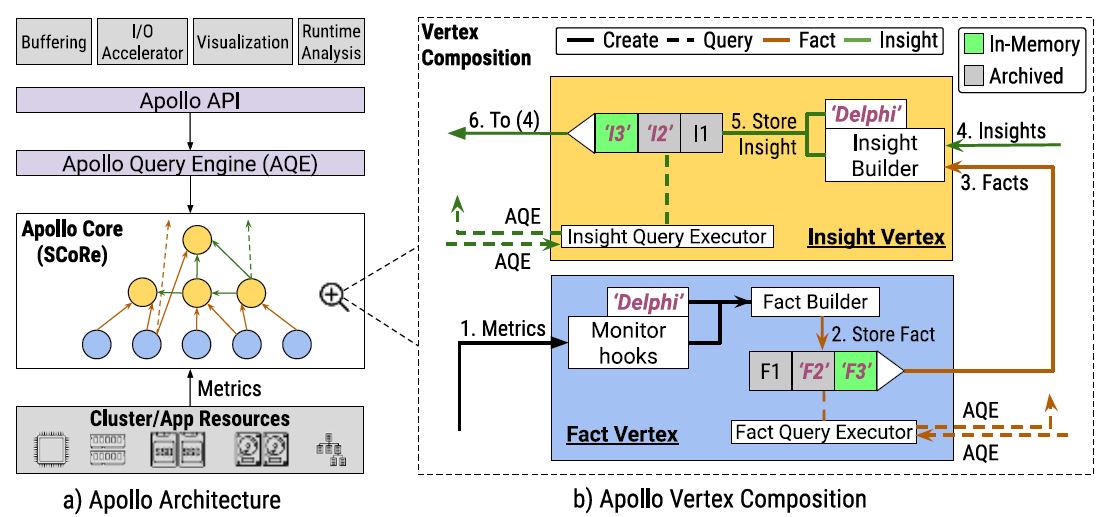}
\caption{Visualisation of the high-level architecture for Apollo. (Source:~\cite{DBLP:conf/hpdc/RajeshDGBLYKS21})}
\label{apollo-architecture}
% \vspace{-4mm}%Put here to reduce too much white space after your table 
\end{figure*}
Apollo's design is based on two principles: (i) reducing telemetry data access latency while increasing I/O throughput, and (ii) reducing the overall cost of resource monitoring while increasing accuracy. Apollo uses a machine-learning model called Delphi to improve monitoring resolution by predicting intermediate values, which can further reduce the overall cost of monitoring. It aims to use an adaptive and dynamic interval to match the dynamic nature of the cluster. The high-level architecture of Apollo is visualised in Fig. \ref{apollo-architecture}.
\begin{itemize}
    \item \textit{Overall architecture of Apollo}: This is depicted in Fig. \ref{apollo-architecture}-(a). Apollo uses SCoRe to enable access to telemetry data in a low-latency manner. The SCoRe data structure is represented as a Directed Acyclic Graph (DAG) of vertices, where each vertex collects information from Apollo. Information is categorised into Facts and/or Insights, and is stored as a tuple along with (timestamp, fact/insight, predicted/measured). The Fact vertex is the source vertex that represents the value of a given metric extracted from a particular hardware or software resource. An Insight vertex forms the sink and inner vertex of SCoRe and is a high-level combination of one or more Facts and/or Insights. Apollo has an Apollo Query Engine (AQE) which resolves queries into multiple accesses within SCoRe.
    \item \textit{Storage Condition Report (SCoRe)}: It is a distributed data store based on a graph data structure. Other than the telemetry data collection, its responsibility includes generating insights, maintaining Facts, and providing service to various middleware libraries of clients. It uses a Pub-Sub communication fabric to meet low-latency requirements, where it uses \textit{libuv}\footnote{libuv - \url{https://libuv.org/}} library for asynchronous I/O and \textit{Redis Streams}\footnote{Redis Streams - \url{https://redis.io/docs/data-types/streams/}} for maintaining telemetry data in a queue and providing the Pub-Sub communication paradigm. The Fact vertices and Insight vertices (two key components of SCoRe) are implemented using concurrent lock-free queues, and they are added only if there is a change from their previous value. Once they are added to the queue, they can be serviced immediately.
    \item \textit{Flow of information through SCoRe}: This is depicted in Fig. \ref{apollo-architecture}-(b). The information flow starts from the Fact vertices (labelled as 'Create') that capture metrics from cluster/application resources using the Monitor Hook. The Monitor Hook sends the metric to the Fact Builder, which converts the metric into a Fact that is later linearised and published onto the Fact Queue. The Facts are ordered by timestamp and can be consumed by an Insight vertex, which generates new Insights in the Insight Builder and pushes them into Insight Queue for immediate consumption. Both Fact and Insight vertices hold dedicated, in-memory queues and Archiver (that stores the queue in a log, and is efficient and scalable). The Monitor Hooks and Insight Builder are complemented with an ML inference model called Delphi, which predicts Facts for Fact vertices and Insights for Insight vertices between the monitoring intervals, that gradually increases the resolution of the telemetry data. The middleware services query Apollo via the AQE, which uses state-of-the-art intelligent algorithms and converts a client query into multiple information access calls to be served by the Query Executor of that vertex.
\end{itemize}
\subsubsection*{Evaluation Environment overview}
The authors evaluate Apollo on the Ares cluster at the Illinois Institute of Technology, which consists of 32 compute nodes and 32 storage nodes interconnected by a 40 Gb/s Ethernet network with RoCE (RDMA over Converged Ethernet) enabled. Each compute node consists of a dual Intel(R) Xeon Scalable Silver 4114 (i.e.,40 cores per node), 96GB RAM and a local 250GBNVMe. Each storage node has a dual AMD Opteron 2384 (i.e.,8 cores per node), 32GB RAM, a 150GB SATA SSD and 1TB HDD. The Ares cluster runs on CentOS 7.1 and the MPI library version is MPICH3.3.2, and uses TensorFlow 2.3.1 for training the machine learning models. As each compute node possesses an NVMe and SSD device and each storage node contains an HDD, two Fact vertices are deployed in every compute node and one in each storage node. Additionally, four Insight vertices are deployed in the cluster where three of them manage the individual streams of all devices in the same node and aggregate data for their Insights into respective Insight vertices, and the final Insight vertex subscribes to those three Insight vertices for generating a combined view of total available space (in the cluster).
%----------------COMPARISON--------------
\begin{table*}[!ht]
\begin{tabular}{|p{0.35in}|p{1.2in}|p{0.3in}|p{0.3in}|p{0.3in}|p{0.3in}|p{0.3in}|p{0.3in}|p{0.3in}|p{0.3in}|p{0.3in}|p{0.3in}|} \hline
% \begin{tabular}[c]{@{}l@{}}Ecosystem\\       Layer\end{tabular} & ODA capability & \begin{tabular}[c]{@{}l@{}}OMNI,\\      Berkeley Lab,\\      USA\end{tabular} & \begin{tabular}[c]{@{}l@{}}Wintermute,DCDB,\\      LRZ,\\      Germany\end{tabular} & \begin{tabular}[c]{@{}l@{}}ExaMon+\\      ExaMon-X,\\      CINECA\end{tabular} & \begin{tabular}[c]{@{}l@{}}Fugaku,\\      RIKEN,\\      Japan\end{tabular} & AutoDiagn & \begin{tabular}[c]{@{}l@{}}Theta,\\      ANL, USA\end{tabular} & \begin{tabular}[c]{@{}l@{}}Summit,\\      OLCF,\\      USA\end{tabular} & \begin{tabular}[c]{@{}l@{}}EAS,\\      IBM Research, UK\end{tabular} & \begin{tabular}[c]{@{}l@{}}Kaleidoscope,\\      Blue Waters,\\      UIUC,USA\end{tabular} & \begin{tabular}[c]{@{}l@{}}Apollo,\\      IIT,\\      USA\end{tabular} \\
\textbf{Layer} & \textbf{ODA Capability} & \textbf{OM} & \textbf{Win} & \textbf{Exa} & \textbf{Fug} & \textbf{Aut} & \textbf{Thet} & \textbf{Sum} & \textbf{EAS} & \textbf{Kal} & \textbf{Apo} \\ \hline
% 0 & \begin{tabular}[c]{@{}l@{}}Infrastructure\\      management\end{tabular} & Y & Y & Y & Y & - & Y &  &  &  & Y \\ \hline
0 & Infrastructure management & Y & Y & Y & Y & - & - & Y & Y & - & - \\ \hline
1 & Runtime tuning & Y & Y & - & Y & X & - & - & Y & - & - \\ \hline
1,2 & Hardware-managed security & - & - & - & - & - & - & - & - & - & - \\ \hline
2 & Fault detection & Y & Y & Y & Y & Y & Y & Y & - & Y & - \\ \hline
2 & Scalability tuning & Y & - & - & - & Y & - & - & - & - & - \\ \hline
3 & Allocation \& scheduling & Y & Y & - & - & - & - & - & X & - & - \\ \hline
3 & Application fingerprinting & Y & Y & Y & Y & Y & Y & Y & Y & - & Y \\ \hline
3 & Workload modelling & - & Y & - & Y & - & - & Y & Y & - & - \\ \hline
4 & Interoperability & - & - & - & - & Y & - & - & - & - & Y \\ \hline
4 & Error handling & Y & - & - & - & - & Y & - & - & Y & - \\ \hline
4,5 & Software-based security & Y & - & - & - & - & - & - & - & - & - \\ \hline
4,5 & Profiling & - & - & - & - & - & - & - & - & - & - \\ \hline
5 & Productivity enhancement & - & Y & - & - & Y & Y & - & - & - & - \\ \hline
7 & Data Reusability & Y & - & Y & - & - & - & - & - & - & - \\ \hline
7 & Experimentation & Y & Y & Y & - & - & - & - & - & - & - \\ \hline
7 & Data governance \& security & Y & - & - & - & - & - & - & - & - & - \\ \hline
0-6 & Monitoring \& alerting & Y & Y & Y & Y & Y & Y & Y & Y & Y & Y \\ \hline
0-6 & Energy modelling & Y & Y & Y & Y & - & - & Y & - & - & - \\ \hline
0-6 & Pipeline optimisation & Y & Y & - & - & - & - & - & - & - & - \\ \hline
\end{tabular}
\caption{\label{sota-comparison-with-oda-framework} Comparison of the state-of-the-art ODA frameworks (represented horizontally in the table with an alias) capabilities with the functionalities proposed in our reference architecture (\textbf{Y → Yes}, \textbf{X → Not available yet}, \textbf{- → Unknown}). Column headings representing systems: OM = OMNI (LBNL), Win = Wintermute (LRZ), Exa = ExaMon/ExaMon-X (CINECA), Fug = Fugaku (RIKEN), Aut = AutoDiagn, Thet = Theta (ANL), Sum = Summit (OLCF), EAS = EAS (IBM Research), Kal = Kaleidoscope (NCSA, UIUC), Apo = Apollo (IIT).}
% \vspace{-276pt}
\end{table*}
%----------------COMPARISON-------------- // Apollo section below for adjustment
\subsubsection*{Additional information}
The authors performed two sets of evaluations: (i) a set of evaluations of the internal components of Apollo exploring the three major components (SCoRe, Adaptive Internal Module, and Delphi), (ii) an evaluation of capabilities of the data structures and the communication layer. The middleware services need I/O-specific insights for making data placement, computation placement, and resource allocation decisions, which can be organised based on performance, energy, access, or workflow info. A few of the insights provided are interference factor (to indicate the degree to which I/O is being interfered with), node availability list (to facilitate leader election algorithms), energy consumption per transfer (the amount of power used vs the amount of work done by the node), allocation characteristics (metrics provided by Slurm) etc. The optimal polling interval for any metric could change (for example, when there is little change in the metric), so the monitoring service should be able to that change when it occurs. Delphi utilises machine learning techniques to reduce the monitoring interval by predicting intermediate values between two measurements.
%----------------COMPARISON--------------
\subsection{Comparing proposed ODA framework's functionalities with the state-of-the-art}
In this subsection, we briefly summarise how various state-of-the-art frameworks discussed in previous subsections compare to our proposed ODA framework architecture (involving various ODA functionalities). We consider the functionalities to be present if the state-of-the-art (SOTA) frameworks have implemented even a simplified version of the component functionality described in the proposed architecture (taking an open stance, to begin with). Some of the SOTA frameworks have a few components that are not fully controlled using ML techniques, but rely on manual interventions by the operations team based on the alerts generated in the visualisation or through other means; we have still considered those frameworks as feature-compliant (as they could optimise the same in the future). Also, many of these frameworks only have descriptive and/or diagnostic analytics being performed on the data collected in the pipeline (and lack predictive and prescriptive analytics), but we don't differentiate them on the level of the type of analysis performed in our comparison, and more on whether the functionality exists in the framework. Frameworks like Kaleidoscope are failure detection and diagnosis frameworks, so it is expected that they do not have advanced capabilities at various layers of the large-scale infrastructure ecosystem, but we have studied them here because they were unique in their own way in researching the diagnosis of the hardware resources.\\
The resultant comparison is shown in Table \ref{sota-comparison-with-oda-framework}, which has been carefully labelled following various references cited in this literature. We have marked the functionality which is present in the SOTA frameworks with a 'Y' representing that it is available, with an 'X' if the authors explicitly state that the functionality is not available and/or proposed as future work, and with a '-' if such a functionality is not discussed (we cannot claim that it is missing, as the respective authors of those works of literature might have refrained from disclosing those details for various possible reasons like privacy concerns besides others). As can be in from the table, the ODA frameworks appearing on the left are more advanced with more functionalities included whereas the right ones have limited functionalities, and they need to evolve their advanced ODA/monitoring frameworks for reaping the benefits realised as part of enabling those advanced ODA functionalities (which we have covered in the next section).

\section{Improvements observed after deployment of ODA techniques}
In this section, we discuss the impact analysis considering the system's energy-related aspects, after deploying the ODA framework techniques in a data centre environment for answering the third research question (RQ3). We characterise the efficiencies achieved in two categories broadly, namely the quantitative gains and the qualitative gains, which are further discussed below.
\subsection{Quantitative Gains}
In this sub-section, we mention the significant quantitative impacts observed in various production or experimental environments after enabling the operational data analytics capabilities. These types of impact are associated mainly with various savings like cost, power, water, etc., or provide highly significant analytical benefits in terms of percentages like model accuracy, etc. A summary of the prominent quantitative gains realised after implementing ODA functionalities in a large-scale computing infrastructure is presented in Table \ref{oda-quantified} for quick reference.
\subsubsection*{OMNI, NERSC, LBNL, USA}
\begin{figure}
\centering
\includegraphics[width=0.5\textwidth]{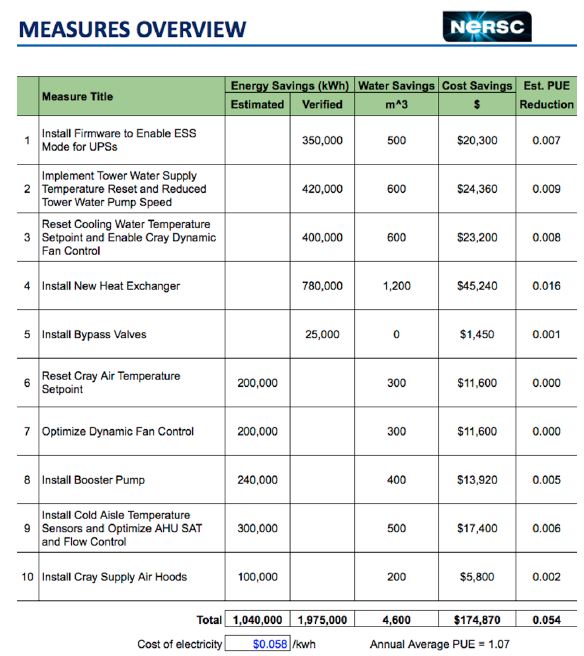}
\caption{Energy Efficiency Measures Identified at NERSC \cite{DBLP:conf/icppw/BourassaJBCJVS19}}
\label{omni-results}
\vspace{-4mm}%Put here to reduce too much white space after your table 
\end{figure}
NERSC utilised the operational data analytics capabilities of OMNI for their business decisions related to the expansion of electrical supply needs. They had an estimated \$23k of annual energy cost savings resulting from the optimisation of cooling plant operations \cite{DBLP:conf/cluster/OttSBWCRB20}. A major bold decision has been the elimination of an electrical substation for the large Perlmutter project planning, which helped them save an estimated \$2 million \cite{DBLP:conf/cluster/OttSBWCRB20}. They mention that the new HPE/Cray Perlmutter system would be 3-4x more performance efficient than the current NERSC Cori system, and the HPC footprint would also be actually smaller than that for Cori.\\
In addition to this, they claimed that OMNI has realised over 1.8 gigawatt-hours of energy savings and 0.56 million gallons (2.1 million Litres) of water savings annually \cite{liu2020continuously}. As per the literature, the current Level 2 PUE annual average is already a very efficient value of 1.08, and the team is working to optimise (lower) it further. Fig. \ref{omni-results} summarises the energy efficiency measures identified at NERSC, Berkeley Lab using OMNI's operational data analytics capabilities with a focus on reducing the annual average PUE value. They have already implemented various projects that had higher impacts like adding submetering and PUE tracking, air management with improved containment, enhancing the cooling system by adding a second heat exchanger and deep collaboration with IT vendors for access to HPC system performance data \cite{liu2020continuously}.
\subsubsection*{Wintermute, DCDB, LRZ, Germany}
The LRZ team uses an "Energy Aware Scheduling" (EAS) scheme for the reduction in the application's power usage without impacting its performance or runtime which uses dynamic voltage and frequency scaling (DVFS). They have estimated that its lifetime savings value is around 1.8M€ \cite{DBLP:conf/cluster/OttSBWCRB20}. They have studied the performance of Wintermute and found that it has a very low resource footprint (memory usage not exceeding 25MB and the per-core CPU load of the Pusher component is mostly uniform and peaks at 1.2\%) in the experimental evaluation, however, it would be different when measured in a deployed production environment. They have already worked on power consumption prediction with a highly accurate prediction (where the error is close to 5\% in the area of dense concentration values), analysis of job behaviour (by providing online visualisation), identification of performance anomalies (using unsupervised learning techniques), and performance and scalability characterisation (within the experimental evaluation) \cite{DBLP:conf/hpdc/NettiMGOTO020}. They also studied the deployment of Wintermute and DCDB in production HPC environments like SuperMUC-NG (for job-data visualisation) and modular DEEP-EST systems (for predictive cooling control), but could only loosely estimate the respective deployment benefits \cite{DBLP:journals/pc/NettiOGTS22}.
% --------- Table -------------
\begin{table}[!ht]
\begin{tabular}{|p{1in}|p{1.9in}|}
\hline
\textbf{Framework/\newline System\newline (Organisation)} & \textbf{Leveraged Benefits of ODA} \\ \hline
OMNI (LBNL)       & Estimated cost savings of \$23k annually from the optimisation of cooling plan operations \\ \hline
OMNI (LBNL)       & Estimated cost savings of \$2 million as a result of the elimination of an electrical substation \\ \hline
OMNI (LBNL)       & Realisation of 1.8 gigawatt-hours of energy savings and 0.56 million gallons (2.1 million Litres) of  water savings annually \\ \hline
OMNI (LBNL)       & Berkeley Lab's PUE value is already at 1.08, with further optimisation already planned \\ \hline
Wintermute / DCDB (LRZ)       & Lifetime cost savings estimated around 1.8 million € due to "Energy Aware Scheduling" scheme \\ \hline
Theta (ANL) &  Job exit status prediction accuracy of 92.3\% and identification of 92.65\% of the correctly-predicted jobs (at least 30-minutes) before failure \\ \hline
Kaleidoscope (NCSA) & Successful reporting of 98.3\% of reliability failures and 94.2\% of resource overload concerns, and a couple of additional issues later manually validated as genuine \\ \hline
\end{tabular}
\caption{\label{oda-quantified}Some of the prominent quantitative benefits gained after implementing ODA functionalities in various large-scale production environments.}
\end{table}
% --------- Table -------------
\subsubsection*{Theta HPC System, ANL, USA}
The ANL team have claimed various contributions as part of their research namely, (i) the system prediction accuracy of job exit status being 92.3\%, (ii) the system prediction accuracy of job exit status in case of a component failure being 88\%, (iii) visual representation representing correlations between environment (SEDC) logs and hardware errors with synchronised timelines, and (iv) identification of 92.65\% of the correctly-predicted jobs before failure (at least 30 minutes before) \cite{DBLP:conf/ccgrid/ShilpikaLESVPM22}. They further want to analyse environment logs to help pinpoint faulty nodes for further investigation, but they are currently ignored due to their size (growth to gigabyte-terabyte in a few weeks).
\subsubsection*{Kaleidoscope, Blue Waters, NCSA, USA}
The NCSA team deployed Store Ping monitors on the Cray Sonexion storage subsystem for two years and Kaleidoscope’s live forensics on Cray Sonexion for more than three months to evaluate the effectiveness of the framework \cite{DBLP:conf/sc/JhaCBXESKI20}. They identified a baseline of 843 production issues that were resolved by the Cray Sonexion operators over two years to measure the true positives and false negatives of Kaleidoscope, which is the first study to consider the impact of the kind (discussed further) as per the authors. They found that Kaleidoscope was able to localise the failed components for 99.3\% of the production issues (837 out of 843), and only 6 were not detected that belonged to disk drive failures (which also had no impact on the I/O completion time). Among those 843 production issues, 346 were caused by reliability failures and 497 were caused by resource overload, and applying relevant heuristics on Kaleidoscope resulted in it reporting 340 reliability failures (accounting for 98.3\%) and 468 overloads (accounting for 94.2\%), and a couple of additional reliability failures and resource overload issues that were later manually validated and turned out to be true. Though its diagnosis module did miss 35 production issues either by the localisation module or due to random noise in the logs that confused it. It also found four one-off, unique failures per month on average that were previously unknown, and that can hardly be anticipated based on historical datasets.

\subsection{Qualitative Gains}
In this sub-section, we list the significant qualitative impacts observed in various production or experimental environments after enabling the operational data analytics capabilities. These types of impact might also represent quantitative improvements, but they do not associate directly with cost, power or energy benefits in a numerical sense, thus, they are categorised under qualitative gains. A summary of the prominent qualitative gains realised after implementing ODA functionalities in a large-scale computing infrastructure is presented in Table \ref{oda-qualified} for quick reference.
% --------- Table -------------
\begin{table}[!ht]
\begin{tabular}{|p{1in}|p{1.9in}|}
\hline
\textbf{Framework/\newline System\newline (Organisation)} & \textbf{Leveraged Benefits of ODA} \\ \hline
Wintermute (LRZ), ExaMon (CINECA) & Anomaly detection module predicts the hardware-level anomalies \\ \hline
Wintermute (LRZ), Fugaku (RIKEN), AutoDiagn  & Performance and scalability improvements \\ \hline
ExaMon (CINECA)  & Improvement in strategic planning in the facility cooling system \\ \hline
AutoDiagn & Modelling memory footprint reduction \\ \hline
Wintermute (LRZ), ExaMon (CINECA) & Electrical power consumption prediction \\ \hline
\end{tabular}
\caption{\label{oda-qualified}Some of the prominent qualitative benefits gained after implementing ODA functionalities in various large-scale production environments.}
\vspace{-5mm}%Put here to reduce too much white space after your table 
\end{table}
% --------- Table -------------
\subsubsection*{ExaMon \& ExaMon-X, CINECA, Italy}
ExaMon \& ExaMon-X team (comprising of members from both CINECA and the University of Bologna, Unibo) have been exploring automatically annotated data and deep-learning models for anomaly detection and prediction in HPC systems, and their experimental results appear promising. Their proposed approach shows the possibility of anticipating the insurgence of faults in advance, on average between 40 and 50 minutes \cite{DBLP:journals/tpds/BorghesiMMB22}. Another literature discusses several use cases of these frameworks with functionalities like anomaly detection (where the semi-supervised approach is highly accurate with a 90\%-95\% range of accuracy), failure prediction of hard disks (with a failure detection rate of 89.1\%), job power prediction (with the vast majority of jobs having a percentage error smaller than 5\%), and thermal prediction in HPC systems \cite{DBLP:journals/iotj/BorghesiBB23}. In another literature, they present RUAD (Recurrent Unsupervised Anomaly Detection) model which has better results than the semi-supervised and unsupervised approaches in anomaly detection, if accurate anomaly labels are available \cite{DBLP:journals/fgcs/MolanBCBB23}. CINECA also benefitted from the use of historical power management data, where they developed several models of the facility cooling system under different possible designs, and thus decided to increase the free-cooling capacity in the data centre (instead of using rear door heat exchangers for that purpose) \cite{DBLP:conf/cluster/OttSBWCRB20}.
\subsubsection*{Fugaku, RIKEN, Japan}
The combined dashboard (for both the IT equipment information and the facility information) can visualise detailed power consumption for each group of racks and can show electricity supplied by the two CGSs as well as the public utility company. RIKEN's monitoring system could automatically detect and kill a running job if it overruns its estimated value which might result in the facility drawing more power and approaching the upper limit of its allowed power consumption, which helps RIKEN avoid utility penalties \cite{DBLP:conf/cluster/OttSBWCRB20}. The authors claim that the switching operations with two CGSs could lead to the extension of the lifetime of the equipment as well as make it easier to maintain. The average power usage effectiveness (PUE) which is commonly used as a comparable metric to evaluate energy efficiency operations in data centres is about 1.2.
\begin{figure}
\centering
\includegraphics[width=0.5\textwidth]{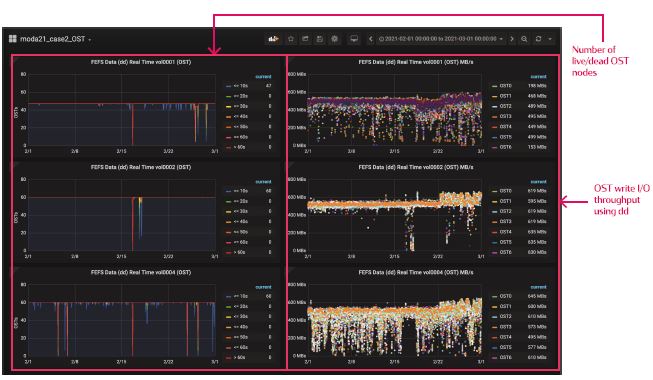}
\caption{An example of the Lustre-based file system state, presenting OST write I/O performance. (Source:~\cite{DBLP:conf/supercomputer/TeraiYMS21})}
\label{fugaku-results}
\vspace{-4mm}%Put here to reduce too much white space after your table 
\end{figure}
The authors have also mentioned that the operators are expecting an ability to be developed to detect certain workload behaviours when a number of workloads write to a single file, as it results in flooding I/Os which could result in system-wide failures. The dashboard showing the Lustre-based filesystem state (that presents MGS/MDS/OSS nodes performance) appears to be very useful for finding and isolating the problem in operations, thus helping to avoid overloading (visualisation show in Fig. \ref{fugaku-results}). They are continuing the efforts to resolve the ODA issues found with the prior K computer.
\subsubsection*{AutoDiagn}
AutoDiagn has mainly been evaluated in an experimental cloud-setup-based environment, therefore its benefits are characterised here under the qualitative gains subsection. The system overhead induced by the AutoDiagn Monitoring agent consumed approximately 2.52\% memory and 4.69\% CPU, while AutoDiagn Diagnosis consumed approximately 2.08\% memory and 3.49\% CPU \cite{DBLP:journals/tc/DemirbagaWNMAGZ22}. The storage overhead during the experimental evaluation was only 3.75 MB disk space, though increasing the types of symptom detection and root cause analysis will add to the consumption of more storage resources.
\subsubsection*{EAS, IBM Research, UK}
The IBM research team implementing a holistic EAS system performed clustering and prediction analysis, and found that CPU utilisation and Fan power were the main factors responsible for leading energy consumption of the rack \cite{DBLP:conf/supercomputer/TraceyHSE20}. The clustering analysis based on the time series variables helps pinpoint groups of workloads like ML series of jobs with high power consumption, and low CPU utilisation, in turn meaning heavy usage of the GPUs. The LSTM prediction model predicts the rack power with high precision, based on corresponding CPU utilisation and Fan power of the current and previous timesteps.
\subsubsection*{Summit, OLCF, USA}
The OLCF team studied the system's overall power and energy consumption, and observed that the behaviour of HPC applications was the best way to explain the resulting power consumption. Under the average power consumption between 5MW and 6MW, the average PUE of the data centre turned out to be 1.11 (thankfully due to the cheap evaporative-based cooling), but it increased to an average of 1.22 during the summer (due to the use of chilled water to trim down the supply temperature) \cite{DBLP:conf/sc/ShinOKEW21}. They categorised jobs into five classes (running on a determined range of nodes) and learnt that variations in peak power values are due to different algorithms across various disciplines that outweigh contributions, or depending on the degree of parallelism, etc. The decrease in per job node count highlighted an interesting finding that GPUs were the main workhorses defining the peak, and CPUs mostly defined the average power consumption. They tweaked the CPU and GPU temperatures just under the threshold of throttling to help in increasing the free cooling even on days with adverse wet bulb temperatures, resulting in improvement in the site's energy efficiency and OPEX \cite{DBLP:conf/cluster/OttSBWCRB20}.\\
Their finding also revealed that given the highly dynamic nature of HPC workloads, the temporal characteristics in temperature changes may have a higher impact on the GPU failures, however, the errors associated with lower-temperature GPUs remained inconclusive. Lastly, they conclude that there is temperature variability in the nodes that are introduced by spatial features (like manufacturing variation in the chips) within the node and on the actual Summit floor (due to the uneven impact of the cooling system due to their location), but the tight thermal response of components to power consumption dynamics can be studied well with the advance in modern HPC data collection systems.
\subsubsection*{Apollo, Illinois Institute of Technology, USA}
The authors experimented with Apollo and identified that the SCoRe data structure had high performance while providing low latency (they also tested overhead analysis and throughput analysis) \cite{DBLP:conf/hpdc/RajeshDGBLYKS21}. Apollo was compared with LDMS monitoring service in terms of performance as part of their experimentation, where they found that its latency was 3.5x lower than LDMS while having an overhead of only 7\% (which could be considered negligible given the lower latency and the adaptive interval feature). They also observed that Apollo could aid various middleware libraries by boosting their performance between 10\% and 20\% (for example, Hierarchical Data Placement Engine (HDPE) performance improvement was 18\% over the round-robin policy).
\subsection{Other considerable gains}
Some of the additional significant improvements realised in other environments after enabling the ODA framework (which are not discussed in the state-of-the-art ODA frameworks for brevity), as reported in the global experiences survey are briefly listed below \cite{DBLP:conf/cluster/OttSBWCRB20}:
\begin{itemize}
    \item \textit{National Center for Atmospheric Research (NCAR)}: The NCAR team were able to correlate site-wide power quality issues with the failure in the Variable Frequency Drive (VFD) of the HPC system Cooling Distribution Units (CDU) pumps. They were able to avoid multiple such incidents by providing clean power to the CDUs, which saved them an estimated \$20k for each incident (around two per year).
    \item \textit{Lawrence Livermore National Lab (LLNL)}: LLNL team was able to analyse three years of data to identify power spike patterns, for providing the electricity provider with a forecasted value. They could attribute such occasional large spike patterns to employee schedules, but also to scheduled and random events.
\end{itemize}

\section{Trending Research and Possibilities for future work}
We discuss the trending research in the field of ODA being conducted at various state-of-the-art frameworks in this section, and we present our view on what could possibly be additionally explored in this field. We also discuss how this literature study could be improved by adding more value to this work. This section serves as a response to our final research question (RQ4).
\subsection*{Ongoing research with(in) various ODA frameworks}
There is a lot of scope for the improvement of existing advanced ODA/monitoring frameworks, as we already presented an overview of ODA functionalities that do not appear in them (although they might not have been discussed in those works of literature for various reasons like data governance). In this section, we first go through the future work planned or mentioned in the various works of literature corresponding to the state-of-the-art ODA/monitoring frameworks discussed in section \ref{state-of-the-art}. Lastly, we present our understanding of the research opportunities in the field of ODA in large-scale distributed infrastructures, and also suggest future work for refinement of this study.\\~\\
% -----------------------OMNI, LBNL-----------------------
The OMNI infrastructure is undergoing an upgrade to support the increased size and scale of metrics from the Perlmutter HPC system which might require installing more edge services for more efficient real-time analysis, and at the same time supporting existing metric sets on the HPE/Cray Cori system \cite{bautista2022omni}. The authors acknowledged that their future work includes an investigation of an alerting framework at the edge like machines self-alerting themselves and taking mitigative actions based on their localised data in case when OMNI is not reachable or its unreliability. Additionally, they are also looking to make OMNI available externally to DOE (U.S. Department of Energy) collaboration users using blockchain, where users can request a dataset of analysis using a front-end graphical interface.\\
% -----------------------Wintermute, LRZ-----------------------
Wintermute team is planning to identify additional production use cases (that are discussed in the impact analysis section), and also explore solutions for ease in management for the operators and providing high availability \cite{DBLP:conf/hpdc/NettiMGOTO020}. They are also working on a machine-learning-based framework "Infrastructure Data Analyzer and Forecaster (IDAF)" that would enable forecasting various data centre energy/power consumption relevant key performance indicators (KPIs) \cite{DBLP:conf/cluster/OttSBWCRB20}. They have planned to extend Wintermute's ODA capabilities with the DCDB event data (like application-level instrumentation and telemetry about data migrations in hierarchical storage systems) \cite{DBLP:journals/pc/NettiOGTS22}. They (along with the TUM team) argued that it is a critical time to re-evaluate the co-scheduling capabilities for full utilisation of heterogeneous resources, manage dynamic addition or removal of resources from the application itself for efficient reaction to global system events, manage power/energy dynamically within the software, and ensuring much closer interaction among applications, systems and facilities by blurring the line \cite{DBLP:conf/heart/0001KSTW21}. In another work, they have planned to further characterise the properties and performance of their correlation-wise smoothing (CS) method by deploying it in a production HPC system, characterising its overhead and scalability in the production environment, testing its effectiveness on accelerator sensor data (e.g., GPUs), and exploring use case outside of the data centre domain \cite{DBLP:conf/ipps/NettiTO021}.\\
% -----------------------Examon, CINECA-----------------------
The ExaMon/ExaMon-X team is working on adding an ability in their ODA system to actively control frequency settings with phases of the application, depending on whether the phases are compute-bound or communication-bound \cite{DBLP:conf/cluster/OttSBWCRB20}. As part of future work for ExaMon, the authors started with a vision to analyse the data sets of the Marconi100 HPC system for a detailed investigation of the cause of the anomalies and classifying them in different categories, with additional plans to store and analyse historical log traces (providing insights related to supercomputing node failures) \cite{DBLP:journals/tpds/BorghesiMMB22}. In ExaMon-x literature, they have planned to enhance and extend the Deep learning (DL) models, and move from predictive forecasting to prescriptive maintenance to assist system owners and administrators \cite{DBLP:journals/iotj/BorghesiBB23}. Their other work focuses on thermal anomaly detection, which could be extended to other kinds of anomalies (e.g., application-level anomaly detection, etc.), and they are also exploring a semi-supervised ML-based approach for improving the anomaly detection performance \cite{DBLP:conf/supercomputer/ArdebiliBAB22}. In another work, they have acknowledged exploring the root cause of the anomalies in the first place in HPC systems, along with detecting anomalies involving multiple nodes at the same time \cite{DBLP:journals/fgcs/MolanBCBB23}. 
% -----------------------Fugaku, RIKEN-----------------------
A portion of the data centre infrastructure has been upgraded to handle any unprecedented power consumption (of 30 MW) since Fugaku has been brought into official service \cite{DBLP:conf/supercomputer/TeraiYMS21}. They have further planned to handle and resolve the ODA issue found in the K computer.\\
% -----------------------AutoDiagn-----------------------
The AutoDiagn authors mention that they focused mainly on evaluating the big data applications for performance reduction issues' outliers, and that newer features like symptom detection and root-cause analysis were required to be added to their system \cite{DBLP:journals/tc/DemirbagaWNMAGZ22}. They also mention that the current AutoDiagn implementation involves storage overhead and network overhead with scalability, which they want to reduce in future by exploring caching methods to aggregate the information before sending it to the destination nodes). Lastly, they have acknowledged an open research direction of building a system that could react to the analysis results, to help improve the performance of big data systems.\\
% -----------------------Theta HPC System, ANL, USA-----------------------
The ANL team highlight that their visualisation corresponding to machine-learning and topic-generation results provides a nice overview of how errors spread across the system. They have planned to further incorporate an end-to-end interactive visual analytics system with the functionality of a built-in user-feedback mechanism that would enhance the root cause analysis process \cite{DBLP:conf/ccgrid/ShilpikaLESVPM22}.\\
% -----------------------Summit, OLCF-----------------------
The OLCF team acknowledged that the analysis of the collected 2020 Summit metrics resulted in a deeper understanding of the workload and the system's response. They confirmed the well-known swinging behaviour of HPC applications after gaining insight into the magnitude, swing frequency and occurrence throughout the year. Large power swings paved the way for analysis of long-term cooling system response analysis, which further revealed the huge scope for facility improvements. The team has planned to research further about combined user and job power-profile fingerprinting capability, which could help in the predictive analysis of node-level power consumption \cite{DBLP:conf/sc/ShinOKEW21}. They have also planned to focus in future on analysing the power usage patterns of AI/ML applications (as to how they differ from traditional modelling) due to their higher demand lately, and correlate the AI/ML jobs and GPU power consumption as to how these jobs affect the HPC power profile.\\
%%%% ------------  EAS, IBM Research, UK  ------------ %%%%
The IBM research team have planned to evolve their system by extending the number and type of IoT devices to better understand the data centre environment behaviour, which would include Edge devices \cite{DBLP:conf/supercomputer/TraceyHSE20}. They also plan to integrate in-band data collection from the applications into the existing data collection framework. Other plans include feeding the results of the models into schedulers like LSF and cooling system controlling software. The directions for their future research in the field of ML include the implementation of AI-driven control techniques combining Predictive Analytics and Deep Reinforcement Learning.\\
% -----------------------Kaleidoscope, NCSA----------------------
The Kaleidoscope team have clarified that their framework is not tied to a specific storage architecture and can be extended to different system topologies and storage system architectures, as it automatically creates/changes the ML models with appropriate parameters using the system topology and file system I/O protocols. However, it cannot detect micro-burst performance anomalies and transient failures shorter than the dataset collection interval, which is an active area of research \cite{DBLP:conf/sc/JhaCBXESKI20}.\\
% -----------------------Apollo, IIT-----------------------
The experiments conducted using Apollo indicate the potential in optimising the collection of telemetry data. The authors acknowledge that some of the I/O curators need tweaking such that the metrics reflect what is needed by the middleware library. They suggest improving the adaptive internal heuristic by using a more intricate heuristic metric, and also suggest that the monitoring can further be improved using KProbes (a dynamic instrumentation mechanism) \cite{DBLP:conf/hpdc/RajeshDGBLYKS21}.\\~\\
% -----------------------Other noticeable concerns-----------------------
Jakobsche et al. have highlighted the challenges and opportunities  of machine learning for monitoring and operational data analytics in the context of quantitative codesign of supercomputers (QCS) in \cite{DBLP:journals/corr/abs-2209-07164}. They have described challenges related to a lack of guidelines for HPC researchers about what appropriate data to collect, no single best way to filter and prepare data for ML, no clarity about standard models for HPC centre-collected data, disjoint perspectives and approaches to data analysis, making the data FAIR resolving privacy and sharing concerns \cite{DBLP:journals/datasci/Dumontier22}, and missing knowledge behind a specific ML-based action, besides others. They suggest some opportunities like inviting ML experts to interact with the QCS communities to guide them about identifying appropriate ML models and suggesting the use of explainable ML models to help optimise system operations, and initiating an "Open-Data-Challenge" (ODC) to enrich the ML guidelines for QCS, targeting several topics like anomaly detection. They have suggested that HPC researchers and ML communities connect and collaboratively build and deploy the standard practices of machine learning for QCS.\\
% -----------------------Our take-----------------------
We envision that there is a huge scope for improvement of power usage and performance of the data centre environments given the extensiveness of our proposed holistic ODA framework, and suggest a geared research field to make standard approaches for analysing the data centre-wide issues, such that the scientific (or even other) community(-ies) benefits from those well-known practices. Another possible research work we envision could be to explore how to utilise/port the learning from one data centre environment to another (referred to as the "inter-ODA" framework), and earn the benefits with a comparatively lesser resource footprint. This idea is an extension of the concept of hybrid cloud evolution for cloud interoperability \cite{barhate2018hybrid}.

\subsection*{Future work for the study}
There are certain limitations of this literature study where mainly those ODA frameworks have been taken into consideration that was published in scientific pieces of literature and have the keyword "operational data analytics" (or ODA) within the literature. But, this leads to skipping several existing monitoring and analysis frameworks used in the HPC environment. We included the MonALISA HPC supercomputer in our study under the monitoring framework though this is one such framework which is widely used and has great analytical capabilities which has been evolving ever, even before the term "operational data analytics" was officially used in this regard. We think that a better approach would be to consider the top 100 supercomputers from the Green 500 supercomputers list, identify the relevant ones (based on criteria like the data centres and large HPC supercomputers with optimal PUE value below a certain threshold) and publish another literature study including the highly rated ones (in overall aspect) in the state-of-the-art ODA frameworks.\\
Other aspects of improving this study could be to assess the reference architecture for the ODA framework holistically proposed by us with respect to the existing large-scale data centre topologies, and add or modify the components of the reference architecture suitably. It would be valuable to research some notable ones among the top 100 HPC supercomputing environments and add to the state-of-the-art frameworks for a more holistic version, and also include the unique components into the reference architecture of the ODA framework such that it encompasses the ODA features from all of them.

\section{Conclusion} 
We have studied different topics throughout the literature. Here we summarise our findings, re-iterating briefly the research questions and their brief response to conclude the literature survey.
\begin{itemize}
    % \item \textbf{RQ1: What is the overall process flow of enabling the ODA capabilities in an HPC environment?} This section will highlight the overall process which allows the deployment of the ODA framework in an HPC environment. We highlight the characteristics of data which allow one to deploy the ODA framework. We will see the main source of data contributors for ODA in an HPC environment, and also go through some of the popular monitoring frameworks. Overall, we will see various components involved in the implementation of an ODA framework to provide intelligent insights into an HPC environment.
    \item \textbf{RQ1: How to lay out various components which are geared towards the performance and energy-efficiency of a large-scale computing infrastructure in a reference architecture for an ODA framework design?}: We propose a reference architecture that inherits from an existing ODA reference architecture and introduces novelty by combing the multi-layer architecture of large-scale computing infrastructure and ODA functionalities at the respective layers, along with drilled-down data collection, characterisation and learning mechanism to facilitate the reassessment and redeployment at each component of the reference architecture.
    \item \textbf{RQ2: What are the state-of-the-art ODA frameworks published in scientific pieces of literature for a large-scale distributed computing data centre?}: We briefly discussed the architecture of several state-of-the-art data centre infrastructures which have deployed their novel ODA frameworks either in their production environment or performed an experimental evaluation in a test environment.
    \item \textbf{RQ3: What are the various (quantitative/qualitative) energy or performance benefits that have been realised after implementing ODA techniques in a large-scale distributed data centre?}: We iterated through the realised gains on deploying the ODA framework in a large-scale distributed data centre (as published in standard pieces of literature). We walked through some interesting quantitative gains (related to energy efficiency) and qualitative gains (in terms of energy efficiency or performance improvements), wherever applicable.
    \item \textbf{RQ4: What are the ongoing ODA-related research works in a large-scale distributed environment?}: We explored the ongoing trends and the associated challenges in this field, as this field is gaining momentum gradually. The major research is geared towards involving more machine-learning-based analytics on the various components to know about the unknown behaviours of these large-scale computing infrastructures.
\end{itemize}
We hope that this literature study contributes to active research in the field of operational data analytics (ODA) within the large-scale computing infrastructure-related communities, and that we create an awareness of sustainability practices in the academia as well as the industry promoting energy and performance-efficient measures.
\medskip
% \cleardoublepage
\phantomsection

\addcontentsline{toc}{section}{References}
\printbibliography

@misc{green500,
  author = {TOP500.org},
  title = {{Green500 List - November 2022 | TOP500}},
  howpublished = "\url{https://www.top500.org/lists/green500/list/2022/11/}",
  year = {2022}, 
  note = "[Online; accessed 11-June-2023]"
}

@article{cappello2014toward,
  title={Toward exascale resilience: 2014 update},
  author={Cappello, Franck and Al, Geist and Gropp, William and Kale, Sanjay and Kramer, Bill and Snir, Marc},
  journal={Supercomputing Frontiers and Innovations: an International Journal},
  volume={1},
  number={1},
  pages={5--28},
  year={2014},
  publisher={South Ural State University Chelyabinsk, Russia, Russia}
}

@inproceedings{villa2014scaling,
  title={Scaling the power wall: a path to exascale},
  author={Villa, Oreste and Johnson, Daniel R and Oconnor, Mike and Bolotin, Evgeny and Nellans, David and Luitjens, Justin and Sakharnykh, Nikolai and Wang, Peng and Micikevicius, Paulius and Scudiero, Anthony and others},
  booktitle={SC'14: Proceedings of the International Conference for High Performance Computing, Networking, Storage and Analysis},
  pages={830--841},
  year={2014},
  organization={IEEE}
}

@book{eckerson2010performance,
  title={Performance dashboards: measuring, monitoring, and managing your business},
  author={Eckerson, Wayne W},
  year={2010},
  publisher={John Wiley \& Sons}
}

@inproceedings{DBLP:conf/cluster/OttSBWCRB20,
  author       = {Michael Ott and
                  Woong Shin and
                  Norman Bourassa and
                  Torsten Wilde and
                  Stefan Ceballos and
                  Melissa Romanus and
                  Natalie J. Bates},
  title        = {Global Experiences with {HPC} Operational Data Measurement, Collection
                  and Analysis},
  booktitle    = {{IEEE} International Conference on Cluster Computing, {CLUSTER} 2020,
                  Kobe, Japan, September 14-17, 2020},
  pages        = {499--508},
  publisher    = {{IEEE}},
  year         = {2020},
  url          = {https://doi.org/10.1109/CLUSTER49012.2020.00071},
  doi          = {10.1109/CLUSTER49012.2020.00071},
  bibsource    = {dblp computer science bibliography, https://dblp.org}
}

@inproceedings{DBLP:conf/ipps/NettiTO021,
  author       = {Alessio Netti and
                  Daniele Tafani and
                  Michael Ott and
                  Martin Schulz},
  title        = {Correlation-wise Smoothing: Lightweight Knowledge Extraction for {HPC}
                  Monitoring Data},
  booktitle    = {35th {IEEE} International Parallel and Distributed Processing Symposium,
                  {IPDPS} 2021, Portland, OR, USA, May 17-21, 2021},
  pages        = {2--12},
  publisher    = {{IEEE}},
  year         = {2021},
  url          = {https://doi.org/10.1109/IPDPS49936.2021.00010},
  doi          = {10.1109/IPDPS49936.2021.00010},
  bibsource    = {dblp computer science bibliography, https://dblp.org}
}

@inproceedings{DBLP:conf/sle/Pereira0RRCFS17,
  author       = {Rui Pereira and
                  Marco Couto and
                  Francisco Ribeiro and
                  Rui Rua and
                  J{\'{a}}come Cunha and
                  Jo{\~{a}}o Paulo Fernandes and
                  Jo{\~{a}}o Saraiva},
  editor       = {Beno{\^{i}}t Combemale and
                  Marjan Mernik and
                  Bernhard Rumpe},
  title        = {Energy efficiency across programming languages: how do energy, time,
                  and memory relate?},
  booktitle    = {Proceedings of the 10th {ACM} {SIGPLAN} International Conference on
                  Software Language Engineering, {SLE} 2017, Vancouver, BC, Canada,
                  October 23-24, 2017},
  pages        = {256--267},
  publisher    = {{ACM}},
  year         = {2017},
  url          = {https://doi.org/10.1145/3136014.3136031},
  doi          = {10.1145/3136014.3136031},
  bibsource    = {dblp computer science bibliography, https://dblp.org}
}

@article{kitchenham2007guidelines,
  title={Guidelines for performing systematic literature reviews in software engineering version 2.3},
  author={Kitchenham, Barbara and Charters, Stuart and others},
  journal={Engineering},
  volume={45},
  number={4ve},
  pages={1051},
  year={2007}
}

@article{DBLP:journals/corr/abs-2206-03259,
  author       = {Alexandru Iosup and
                  Fernando Kuipers and
                  Ana Lucia Varbanescu and
                  Paola Grosso and
                  Animesh Trivedi and
                  Jan S. Rellermeyer and
                  Lin Wang and
                  Alexandru Uta and
                  Francesco Regazzoni},
  title        = {Future Computer Systems and Networking Research in the Netherlands:
                  {A} Manifesto},
  journal      = {CoRR},
  volume       = {abs/2206.03259},
  year         = {2022},
  url          = {https://doi.org/10.48550/arXiv.2206.03259},
  doi          = {10.48550/arXiv.2206.03259},
  eprinttype    = {arXiv},
  eprint       = {2206.03259},
  bibsource    = {dblp computer science bibliography, https://dblp.org}
}

@article{DBLP:journals/ife/WildeAS14,
  author    = {Torsten Wilde and
               Axel Auweter and
               Hayk Shoukourian},
  title     = {The 4 Pillar Framework for energy efficient {HPC} data centers},
  journal   = {Comput. Sci. Res. Dev.},
  volume    = {29},
  number    = {3-4},
  pages     = {241--251},
  year      = {2014},
  url       = {https://doi.org/10.1007/s00450-013-0244-6},
  doi       = {10.1007/s00450-013-0244-6},
  bibsource = {dblp computer science bibliography, https://dblp.org}
}

@article{DBLP:journals/cacm/SakrBVIAAAABBDV21,
  author       = {Sherif Sakr and
                  Angela Bonifati and
                  Hannes Voigt and
                  Alexandru Iosup and
                  Khaled Ammar and
                  Renzo Angles and
                  Walid G. Aref and
                  Marcelo Arenas and
                  Maciej Besta and
                  Peter A. Boncz and
                  Khuzaima Daudjee and
                  Emanuele Della Valle and
                  Stefania Dumbrava and
                  Olaf Hartig and
                  Bernhard Haslhofer and
                  Tim Hegeman and
                  Jan Hidders and
                  Katja Hose and
                  Adriana Iamnitchi and
                  Vasiliki Kalavri and
                  Hugo Kapp and
                  Wim Martens and
                  M. Tamer {\"{O}}zsu and
                  Eric Peukert and
                  Stefan Plantikow and
                  Mohamed Ragab and
                  Matei Ripeanu and
                  Semih Salihoglu and
                  Christian Schulz and
                  Petra Selmer and
                  Juan F. Sequeda and
                  Joshua Shinavier and
                  G{\'{a}}bor Sz{\'{a}}rnyas and
                  Riccardo Tommasini and
                  Antonino Tumeo and
                  Alexandru Uta and
                  Ana Lucia Varbanescu and
                  Hsiang{-}Yun Wu and
                  Nikolay Yakovets and
                  Da Yan and
                  Eiko Yoneki},
  title        = {The future is big graphs: a community view on graph processing systems},
  journal      = {Commun. {ACM}},
  volume       = {64},
  number       = {9},
  pages        = {62--71},
  year         = {2021},
  url          = {https://doi.org/10.1145/3434642},
  doi          = {10.1145/3434642},
  bibsource    = {dblp computer science bibliography, https://dblp.org}
}

@inproceedings{DBLP:conf/hpdc/NettiMGOTO020,
  author    = {Alessio Netti and
               Micha M{\"{u}}ller and
               Carla Guill{\'{e}}n and
               Michael Ott and
               Daniele Tafani and
               Gence Ozer and
               Martin Schulz},
  editor    = {Manish Parashar and
               Vladimir Vlassov and
               David E. Irwin and
               Kathryn Mohror},
  title     = {{DCDB} Wintermute: Enabling Online and Holistic Operational Data Analytics
               on {HPC} Systems},
  booktitle = {{HPDC} '20: The 29th International Symposium on High-Performance Parallel
               and Distributed Computing, Stockholm, Sweden, June 23-26, 2020},
  pages     = {101--112},
  publisher = {{ACM}},
  year      = {2020},
  url       = {https://doi.org/10.1145/3369583.3392674},
  doi       = {10.1145/3369583.3392674},
  bibsource = {dblp computer science bibliography, https://dblp.org}
}

@article{moschny2021deep,
  title={DEEP-EST},
  author={Moschny, Th and Clau{\ss}, C and Huda, Z Ul and Krempel, S and Netti, A and Nuessle, M and Groschup, T and Ott, M and Pickartz, S},
  year={2021}
}

@inproceedings{DBLP:conf/sc/NettiMAGOT019,
  author       = {Alessio Netti and
                  Micha M{\"{u}}ller and
                  Axel Auweter and
                  Carla Guill{\'{e}}n and
                  Michael Ott and
                  Daniele Tafani and
                  Martin Schulz},
  editor       = {Michela Taufer and
                  Pavan Balaji and
                  Antonio J. Pe{\~{n}}a},
  title        = {From facility to application sensor data: modular, continuous and
                  holistic monitoring with {DCDB}},
  booktitle    = {Proceedings of the International Conference for High Performance Computing,
                  Networking, Storage and Analysis, {SC} 2019, Denver, Colorado, USA,
                  November 17-19, 2019},
  pages        = {64:1--64:27},
  publisher    = {{ACM}},
  year         = {2019},
  url          = {https://doi.org/10.1145/3295500.3356191},
  doi          = {10.1145/3295500.3356191},
  bibsource    = {dblp computer science bibliography, https://dblp.org}
}

@techreport{mqtt31,
  author = {Dave Locke},
  title = {MQ Telemetry Transport (MQTT) V3.1 Protocol Specification},
  institution = {{IBM}},
  year = {2010},
  month = {August},
  uri = {http://www.ibm.com/developerworks/webservices/library/ws-mqtt/index.html}
}

@misc{gartner2013,
  title = {Gartner IT glossary},
  howpublished = {\url{https://www.gartner.com/en/information-technology/glossary}},
  note = {Accessed: 2023-07-25}
}

@inproceedings{DBLP:conf/icppw/BartoliniBBCLBC19,
  author       = {Andrea Bartolini and
                  Francesco Beneventi and
                  Andrea Borghesi and
                  Daniele Cesarini and
                  Antonio Libri and
                  Luca Benini and
                  Carlo Cavazzoni},
  title        = {Paving the Way Toward Energy-Aware and Automated Datacentre},
  booktitle    = {48th International Conference on Parallel Processing, {ICPP} 2019
                  Workshop Proceedings, Kyoto, Japan, August 05-08, 2019},
  pages        = {8:1--8:8},
  publisher    = {{ACM}},
  year         = {2019},
  url          = {https://doi.org/10.1145/3339186.3339215},
  doi          = {10.1145/3339186.3339215},
  bibsource    = {dblp computer science bibliography, https://dblp.org}
}

@article{DBLP:journals/tpds/BorghesiMMB22,
  author       = {Andrea Borghesi and
                  Martin Molan and
                  Michela Milano and
                  Andrea Bartolini},
  title        = {Anomaly Detection and Anticipation in High Performance Computing Systems},
  journal      = {{IEEE} Trans. Parallel Distributed Syst.},
  volume       = {33},
  number       = {4},
  pages        = {739--750},
  year         = {2022},
  url          = {https://doi.org/10.1109/TPDS.2021.3082802},
  doi          = {10.1109/TPDS.2021.3082802},
  bibsource    = {dblp computer science bibliography, https://dblp.org}
}

@article{DBLP:journals/iotj/BorghesiBB23,
  author       = {Andrea Borghesi and
                  Alessio Burrello and
                  Andrea Bartolini},
  title        = {ExaMon-X: {A} Predictive Maintenance Framework for Automatic Monitoring
                  in Industrial IoT Systems},
  journal      = {{IEEE} Internet Things J.},
  volume       = {10},
  number       = {4},
  pages        = {2995--3005},
  year         = {2023},
  url          = {https://doi.org/10.1109/JIOT.2021.3125885},
  doi          = {10.1109/JIOT.2021.3125885},
  bibsource    = {dblp computer science bibliography, https://dblp.org}
}

@article{DBLP:journals/fgcs/MolanBCBB23,
  author       = {Martin Molan and
                  Andrea Borghesi and
                  Daniele Cesarini and
                  Luca Benini and
                  Andrea Bartolini},
  title        = {{RUAD:} Unsupervised anomaly detection in {HPC} systems},
  journal      = {Future Gener. Comput. Syst.},
  volume       = {141},
  pages        = {542--554},
  year         = {2023},
  url          = {https://doi.org/10.1016/j.future.2022.12.001},
  doi          = {10.1016/j.future.2022.12.001},
  bibsource    = {dblp computer science bibliography, https://dblp.org}
}

@inproceedings{DBLP:conf/europar/MolanBBB22,
  author       = {Martin Molan and
                  Andrea Borghesi and
                  Luca Benini and
                  Andrea Bartolini},
  editor       = {Jos{\'{e}} Cano and
                  Phil Trinder},
  title        = {Analysing Supercomputer Nodes Behaviour with the Latent Representation
                  of Deep Learning Models},
  booktitle    = {Euro-Par 2022: Parallel Processing - 28th International Conference
                  on Parallel and Distributed Computing, Glasgow, UK, August 22-26,
                  2022, Proceedings},
  series       = {Lecture Notes in Computer Science},
  volume       = {13440},
  pages        = {171--185},
  publisher    = {Springer},
  year         = {2022},
  url          = {https://doi.org/10.1007/978-3-031-12597-3\_11},
  doi          = {10.1007/978-3-031-12597-3\_11},
  bibsource    = {dblp computer science bibliography, https://dblp.org}
}

@inproceedings{DBLP:conf/cf/MolanBBB22,
  author       = {Martin Molan and
                  Andrea Borghesi and
                  Luca Benini and
                  Andrea Bartolini},
  editor       = {Luca Sterpone and
                  Andrea Bartolini and
                  Anastasiia Butko},
  title        = {Semi-supervised anomaly detection on a Tier-0 {HPC} system},
  booktitle    = {{CF} '22: 19th {ACM} International Conference on Computing Frontiers,
                  Turin, Italy, May 17 - 22, 2022},
  pages        = {203--204},
  publisher    = {{ACM}},
  year         = {2022},
  url          = {https://doi.org/10.1145/3528416.3530867},
  doi          = {10.1145/3528416.3530867},
  bibsource    = {dblp computer science bibliography, https://dblp.org}
}

@inproceedings{DBLP:conf/europar/MolanBBB22a,
  author       = {Martin Molan and
                  Andrea Borghesi and
                  Luca Benini and
                  Andrea Bartolini},
  editor       = {Jeremy Singer and
                  Yehia Elkhatib and
                  Dora Blanco Heras and
                  Patrick Diehl and
                  Nick Brown and
                  Aleksandar Ilic},
  title        = {Machine Learning Methodologies to Support {HPC} Systems Operations:
                  Anomaly Detection},
  booktitle    = {Euro-Par 2022: Parallel Processing Workshops - Euro-Par 2022 International
                  Workshops, Glasgow, UK, August 22-26, 2022, Revised Selected Papers},
  series       = {Lecture Notes in Computer Science},
  volume       = {13835},
  pages        = {294--298},
  publisher    = {Springer},
  year         = {2022},
  url          = {https://doi.org/10.1007/978-3-031-31209-0\_24},
  doi          = {10.1007/978-3-031-31209-0\_24},
  bibsource    = {dblp computer science bibliography, https://dblp.org}
}

@inproceedings{DBLP:conf/supercomputer/TeraiYMS21,
  author       = {Masaaki Terai and
                  Keiji Yamamoto and
                  Shin'ichi Miura and
                  Fumiyoshi Shoji},
  editor       = {Heike Jagode and
                  Hartwig Anzt and
                  Hatem Ltaief and
                  Piotr Luszczek},
  title        = {An Operational Data Collecting and Monitoring Platform for Fugaku:
                  System Overviews and Case Studies in the Prelaunch Service Period},
  booktitle    = {High Performance Computing - {ISC} High Performance Digital 2021 International
                  Workshops, Frankfurt am Main, Germany, June 24 - July 2, 2021, Revised
                  Selected Papers},
  series       = {Lecture Notes in Computer Science},
  volume       = {12761},
  pages        = {365--377},
  publisher    = {Springer},
  year         = {2021},
  url          = {https://doi.org/10.1007/978-3-030-90539-2\_24},
  doi          = {10.1007/978-3-030-90539-2\_24},
  bibsource    = {dblp computer science bibliography, https://dblp.org}
}

@inproceedings{DBLP:conf/cluster/TeraiSTY20,
  author       = {Masaaki Terai and
                  Fumiyoshi Shoji and
                  Toshiyuki Tsukamoto and
                  Yukihiro Yamochi},
  title        = {A Study of Operational Impact on Power Usage Effectiveness using Facility
                  Metrics and Server Operation Logs in the {K} Computer},
  booktitle    = {{IEEE} International Conference on Cluster Computing, {CLUSTER} 2020,
                  Kobe, Japan, September 14-17, 2020},
  pages        = {509--513},
  publisher    = {{IEEE}},
  year         = {2020},
  url          = {https://doi.org/10.1109/CLUSTER49012.2020.00072},
  doi          = {10.1109/CLUSTER49012.2020.00072},
  bibsource    = {dblp computer science bibliography, https://dblp.org}
}

@inproceedings{DBLP:conf/asiasim/FujitaSFNT21,
  author       = {Keijiro Fujita and
                  Naohisa Sakamoto and
                  Takanori Fujiwara and
                  Jorji Nonaka and
                  Toshiyuki Tsukamoto},
  editor       = {Byeong{-}Yun Chang and
                  Changbeom Choi},
  title        = {A Visual Analytics Method for Time-Series Log Data Using Multiple
                  Dimensionality Reduction},
  booktitle    = {Methods and Applications for Modeling and Simulation of Complex Systems
                  - 20th Asian Simulation Conference, AsiaSim 2021, Virtual Event, November
                  17-20, 2021, Proceedings},
  series       = {Communications in Computer and Information Science},
  volume       = {1636},
  pages        = {19--27},
  publisher    = {Springer},
  year         = {2021},
  url          = {https://doi.org/10.1007/978-981-19-6857-0\_3},
  doi          = {10.1007/978-981-19-6857-0\_3},
  bibsource    = {dblp computer science bibliography, https://dblp.org}
}

@inproceedings{DBLP:conf/icppw/BourassaJBCJVS19,
  author    = {Norman Bourassa and
               Walker Johnson and
               Jeff Broughton and
               Deirdre McShane Carter and
               Sadie Joy and
               Raphael Vitti and
               Peter Seto},
  title     = {Operational Data Analytics: Optimizing the National Energy Research
               Scientific Computing Center Cooling Systems},
  booktitle = {48th International Conference on Parallel Processing, {ICPP} 2019
               Workshop Proceedings, Kyoto, Japan, August 05-08, 2019},
  pages     = {5:1--5:7},
  publisher = {{ACM}},
  year      = {2019},
  url       = {https://doi.org/10.1145/3339186.3339210},
  doi       = {10.1145/3339186.3339210},
  bibsource = {dblp computer science bibliography, https://dblp.org}
}

@inproceedings{DBLP:conf/icppw/BautistaRDWK19,
  author       = {Elizabeth Bautista and
                  Melissa Romanus and
                  Thomas Davis and
                  Cary Whitney and
                  Theodore Kubaska},
  title        = {Collecting, Monitoring, and Analyzing Facility and Systems Data at
                  the National Energy Research Scientific Computing Center},
  booktitle    = {48th International Conference on Parallel Processing, {ICPP} 2019
                  Workshop Proceedings, Kyoto, Japan, August 05-08, 2019},
  pages        = {10:1--10:9},
  publisher    = {{ACM}},
  year         = {2019},
  url          = {https://doi.org/10.1145/3339186.3339213},
  doi          = {10.1145/3339186.3339213},
  bibsource    = {dblp computer science bibliography, https://dblp.org}
}

@incollection{bautista2022omni,
  title={OMNI at the Edge},
  author={Bautista, Elizabeth and Sukhija, Nitin and Romanus, Melissa and Davis, Thomas and Whitney, Cary},
  booktitle={Cybersecurity and High-Performance Computing Environments},
  pages={63--84},
  year={2022},
  publisher={Chapman and Hall/CRC}
}

@article{liu2020continuously,
  title={Continuously improving energy and water management},
  author={Liu, Jingjing and Bourassa, Norman},
  journal={ASHRAE Journal},
  volume={62},
  number={12},
  pages={30--36},
  year={2020}
}

@inproceedings{DBLP:conf/cluster/NettiSOWB21,
  author    = {Alessio Netti and
               Woong Shin and
               Michael Ott and
               Torsten Wilde and
               Natalie J. Bates},
  title     = {A Conceptual Framework for {HPC} Operational Data Analytics},
  booktitle = {{IEEE} International Conference on Cluster Computing, {CLUSTER} 2021,
               Portland, OR, USA, September 7-10, 2021},
  pages     = {596--603},
  publisher = {{IEEE}},
  year      = {2021},
  url       = {https://doi.org/10.1109/Cluster48925.2021.00086},
  doi       = {10.1109/Cluster48925.2021.00086},
  bibsource = {dblp computer science bibliography, https://dblp.org}
}

@article{DBLP:journals/misq/WebsterW02,
  author    = {Jane Webster and
               Richard T. Watson},
  title     = {Analyzing the Past to Prepare for the Future: Writing a Literature
               Review},
  journal   = {{MIS} Q.},
  volume    = {26},
  number    = {2},
  year      = {2002},
  bibsource = {dblp computer science bibliography, https://dblp.org}
}

@inproceedings{DBLP:conf/ease/Wohlin14,
  author    = {Claes Wohlin},
  editor    = {Martin J. Shepperd and
               Tracy Hall and
               Ingunn Myrtveit},
  title     = {Guidelines for snowballing in systematic literature studies and a
               replication in software engineering},
  booktitle = {18th International Conference on Evaluation and Assessment in Software
               Engineering, {EASE} '14, London, England, United Kingdom, May 13-14,
               2014},
  pages     = {38:1--38:10},
  publisher = {{ACM}},
  year      = {2014},
  url       = {https://doi.org/10.1145/2601248.2601268},
  doi       = {10.1145/2601248.2601268},
  bibsource = {dblp computer science bibliography, https://dblp.org}
}

@inproceedings{DBLP:conf/sc/ShinOKEW21,
  author       = {Woong Shin and
                  Vladyslav Oles and
                  Ahmad Maroof Karimi and
                  J. Austin Ellis and
                  Feiyi Wang},
  editor       = {Bronis R. de Supinski and
                  Mary W. Hall and
                  Todd Gamblin},
  title        = {Revealing power, energy and thermal dynamics of a 200PF pre-exascale
                  supercomputer},
  booktitle    = {International Conference for High Performance Computing, Networking,
                  Storage and Analysis, {SC} 2021, St. Louis, Missouri, USA, November
                  14-19, 2021},
  pages        = {12},
  publisher    = {{ACM}},
  year         = {2021},
  url          = {https://doi.org/10.1145/3458817.3476188},
  doi          = {10.1145/3458817.3476188},
  bibsource    = {dblp computer science bibliography, https://dblp.org}
}

@article{fischer2020metrics,
  title={Metrics for Job Similarity Based on Hardware Performance Data},
  author={Fischer, Felix},
  year={2020}
}

@article{wood2021online,
  title={Online Monitoring for High-Performance Computing Systems},
  author={Wood, Chad},
  year={2021}
}

@phdthesis{wood2022scalable,
  title={Scalable Observation, Analysis, and Tuning for Parallel Portability in HPC},
  author={Wood, Chad},
  year={2022},
  school={University of Oregon}
}

@phdthesis{DBLP:phd/dnb/Netti22,
  author       = {Alessio Netti},
  title        = {Holistic and Portable Operational Data Analytics on Production {HPC}
                  Systems},
  school       = {Technical University of Munich, Germany},
  year         = {2022},
  url          = {https://nbn-resolving.org/urn:nbn:de:bvb:91-diss-20220222-1620116-1-0},
  urn          = {urn:nbn:de:bvb:91-diss-20220222-1620116-1-0},
  bibsource    = {dblp computer science bibliography, https://dblp.org}
}

@inproceedings{DBLP:conf/cluster/SottileM02,
  author       = {Matthew J. Sottile and
                  Ronald G. Minnich},
  title        = {Supermon: {A} High-Speed Cluster Monitoring System},
  booktitle    = {2002 {IEEE} International Conference on Cluster Computing {(CLUSTER}
                  2002), 23-26 September 2002, Chicago, IL, {USA}},
  pages        = {39--46},
  publisher    = {{IEEE} Computer Society},
  year         = {2002},
  url          = {https://doi.org/10.1109/CLUSTR.2002.1137727},
  doi          = {10.1109/CLUSTR.2002.1137727},
  bibsource    = {dblp computer science bibliography, https://dblp.org}
}

@article{DBLP:journals/corr/cs-DC-0306096,
  author       = {Harvey B. Newman and
                  Iosif Legrand and
                  Philippe Galvez and
                  Ramiro Voicu and
                  Catalin Cirstoiu},
  title        = {MonALISA : {A} Distributed Monitoring Service Architecture},
  journal      = {CoRR},
  volume       = {cs.DC/0306096},
  year         = {2003},
  url          = {http://arxiv.org/abs/cs/0306096},
  bibsource    = {dblp computer science bibliography, https://dblp.org}
}

@inproceedings{DBLP:conf/cluster/SacerdotiKMC03,
  author       = {Federico D. Sacerdoti and
                  Mason J. Katz and
                  Matthew L. Massie and
                  David E. Culler},
  title        = {Wide Area Cluster Monitoring with Ganglia},
  booktitle    = {2003 {IEEE} International Conference on Cluster Computing {(CLUSTER}
                  2003), 1-4 December 2003, Kowloon, Hong Kong, China},
  pages        = {289},
  publisher    = {{IEEE} Computer Society},
  year         = {2003},
  url          = {https://doi.org/10.1109/CLUSTR.2003.1253327},
  doi          = {10.1109/CLUSTR.2003.1253327},
  bibsource    = {dblp computer science bibliography, https://dblp.org}
}

@article{DBLP:journals/pc/MassieCC04,
  author       = {Matthew L. Massie and
                  Brent N. Chun and
                  David E. Culler},
  title        = {The ganglia distributed monitoring system: design, implementation,
                  and experience},
  journal      = {Parallel Comput.},
  volume       = {30},
  number       = {5-6},
  pages        = {817--840},
  year         = {2004},
  url          = {https://doi.org/10.1016/j.parco.2004.04.001},
  doi          = {10.1016/j.parco.2004.04.001},
  bibsource    = {dblp computer science bibliography, https://dblp.org}
}

@inproceedings{DBLP:conf/IEEEscc/KatsarosKG11,
  author       = {Gregory Katsaros and
                  Roland K{\"{u}}bert and
                  Georgina Gallizo},
  editor       = {Hans{-}Arno Jacobsen and
                  Yan Wang and
                  Patrick Hung},
  title        = {Building a Service-Oriented Monitoring Framework with {REST} and Nagios},
  booktitle    = {{IEEE} International Conference on Services Computing, {SCC} 2011,
                  Washington, DC, USA, 4-9 July, 2011},
  pages        = {426--431},
  publisher    = {{IEEE} Computer Society},
  year         = {2011},
  url          = {https://doi.org/10.1109/SCC.2011.53},
  doi          = {10.1109/SCC.2011.53},
  bibsource    = {dblp computer science bibliography, https://dblp.org}
}

@inproceedings{mongkolluksamee2010strengths,
  title={Strengths and limitations of Nagios as a network monitoring solution},
  author={Mongkolluksamee, Sophon and Pongpaibool, Panita and Issariyapat, Chavee},
  booktitle={Proceedings of the 7th International Joint Conference on Computer Science and Software Engineering (JCSSE 2010). Bangkok, Thailand},
  pages={96--101},
  year={2010}
}

@inproceedings{DBLP:conf/lisa/Oetiker98,
  author       = {Tobias Oetiker},
  editor       = {Xev Gittler and
                  Rob Kolstad},
  title        = {{MRTG:} The Multi Router Traffic Grapher},
  booktitle    = {Proceedings of the 12th Conference on Systems Administration (LISA-98),
                  Boston, MA, USA, December 6-11, 1998},
  pages        = {141--148},
  publisher    = {{USENIX}},
  year         = {1998},
  url          = {http://www.usenix.org/publications/library/proceedings/lisa98/oetiker.html},
  bibsource    = {dblp computer science bibliography, https://dblp.org}
}

@inproceedings{DBLP:conf/sc/EvansBBDFGJP14,
  author       = {R. Todd Evans and
                  William L. Barth and
                  James C. Browne and
                  Robert L. DeLeon and
                  Thomas R. Furlani and
                  Steven M. Gallo and
                  Matthew D. Jones and
                  Abani K. Patra},
  editor       = {Christopher Bording and
                  Andy Georges},
  title        = {Comprehensive resource use monitoring for {HPC} systems with {TACC}
                  stats},
  booktitle    = {Proceedings of the First International Workshop on {HPC} User Support
                  Tools, {HUST} '14, New Orleans, Louisiana, USA, November 16-21, 2014},
  pages        = {13--21},
  publisher    = {{IEEE} Computer Society},
  year         = {2014},
  url          = {https://doi.org/10.1109/HUST.2014.7},
  doi          = {10.1109/HUST.2014.7},
  bibsource    = {dblp computer science bibliography, https://dblp.org}
}

@inproceedings{DBLP:conf/sc/AgelastosABCEFGMNORSSTT14,
  author       = {Anthony M. Agelastos and
                  Benjamin A. Allan and
                  Jim M. Brandt and
                  Paul Cassella and
                  Jeremy Enos and
                  Joshi Fullop and
                  Ann C. Gentile and
                  Steve Monk and
                  Nichamon Naksinehaboon and
                  Jeff Ogden and
                  Mahesh Rajan and
                  Michael T. Showerman and
                  Joel Stevenson and
                  Narate Taerat and
                  Thomas W. Tucker},
  editor       = {Trish Damkroger and
                  Jack J. Dongarra},
  title        = {The Lightweight Distributed Metric Service: {A} Scalable Infrastructure
                  for Continuous Monitoring of Large Scale Computing Systems and Applications},
  booktitle    = {International Conference for High Performance Computing, Networking,
                  Storage and Analysis, {SC} 2014, New Orleans, LA, USA, November 16-21,
                  2014},
  pages        = {154--165},
  publisher    = {{IEEE} Computer Society},
  year         = {2014},
  url          = {https://doi.org/10.1109/SC.2014.18},
  doi          = {10.1109/SC.2014.18},
  bibsource    = {dblp computer science bibliography, https://dblp.org}
}

@inproceedings{DBLP:conf/cluster/AgelastosABGLMO15,
  author       = {Anthony M. Agelastos and
                  Benjamin A. Allan and
                  Jim M. Brandt and
                  Ann C. Gentile and
                  Sophia Lefantzi and
                  Steve Monk and
                  Jeff Ogden and
                  Mahesh Rajan and
                  Joel Stevenson},
  title        = {Toward Rapid Understanding of Production {HPC} Applications and Systems},
  booktitle    = {2015 {IEEE} International Conference on Cluster Computing, {CLUSTER}
                  2015, Chicago, IL, USA, September 8-11, 2015},
  pages        = {464--473},
  publisher    = {{IEEE} Computer Society},
  year         = {2015},
  url          = {https://doi.org/10.1109/CLUSTER.2015.71},
  doi          = {10.1109/CLUSTER.2015.71},
  bibsource    = {dblp computer science bibliography, https://dblp.org}
}

@article{DBLP:journals/fgcs/FortiGB21a,
  author       = {Stefano Forti and
                  Marco Gaglianese and
                  Antonio Brogi},
  title        = {Corrigendum to "Lightweight self-organising distributed monitoring
                  of Fog infrastructures" [Future Gener. Comput. Syst. 114 {(2020)}
                  605-618]},
  journal      = {Future Gener. Comput. Syst.},
  volume       = {118},
  pages        = {495},
  year         = {2021},
  url          = {https://doi.org/10.1016/j.future.2021.01.009},
  doi          = {10.1016/j.future.2021.01.009},
  bibsource    = {dblp computer science bibliography, https://dblp.org}
}

@inproceedings{DBLP:conf/uic/SukhijaB19,
  author       = {Nitin Sukhija and
                  Elizabeth Bautista},
  title        = {Towards a Framework for Monitoring and Analyzing High Performance
                  Computing Environments Using Kubernetes and Prometheus},
  booktitle    = {2019 {IEEE} SmartWorld, Ubiquitous Intelligence {\&} Computing,
                  Advanced {\&} Trusted Computing, Scalable Computing {\&} Communications,
                  Cloud {\&} Big Data Computing, Internet of People and Smart City
                  Innovation, SmartWorld/SCALCOM/UIC/ATC/CBDCom/IOP/SCI 2019, Leicester,
                  United Kingdom, August 19-23, 2019},
  pages        = {257--262},
  publisher    = {{IEEE}},
  year         = {2019},
  url          = {https://doi.org/10.1109/SmartWorld-UIC-ATC-SCALCOM-IOP-SCI.2019.00087},
  doi          = {10.1109/SmartWorld-UIC-ATC-SCALCOM-IOP-SCI.2019.00087},
  bibsource    = {dblp computer science bibliography, https://dblp.org}
}

@inproceedings{DBLP:conf/medes/SukhijaBJGDLQL20,
  author       = {Nitin Sukhija and
                  Elizabeth Bautista and
                  Owen James and
                  Daniel Gens and
                  Siqi Deng and
                  Yulok Lam and
                  Tony Quan and
                  Basil Lalli},
  editor       = {Richard Chbeir and
                  Yannis Manolopoulos and
                  Ernesto Damiani and
                  Djamal Benslimane and
                  Ladjel Bellatreche and
                  Tadeusz Morzy},
  title        = {Event Management and Monitoring Framework for {HPC} Environments using
                  ServiceNow and Prometheus},
  booktitle    = {{MEDES} '20: 12th International Conference on Management of Digital
                  EcoSystems, Virtual Event, United Arab Emirates, 2-4 November, 2020},
  pages        = {149--156},
  publisher    = {{ACM}},
  year         = {2020},
  url          = {https://doi.org/10.1145/3415958.3433046},
  doi          = {10.1145/3415958.3433046},
  bibsource    = {dblp computer science bibliography, https://dblp.org}
}

@inproceedings{DBLP:conf/ucc/GraciaRBA16,
  author       = {Victor Medel Gracia and
                  Omer F. Rana and
                  Jos{\'{e}} {\'{A}}ngel Ba{\~{n}}ares and
                  Unai Arronategui},
  editor       = {Changjun Jiang and
                  Omer F. Rana and
                  Nick Antonopoulos},
  title        = {Modelling performance {\&} resource management in kubernetes},
  booktitle    = {Proceedings of the 9th International Conference on Utility and Cloud
                  Computing, {UCC} 2016, Shanghai, China, December 6-9, 2016},
  pages        = {257--262},
  publisher    = {{ACM}},
  year         = {2016},
  url          = {https://doi.org/10.1145/2996890.3007869},
  doi          = {10.1145/2996890.3007869},
  bibsource    = {dblp computer science bibliography, https://dblp.org}
}

@inproceedings{DBLP:conf/xsede/Chan19,
  author       = {Nicolas Chan},
  editor       = {Thomas R. Furlani},
  title        = {A Resource Utilization Analytics Platform Using Grafana and Telegraf
                  for the Savio Supercluster},
  booktitle    = {Proceedings of the Practice and Experience in Advanced Research Computing
                  on Rise of the Machines (learning), {PEARC} 2019, Chicago, IL, USA,
                  July 28 - August 01, 2019},
  pages        = {31:1--31:6},
  publisher    = {{ACM}},
  year         = {2019},
  url          = {https://doi.org/10.1145/3332186.3333053},
  doi          = {10.1145/3332186.3333053},
  bibsource    = {dblp computer science bibliography, https://dblp.org}
}

@inproceedings{DBLP:conf/jcsse/RattanatamrongB20,
  author       = {Prapaporn Rattanatamrong and
                  Yoottana Boonpalit and
                  Siwakorn Suwanjinda and
                  Ayuth Mangmeesap and
                  Ken Subraties and
                  Vahid Daneshmand and
                  Shava Smallen and
                  Jason Haga},
  title        = {Overhead Study of Telegraf as a Real-Time Monitoring Agent},
  booktitle    = {17th International Joint Conference on Computer Science and Software
                  Engineering, {JCSSE} 2020, Bangkok, Thailand, November 4-6, 2020},
  pages        = {42--46},
  publisher    = {{IEEE}},
  year         = {2020},
  url          = {https://doi.org/10.1109/JCSSE49651.2020.9268333},
  doi          = {10.1109/JCSSE49651.2020.9268333},
  bibsource    = {dblp computer science bibliography, https://dblp.org}
}

@article{naqvi2017time,
  title={Time series databases and influxdb},
  author={Naqvi, Syeda Noor Zehra and Yfantidou, Sofia and Zim{\'a}nyi, Esteban},
  journal={Studienarbeit, Universit{\'e} Libre de Bruxelles},
  volume={12},
  year={2017}
}

@article{DBLP:journals/fgcs/WangXZGZ18,
  author       = {Tao Wang and
                  Jiwei Xu and
                  Wenbo Zhang and
                  Zeyu Gu and
                  Hua Zhong},
  title        = {Self-adaptive cloud monitoring with online anomaly detection},
  journal      = {Future Gener. Comput. Syst.},
  volume       = {80},
  pages        = {89--101},
  year         = {2018},
  url          = {https://doi.org/10.1016/j.future.2017.09.067},
  doi          = {10.1016/j.future.2017.09.067},
  bibsource    = {dblp computer science bibliography, https://dblp.org}
}

@article{simmonds2009scf,
  title={Scf/fef evaluation of nagios and zabbix monitoring systems},
  author={Simmonds, Ed and Harrington, Jason},
  journal={SCF/FEF},
  pages={1--9},
  year={2009}
}

@article{DBLP:journals/mam/CascajoSC22,
  author       = {Alberto Cascajo and
                  David E. Singh and
                  Jes{\'{u}}s Carretero},
  title        = {{LIMITLESS} - LIght-weight MonItoring Tool for LargE Scale Systems},
  journal      = {Microprocess. Microsystems},
  volume       = {93},
  pages        = {104586},
  year         = {2022},
  url          = {https://doi.org/10.1016/j.micpro.2022.104586},
  doi          = {10.1016/j.micpro.2022.104586},
  bibsource    = {dblp computer science bibliography, https://dblp.org}
}

@inproceedings{DBLP:conf/cluster/JakobscheLCC21,
  author       = {Thomas Jakobsche and
                  Nicolas Lachiche and
                  Aur{\'{e}}lien Cavelan and
                  Florina M. Ciorba},
  title        = {An Execution Fingerprint Dictionary for {HPC} Application Recognition},
  booktitle    = {{IEEE} International Conference on Cluster Computing, {CLUSTER} 2021,
                  Portland, OR, USA, September 7-10, 2021},
  pages        = {604--608},
  publisher    = {{IEEE}},
  year         = {2021},
  url          = {https://doi.org/10.1109/Cluster48925.2021.00092},
  doi          = {10.1109/Cluster48925.2021.00092},
  bibsource    = {dblp computer science bibliography, https://dblp.org}
}

@article{kunz2022hpc,
  title={HPC Job-Monitoring with SLURM, Prometheus and Grafana},
  author={Kunz, Pascal},
  year={2022}
}

@phdthesis{DBLP:phd/dnb/Ilsche20,
  author       = {Thomas Ilsche},
  title        = {Energy Measurements of High Performance Computing Systems: From Instrumentation
                  to Analysis},
  school       = {Dresden University of Technology, Germany},
  year         = {2020},
  url          = {https://nbn-resolving.org/urn:nbn:de:bsz:14-qucosa2-716000},
  urn          = {urn:nbn:de:bsz:14-qucosa2-716000},
  bibsource    = {dblp computer science bibliography, https://dblp.org}
}

@article{fernandezenergy,
  title={Energy efficiency in a supercomputing center: a case study},
  author={Fern{\'a}ndez Gonz{\'a}lez, A and Matell{\'a}n, V and Mart{\'\i}nez Garc{\'\i}a, JM and Lorenzana, J and L{\'o}pez, M}
}

@inproceedings{DBLP:conf/hpca/RoyPKARST21,
  author       = {Rohan Basu Roy and
                  Tirthak Patel and
                  Raj Kettimuthu and
                  William E. Allcock and
                  Paul Rich and
                  Adam Scovel and
                  Devesh Tiwari},
  title        = {Operating Liquid-Cooled Large-Scale Systems: Long-Term Monitoring,
                  Reliability Analysis, and Efficiency Measures},
  booktitle    = {{IEEE} International Symposium on High-Performance Computer Architecture,
                  {HPCA} 2021, Seoul, South Korea, February 27 - March 3, 2021},
  pages        = {881--893},
  publisher    = {{IEEE}},
  year         = {2021},
  url          = {https://doi.org/10.1109/HPCA51647.2021.00078},
  doi          = {10.1109/HPCA51647.2021.00078},
  bibsource    = {dblp computer science bibliography, https://dblp.org}
}

@article{markus2021framework,
  title={A framework for a multi-source, data-driven building energy management toolkit},
  author={Markus, Andre A and Hobson, Brodie W and Gunay, H Burak and Bucking, Scott},
  journal={Energy and Buildings},
  volume={250},
  pages={111255},
  year={2021},
  publisher={Elsevier}
}

@inproceedings{DBLP:conf/cluster/SchwallerTTAB20,
  author       = {Benjamin Schwaller and
                  Nick Tucker and
                  Tom Tucker and
                  Benjamin A. Allan and
                  Jim M. Brandt},
  title        = {{HPC} System Data Pipeline to Enable Meaningful Insights through Analysis-Driven
                  Visualizations},
  booktitle    = {{IEEE} International Conference on Cluster Computing, {CLUSTER} 2020,
                  Kobe, Japan, September 14-17, 2020},
  pages        = {433--441},
  publisher    = {{IEEE}},
  year         = {2020},
  url          = {https://doi.org/10.1109/CLUSTER49012.2020.00062},
  doi          = {10.1109/CLUSTER49012.2020.00062},
  bibsource    = {dblp computer science bibliography, https://dblp.org}
}

@inproceedings{DBLP:conf/europar/AksarSALBEC21,
  author       = {Burak Aksar and
                  Benjamin Schwaller and
                  Omar Aaziz and
                  Vitus J. Leung and
                  Jim M. Brandt and
                  Manuel Egele and
                  Ayse K. Coskun},
  editor       = {Leonel Sousa and
                  Nuno Roma and
                  Pedro Tom{\'{a}}s},
  title        = {E2EWatch: An End-to-End Anomaly Diagnosis Framework for Production
                  {HPC} Systems},
  booktitle    = {Euro-Par 2021: Parallel Processing - 27th International Conference
                  on Parallel and Distributed Computing, Lisbon, Portugal, September
                  1-3, 2021, Proceedings},
  series       = {Lecture Notes in Computer Science},
  volume       = {12820},
  pages        = {70--85},
  publisher    = {Springer},
  year         = {2021},
  url          = {https://doi.org/10.1007/978-3-030-85665-6\_5},
  doi          = {10.1007/978-3-030-85665-6\_5},
  bibsource    = {dblp computer science bibliography, https://dblp.org}
}

@article{DBLP:journals/fgcs/NettiKBSBB20,
  author       = {Alessio Netti and
                  Zeynep Kiziltan and
                  {\"{O}}zalp Babaoglu and
                  Alina S{\^{\i}}rbu and
                  Andrea Bartolini and
                  Andrea Borghesi},
  title        = {A machine learning approach to online fault classification in {HPC}
                  systems},
  journal      = {Future Gener. Comput. Syst.},
  volume       = {110},
  pages        = {1009--1022},
  year         = {2020},
  url          = {https://doi.org/10.1016/j.future.2019.11.029},
  doi          = {10.1016/j.future.2019.11.029},
  bibsource    = {dblp computer science bibliography, https://dblp.org}
}

@inproceedings{DBLP:conf/supercomputer/BrownNGBPCMFG22,
  author       = {Nick Brown and
                  Rupert Nash and
                  Gordon Gibb and
                  Evgenij Belikov and
                  Artur Podobas and
                  Wei Der Chien and
                  Stefano Markidis and
                  Markus Flatken and
                  Andreas Gerndt},
  editor       = {Hartwig Anzt and
                  Amanda Bienz and
                  Piotr Luszczek and
                  Marc Baboulin},
  title        = {Workflows to Driving High-Performance Interactive Supercomputing for
                  Urgent Decision Making},
  booktitle    = {High Performance Computing. {ISC} High Performance 2022 International
                  Workshops - Hamburg, Germany, May 29 - June 2, 2022, Revised Selected
                  Papers},
  series       = {Lecture Notes in Computer Science},
  volume       = {13387},
  pages        = {233--244},
  publisher    = {Springer},
  year         = {2022},
  url          = {https://doi.org/10.1007/978-3-031-23220-6\_16},
  doi          = {10.1007/978-3-031-23220-6\_16},
  bibsource    = {dblp computer science bibliography, https://dblp.org}
}

@article{shvets2021endless,
  title={‘‘Endless’’Workload Analysis of Large-Scale Supercomputers},
  author={Shvets, PA and Voevodin, VV},
  journal={Lobachevskii Journal of Mathematics},
  volume={42},
  pages={184--194},
  year={2021},
  publisher={Springer}
}

@article{DBLP:journals/tc/DemirbagaWNMAGZ22,
  author       = {Umit Demirbaga and
                  Zhenyu Wen and
                  Ayman Noor and
                  Karan Mitra and
                  Khaled Alwasel and
                  Saurabh Garg and
                  Albert Y. Zomaya and
                  Rajiv Ranjan},
  title        = {AutoDiagn: An Automated Real-Time Diagnosis Framework for Big Data
                  Systems},
  journal      = {{IEEE} Trans. Computers},
  volume       = {71},
  number       = {5},
  pages        = {1035--1048},
  year         = {2022},
  url          = {https://doi.org/10.1109/TC.2021.3070639},
  doi          = {10.1109/TC.2021.3070639},
  bibsource    = {dblp computer science bibliography, https://dblp.org}
}

@phdthesis{demirbaga2022real,
  title={Real-time performance diagnosis and evaluation of big data systems in cloud datacenters},
  author={Demirbaga, Umit},
  year={2022},
  school={Newcastle University}
}

@Manual{dask2016,
  title = {Dask: Library for dynamic task scheduling},
  author = {{Dask Development Team}},
  year = {2016},
  url = {https://dask.org},
}

@article{DBLP:journals/corr/abs-2209-07164,
  author       = {Thomas Jakobsche and
                  Nicolas Lachiche and
                  Florina M. Ciorba},
  title        = {Challenges and Opportunities of Machine Learning for Monitoring and
                  Operational Data Analytics in Quantitative Codesign of Supercomputers},
  journal      = {CoRR},
  volume       = {abs/2209.07164},
  year         = {2022},
  url          = {https://doi.org/10.48550/arXiv.2209.07164},
  doi          = {10.48550/arXiv.2209.07164},
  eprinttype    = {arXiv},
  eprint       = {2209.07164},
  bibsource    = {dblp computer science bibliography, https://dblp.org}
}

@article{DBLP:journals/datasci/Dumontier22,
  author       = {Michel Dumontier},
  title        = {A formalization of one of the main claims of "The {FAIR} Guiding Principles
                  for scientific data management and stewardship" by Wilkinson et al.
                  20161},
  journal      = {Data Sci.},
  volume       = {5},
  number       = {1},
  pages        = {53--56},
  year         = {2022},
  url          = {https://doi.org/10.3233/ds-210047},
  doi          = {10.3233/ds-210047},
  bibsource    = {dblp computer science bibliography, https://dblp.org}
}

@article{DBLP:journals/fgcs/VersluisCGLPCUI23,
  author       = {Laurens Versluis and
                  Mehmet {\c{C}}etin and
                  Caspar Greeven and
                  Kristian Laursen and
                  Damian Podareanu and
                  Valeriu Codreanu and
                  Alexandru Uta and
                  Alexandru Iosup},
  title        = {Less is not more: We need rich datasets to explore},
  journal      = {Future Gener. Comput. Syst.},
  volume       = {142},
  pages        = {117--130},
  year         = {2023},
  url          = {https://doi.org/10.1016/j.future.2022.12.022},
  doi          = {10.1016/j.future.2022.12.022},
  bibsource    = {dblp computer science bibliography, https://dblp.org}
}

@article{muller2019development,
  title={Development and Evaluation of a Generic Framework for Sensor Data Aquisition, Aggregation and Propagation in HPC Systems},
  author={M{\"u}ller, Micha},
  year={2019}
}

@article{DBLP:journals/corr/abs-2107-11832,
  author       = {Laurens Versluis and
                  Mehmet {\c{C}}etin and
                  Caspar Greeven and
                  Kristian Laursen and
                  Damian Podareanu and
                  Valeriu Codreanu and
                  Alexandru Uta and
                  Alexandru Iosup},
  title        = {A Holistic Analysis of Datacenter Operations: Resource Usage, Energy,
                  and Workload Characterization - Extended Technical Report},
  journal      = {CoRR},
  volume       = {abs/2107.11832},
  year         = {2021},
  url          = {https://arxiv.org/abs/2107.11832},
  eprinttype    = {arXiv},
  eprint       = {2107.11832},
  bibsource    = {dblp computer science bibliography, https://dblp.org}
}

@inproceedings{DBLP:conf/ipps/Das0R21,
  author       = {Anwesha Das and
                  Frank Mueller and
                  Barry Rountree},
  title        = {Systemic Assessment of Node Failures in {HPC} Production Platforms},
  booktitle    = {35th {IEEE} International Parallel and Distributed Processing Symposium,
                  {IPDPS} 2021, Portland, OR, USA, May 17-21, 2021},
  pages        = {267--276},
  publisher    = {{IEEE}},
  year         = {2021},
  url          = {https://doi.org/10.1109/IPDPS49936.2021.00035},
  doi          = {10.1109/IPDPS49936.2021.00035},
  bibsource    = {dblp computer science bibliography, https://dblp.org}
}

@inproceedings{DBLP:conf/supercomputer/SivalingamR20,
  author       = {Karthee Sivalingam and
                  Harvey Richardson},
  editor       = {Heike Jagode and
                  Hartwig Anzt and
                  Guido Juckeland and
                  Hatem Ltaief},
  title        = {Application {IO} Analysis with Lustre Monitoring Using LASSi for {ARCHER}},
  booktitle    = {High Performance Computing - {ISC} High Performance 2020 International
                  Workshops, Frankfurt, Germany, June 21-25, 2020, Revised Selected
                  Papers},
  series       = {Lecture Notes in Computer Science},
  volume       = {12321},
  pages        = {255--266},
  publisher    = {Springer},
  year         = {2020},
  url          = {https://doi.org/10.1007/978-3-030-59851-8\_16},
  doi          = {10.1007/978-3-030-59851-8\_16},
  bibsource    = {dblp computer science bibliography, https://dblp.org}
}

@inproceedings{DBLP:conf/sc/JhaCBXESKI20,
  author       = {Saurabh Jha and
                  Shengkun Cui and
                  Subho S. Banerjee and
                  Tianyin Xu and
                  Jeremy Enos and
                  Mike Showerman and
                  Zbigniew T. Kalbarczyk and
                  Ravishankar K. Iyer},
  editor       = {Christine Cuicchi and
                  Irene Qualters and
                  William T. Kramer},
  title        = {Live forensics for {HPC} systems: a case study on distributed storage
                  systems},
  booktitle    = {Proceedings of the International Conference for High Performance Computing,
                  Networking, Storage and Analysis, {SC} 2020, Virtual Event / Atlanta,
                  Georgia, USA, November 9-19, 2020},
  pages        = {65},
  publisher    = {{IEEE/ACM}},
  year         = {2020},
  url          = {https://doi.org/10.1109/SC41405.2020.00069},
  doi          = {10.1109/SC41405.2020.00069},
  bibsource    = {dblp computer science bibliography, https://dblp.org}
}

@inproceedings{DBLP:conf/hpdc/RajeshDGBLYKS21,
  author       = {Neeraj Rajesh and
                  Hariharan Devarajan and
                  Jaime Cernuda Garcia and
                  Keith Bateman and
                  Luke Logan and
                  Jie Ye and
                  Anthony Kougkas and
                  Xian{-}He Sun},
  editor       = {Erwin Laure and
                  Stefano Markidis and
                  Ana Lucia Verbanescu and
                  Jay F. Lofstead},
  title        = {Apollo: : An ML-assisted Real-Time Storage Resource Observer},
  booktitle    = {{HPDC} '21: The 30th International Symposium on High-Performance Parallel
                  and Distributed Computing, Virtual Event, Sweden, June 21-25, 2021},
  pages        = {147--159},
  publisher    = {{ACM}},
  year         = {2021},
  url          = {https://doi.org/10.1145/3431379.3460640},
  doi          = {10.1145/3431379.3460640},
  bibsource    = {dblp computer science bibliography, https://dblp.org}
}

@inproceedings{DBLP:conf/ccgrid/ShilpikaLESVPM22,
  author       = {Shilpika and
                  Bethany Lusch and
                  Murali Emani and
                  Filippo Simini and
                  Venkatram Vishwanath and
                  Michael E. Papka and
                  Kwan{-}Liu Ma},
  title        = {Toward an In-Depth Analysis of Multifidelity High Performance Computing
                  Systems},
  booktitle    = {22nd {IEEE} International Symposium on Cluster, Cloud and Internet
                  Computing, CCGrid 2022, Taormina, Italy, May 16-19, 2022},
  pages        = {716--725},
  publisher    = {{IEEE}},
  year         = {2022},
  url          = {https://doi.org/10.1109/CCGrid54584.2022.00081},
  doi          = {10.1109/CCGrid54584.2022.00081},
  bibsource    = {dblp computer science bibliography, https://dblp.org}
}

@inproceedings{DBLP:conf/supercomputer/TraceyHSE20,
  author       = {Robert Tracey and
                  Lan Hoang and
                  Felix Subelet and
                  Vadim V. Elisseev},
  editor       = {Heike Jagode and
                  Hartwig Anzt and
                  Guido Juckeland and
                  Hatem Ltaief},
  title        = {AI-Driven Holistic Approach to Energy Efficient {HPC}},
  booktitle    = {High Performance Computing - {ISC} High Performance 2020 International
                  Workshops, Frankfurt, Germany, June 21-25, 2020, Revised Selected
                  Papers},
  series       = {Lecture Notes in Computer Science},
  volume       = {12321},
  pages        = {267--279},
  publisher    = {Springer},
  year         = {2020},
  url          = {https://doi.org/10.1007/978-3-030-59851-8\_17},
  doi          = {10.1007/978-3-030-59851-8\_17},
  bibsource    = {dblp computer science bibliography, https://dblp.org}
}

@inproceedings{DBLP:conf/supercomputer/OzerNT020,
  author       = {Gence Ozer and
                  Alessio Netti and
                  Daniele Tafani and
                  Martin Schulz},
  editor       = {Heike Jagode and
                  Hartwig Anzt and
                  Guido Juckeland and
                  Hatem Ltaief},
  title        = {Characterizing {HPC} Performance Variation with Monitoring and Unsupervised
                  Learning},
  booktitle    = {High Performance Computing - {ISC} High Performance 2020 International
                  Workshops, Frankfurt, Germany, June 21-25, 2020, Revised Selected
                  Papers},
  series       = {Lecture Notes in Computer Science},
  volume       = {12321},
  pages        = {280--292},
  publisher    = {Springer},
  year         = {2020},
  url          = {https://doi.org/10.1007/978-3-030-59851-8\_18},
  doi          = {10.1007/978-3-030-59851-8\_18},
  bibsource    = {dblp computer science bibliography, https://dblp.org}
}

@article{DBLP:journals/pc/NettiOGTS22,
  author       = {Alessio Netti and
                  Michael Ott and
                  Carla Guill{\'{e}}n and
                  Daniele Tafani and
                  Martin Schulz},
  title        = {Operational Data Analytics in practice: Experiences from design to
                  deployment in production {HPC} environments},
  journal      = {Parallel Comput.},
  volume       = {113},
  pages        = {102950},
  year         = {2022},
  url          = {https://doi.org/10.1016/j.parco.2022.102950},
  doi          = {10.1016/j.parco.2022.102950},
  bibsource    = {dblp computer science bibliography, https://dblp.org}
}

@inproceedings{DBLP:conf/heart/0001KSTW21,
  author       = {Martin Schulz and
                  Dieter Kranzlm{\"{u}}ller and
                  Laura Brandon Schulz and
                  Carsten Trinitis and
                  Josef Weidendorfer},
  editor       = {Christian Plessl and
                  Paul Chow and
                  Marco Platzner},
  title        = {On the Inevitability of Integrated {HPC} Systems and How they will
                  Change {HPC} System Operations},
  booktitle    = {{HEART} '21: 11th International Symposium on Highly Efficient Accelerators
                  and Reconfigurable Technologies, Virtual Event, Germany, 21-23 June,
                  2021},
  pages        = {2:1--2:6},
  publisher    = {{ACM}},
  year         = {2021},
  url          = {https://doi.org/10.1145/3468044.3468046},
  doi          = {10.1145/3468044.3468046},
  bibsource    = {dblp computer science bibliography, https://dblp.org}
}

@inproceedings{DBLP:conf/supercomputer/MolanBBGB21,
  author       = {Martin Molan and
                  Andrea Borghesi and
                  Francesco Beneventi and
                  Massimiliano Guarrasi and
                  Andrea Bartolini},
  editor       = {Heike Jagode and
                  Hartwig Anzt and
                  Hatem Ltaief and
                  Piotr Luszczek},
  title        = {An Explainable Model for Fault Detection in {HPC} Systems},
  booktitle    = {High Performance Computing - {ISC} High Performance Digital 2021 International
                  Workshops, Frankfurt am Main, Germany, June 24 - July 2, 2021, Revised
                  Selected Papers},
  series       = {Lecture Notes in Computer Science},
  volume       = {12761},
  pages        = {378--391},
  publisher    = {Springer},
  year         = {2021},
  url          = {https://doi.org/10.1007/978-3-030-90539-2\_25},
  doi          = {10.1007/978-3-030-90539-2\_25},
  bibsource    = {dblp computer science bibliography, https://dblp.org}
}

@inproceedings{DBLP:conf/supercomputer/EganPS22,
  author       = {Hilary Egan and
                  Avi Purkayastha and
                  David Sickinger},
  editor       = {Hartwig Anzt and
                  Amanda Bienz and
                  Piotr Luszczek and
                  Marc Baboulin},
  title        = {Data Center Facility Monitoring with Physics Aware Approach},
  booktitle    = {High Performance Computing. {ISC} High Performance 2022 International
                  Workshops - Hamburg, Germany, May 29 - June 2, 2022, Revised Selected
                  Papers},
  series       = {Lecture Notes in Computer Science},
  volume       = {13387},
  pages        = {251--261},
  publisher    = {Springer},
  year         = {2022},
  url          = {https://doi.org/10.1007/978-3-031-23220-6\_17},
  doi          = {10.1007/978-3-031-23220-6\_17},
  bibsource    = {dblp computer science bibliography, https://dblp.org}
}

@inproceedings{DBLP:conf/supercomputer/ArdebiliBAB22,
  author       = {Mohsen Seyedkazemi Ardebili and
                  Andrea Bartolini and
                  Andrea Acquaviva and
                  Luca Benini},
  editor       = {Hartwig Anzt and
                  Amanda Bienz and
                  Piotr Luszczek and
                  Marc Baboulin},
  title        = {Rule-Based Thermal Anomaly Detection for Tier-0 {HPC} Systems},
  booktitle    = {High Performance Computing. {ISC} High Performance 2022 International
                  Workshops - Hamburg, Germany, May 29 - June 2, 2022, Revised Selected
                  Papers},
  series       = {Lecture Notes in Computer Science},
  volume       = {13387},
  pages        = {262--276},
  publisher    = {Springer},
  year         = {2022},
  url          = {https://doi.org/10.1007/978-3-031-23220-6\_18},
  doi          = {10.1007/978-3-031-23220-6\_18},
  bibsource    = {dblp computer science bibliography, https://dblp.org}
}

@incollection{DBLP:books/sp/14/MavrovouniotisG14,
  author       = {Stathis Mavrovouniotis and
                  Mick Ganley},
  editor       = {Konstantinos Markantonakis and
                  Keith Mayes},
  title        = {Hardware Security Modules},
  booktitle    = {Secure Smart Embedded Devices, Platforms and Applications},
  pages        = {383--405},
  publisher    = {Springer},
  year         = {2014},
  url          = {https://doi.org/10.1007/978-1-4614-7915-4\_17},
  doi          = {10.1007/978-1-4614-7915-4\_17},
  bibsource    = {dblp computer science bibliography, https://dblp.org}
}

@inproceedings{DBLP:conf/itrust/BaldwinS03,
  author       = {Adrian Baldwin and
                  Simon Shiu},
  editor       = {Paddy Nixon and
                  Sotirios Terzis},
  title        = {Hardware Security Appliances for Trust},
  booktitle    = {Trust Management, First International Conference, iTrust 2003, Heraklion,
                  Crete, Greece, May 28-30, 2002, Proceedings},
  series       = {Lecture Notes in Computer Science},
  volume       = {2692},
  pages        = {46--58},
  publisher    = {Springer},
  year         = {2003},
  url          = {https://doi.org/10.1007/3-540-44875-6\_4},
  doi          = {10.1007/3-540-44875-6\_4},
  bibsource    = {dblp computer science bibliography, https://dblp.org}
}

@article{DBLP:journals/jcloudc/IsmaeelKM18,
  author       = {Salam Ismaeel and
                  Raed Karim and
                  Ali Miri},
  title        = {Proactive dynamic virtual-machine consolidation for energy conservation
                  in cloud data centres},
  journal      = {J. Cloud Comput.},
  volume       = {7},
  pages        = {10},
  year         = {2018},
  url          = {https://doi.org/10.1186/s13677-018-0111-x},
  doi          = {10.1186/s13677-018-0111-x},
  bibsource    = {dblp computer science bibliography, https://dblp.org}
}

@inproceedings{DBLP:conf/IEEEcloud/HauserW18,
  author       = {Christopher B. Hauser and
                  Stefan Wesner},
  title        = {Reviewing Cloud Monitoring: Towards Cloud Resource Profiling},
  booktitle    = {11th {IEEE} International Conference on Cloud Computing, {CLOUD} 2018,
                  San Francisco, CA, USA, July 2-7, 2018},
  pages        = {678--685},
  publisher    = {{IEEE} Computer Society},
  year         = {2018},
  url          = {https://doi.org/10.1109/CLOUD.2018.00093},
  doi          = {10.1109/CLOUD.2018.00093},
  bibsource    = {dblp computer science bibliography, https://dblp.org}
}

@article{DBLP:journals/jnca/SinghJP16,
  author       = {Saurabh Singh and
                  Young{-}Sik Jeong and
                  Jong Hyuk Park},
  title        = {A survey on cloud computing security: Issues, threats, and solutions},
  journal      = {J. Netw. Comput. Appl.},
  volume       = {75},
  pages        = {200--222},
  year         = {2016},
  url          = {https://doi.org/10.1016/j.jnca.2016.09.002},
  doi          = {10.1016/j.jnca.2016.09.002},
  bibsource    = {dblp computer science bibliography, https://dblp.org}
}

@inproceedings{DBLP:conf/3pgcic/PretoriusGI10,
  author       = {Maria Pretorius and
                  Mona Ghassemian and
                  C. Ierotheou},
  editor       = {Fatos Xhafa and
                  Leonard Barolli and
                  Hiroaki Nishino and
                  Markus Aleksy},
  title        = {An Investigation into Energy Efficiency of Data Centre Virtualisation},
  booktitle    = {3PGCIC 2010, International Conference on P2P, Parallel, Grid, Cloud
                  and Internet Computing, Fukuoka Institute of Technology, Fukuoka,
                  Japan, 4-6 November 2010},
  pages        = {157--163},
  publisher    = {{IEEE} Computer Society},
  year         = {2010},
  url          = {https://doi.org/10.1109/3PGCIC.2010.28},
  doi          = {10.1109/3PGCIC.2010.28},
  bibsource    = {dblp computer science bibliography, https://dblp.org}
}

@article{avgerinou2017trends,
  title={Trends in data centre energy consumption under the european code of conduct for data centre energy efficiency},
  author={Avgerinou, Maria and Bertoldi, Paolo and Castellazzi, Luca},
  journal={Energies},
  volume={10},
  number={10},
  pages={1470},
  year={2017},
  publisher={MDPI}
}

@article{DBLP:journals/ijcnds/OukfifOBB20,
  author       = {Karima Oukfif and
                  Fatima Oulebsir{-}Boumghar and
                  Samia Bouzefrane and
                  Soumya Banerjee},
  title        = {Workflow scheduling with data transfer optimisation and enhancement
                  of reliability in cloud data centres},
  journal      = {Int. J. Commun. Networks Distributed Syst.},
  volume       = {24},
  number       = {3},
  pages        = {262--283},
  year         = {2020},
  url          = {https://doi.org/10.1504/IJCNDS.2020.106322},
  doi          = {10.1504/IJCNDS.2020.106322},
  bibsource    = {dblp computer science bibliography, https://dblp.org}
}

@article{DBLP:journals/corr/abs-2202-05711,
  author       = {Erica Lin and
                  Luna Xu and
                  Suraj Bramhavar and
                  Marco Montes de Oca and
                  Sean Gorsky and
                  Lingyun Yi and
                  Arianna Groetsema and
                  Jeffrey Chou},
  title        = {Global Optimization of Data Pipelines in Heterogeneous Cloud Environments},
  journal      = {CoRR},
  volume       = {abs/2202.05711},
  year         = {2022},
  url          = {https://arxiv.org/abs/2202.05711},
  eprinttype    = {arXiv},
  eprint       = {2202.05711},
  bibsource    = {dblp computer science bibliography, https://dblp.org}
}

@article{oro2016experimental,
  title={Experimental and numerical analysis of the air management in a data centre in Spain},
  author={Or{\'o}, Eduard and Garcia, Albert and Salom, Jaume},
  journal={Energy and Buildings},
  volume={116},
  pages={553--561},
  year={2016},
  publisher={Elsevier}
}

@article{DBLP:journals/kbs/Chang17,
  author       = {Victor Chang},
  title        = {Towards data analysis for weather cloud computing},
  journal      = {Knowl. Based Syst.},
  volume       = {127},
  pages        = {29--45},
  year         = {2017},
  url          = {https://doi.org/10.1016/j.knosys.2017.03.003},
  doi          = {10.1016/j.knosys.2017.03.003},
  bibsource    = {dblp computer science bibliography, https://dblp.org}
}

@article{DBLP:journals/jcloudc/BaoG22,
  author       = {Guanming Bao and
                  Ping Guo},
  title        = {Federated learning in cloud-edge collaborative architecture: key technologies,
                  applications and challenges},
  journal      = {J. Cloud Comput.},
  volume       = {11},
  pages        = {94},
  year         = {2022},
  url          = {https://doi.org/10.1186/s13677-022-00377-4},
  doi          = {10.1186/s13677-022-00377-4},
  bibsource    = {dblp computer science bibliography, https://dblp.org}
}

@inproceedings{DBLP:conf/kdd/Banitalebi-Dehkordi21,
  author       = {Amin Banitalebi{-}Dehkordi and
                  Naveen Vedula and
                  Jian Pei and
                  Fei Xia and
                  Lanjun Wang and
                  Yong Zhang},
  editor       = {Feida Zhu and
                  Beng Chin Ooi and
                  Chunyan Miao},
  title        = {Auto-Split: {A} General Framework of Collaborative Edge-Cloud {AI}},
  booktitle    = {{KDD} '21: The 27th {ACM} {SIGKDD} Conference on Knowledge Discovery
                  and Data Mining, Virtual Event, Singapore, August 14-18, 2021},
  pages        = {2543--2553},
  publisher    = {{ACM}},
  year         = {2021},
  url          = {https://doi.org/10.1145/3447548.3467078},
  doi          = {10.1145/3447548.3467078},
  bibsource    = {dblp computer science bibliography, https://dblp.org}
}

@article{peng1993software,
  title={Software error analysis},
  author={Peng, Wendy W and Wallace, Dolores R},
  journal={NIST Special Publication},
  volume={500},
  pages={209},
  year={1993}
}

@article{wong2022prescriptive,
  title={A Prescriptive Analytical Logic Model for Error Analysis on Enterprise Level of Software Application},
  author={Wong, Hoo Meng and Amalathas, Sagaya Sabestinal and Ray, Sayan K and Murdani, Harry},
  journal={International Journal of Advanced Research in Technology and Innovation},
  volume={4},
  number={2},
  pages={43--52},
  year={2022}
}

@inproceedings{DBLP:conf/IEEEcloud/DoCWLZZ11,
  author       = {Anh Vu Do and
                  Junliang Chen and
                  Chen Wang and
                  Young Choon Lee and
                  Albert Y. Zomaya and
                  Bing Bing Zhou},
  editor       = {Ling Liu and
                  Manish Parashar},
  title        = {Profiling Applications for Virtual Machine Placement in Clouds},
  booktitle    = {{IEEE} International Conference on Cloud Computing, {CLOUD} 2011,
                  Washington, DC, USA, 4-9 July, 2011},
  pages        = {660--667},
  publisher    = {{IEEE} Computer Society},
  year         = {2011},
  url          = {https://doi.org/10.1109/CLOUD.2011.75},
  doi          = {10.1109/CLOUD.2011.75},
  bibsource    = {dblp computer science bibliography, https://dblp.org}
}

@inproceedings{DBLP:conf/IEEEcloud/ScheunerL18,
  author       = {Joel Scheuner and
                  Philipp Leitner},
  title        = {Estimating Cloud Application Performance Based on Micro-Benchmark
                  Profiling},
  booktitle    = {11th {IEEE} International Conference on Cloud Computing, {CLOUD} 2018,
                  San Francisco, CA, USA, July 2-7, 2018},
  pages        = {90--97},
  publisher    = {{IEEE} Computer Society},
  year         = {2018},
  url          = {https://doi.org/10.1109/CLOUD.2018.00019},
  doi          = {10.1109/CLOUD.2018.00019},
  bibsource    = {dblp computer science bibliography, https://dblp.org}
}

@article{DBLP:journals/ijcim/TzafilkouPK17,
  author       = {Katerina Tzafilkou and
                  Nicolaos Protogeros and
                  Adamantios Koumpis},
  title        = {User-centred cloud service adaptation: an adaptation framework for
                  cloud services to enhance user experience},
  journal      = {Int. J. Comput. Integr. Manuf.},
  volume       = {30},
  number       = {4-5},
  pages        = {472--482},
  year         = {2017},
  url          = {https://doi.org/10.1080/0951192X.2015.1030697},
  doi          = {10.1080/0951192X.2015.1030697},
  bibsource    = {dblp computer science bibliography, https://dblp.org}
}

@inproceedings{DBLP:conf/bigcom/QiuGSL017,
  author       = {Zhijin Qiu and
                  Zhongwen Guo and
                  Yanan Sun and
                  Yingjian Liu and
                  Yu Wang},
  title        = {{POUX:} Performance Optimization Strategy for Cloud Platforms Based
                  on User Experience},
  booktitle    = {3rd International Conference on Big Data Computing and Communications,
                  {BIGCOM} 2017, Chengdu, China, August 10-11, 2017},
  pages        = {200--207},
  publisher    = {{IEEE} Computer Society},
  year         = {2017},
  url          = {https://doi.org/10.1109/BIGCOM.2017.60},
  doi          = {10.1109/BIGCOM.2017.60},
  bibsource    = {dblp computer science bibliography, https://dblp.org}
}

@inproceedings{DBLP:conf/ecis/Otto11,
  author       = {Boris Otto},
  editor       = {Virpi Kristiina Tuunainen and
                  Matti Rossi and
                  Joe Nandhakumar},
  title        = {A morphology of the organisation of data governance},
  booktitle    = {19th European Conference on Information Systems, {ECIS} 2011, Helsinki,
                  Finland, June 9-11, 2011},
  pages        = {272},
  year         = {2011},
  url          = {http://aisel.aisnet.org/ecis2011/272},
  bibsource    = {dblp computer science bibliography, https://dblp.org}
}

@article{DBLP:journals/puc/Al-RuitheBH19,
  author       = {Majid Al{-}Ruithe and
                  Elhadj Benkhelifa and
                  Khawar Hameed},
  title        = {A systematic literature review of data governance and cloud data governance},
  journal      = {Pers. Ubiquitous Comput.},
  volume       = {23},
  number       = {5-6},
  pages        = {839--859},
  year         = {2019},
  url          = {https://doi.org/10.1007/s00779-017-1104-3},
  doi          = {10.1007/s00779-017-1104-3},
  bibsource    = {dblp computer science bibliography, https://dblp.org}
}

@inproceedings{barhate2018hybrid,
  title={Hybrid cloud: A solution to cloud interoperability},
  author={Barhate, Shweta M and Dhore, MP},
  booktitle={2018 Second International Conference on Inventive Communication and Computational Technologies (ICICCT)},
  pages={1242--1247},
  year={2018},
  organization={IEEE}
}
\end{document}